\documentclass[english,aps,prl,amsmath,amssymb,twocolumn,letterpaper,showpacs,superscriptaddress]{revtex4-1}

\usepackage{mathtools} 

\usepackage[normalem]{ulem} 

\usepackage{comment}
\usepackage{amsmath}
\usepackage{physics}
\usepackage{graphicx}
\usepackage{makecell,multirow}
\usepackage{babel}
\usepackage{float}
\usepackage{amstext}
\usepackage{cancel}
\usepackage{esint}
\usepackage[unicode=true,pdfusetitle,
bookmarks=true,bookmarksnumbered=false,bookmarksopen=false, breaklinks=false,pdfborder={0 0 1},backref=false,colorlinks=false]
 {hyperref}
 \hypersetup{
    colorlinks=true,
    linkcolor=red,
    citecolor=blue,
    filecolor=magenta,      
    urlcolor=blue
    }
\usepackage{pdfpages}	
\usepackage{pgffor}		
\usepackage{upgreek}
\usepackage{braket}
\usepackage[caption=false]{subfig}
\usepackage{verbatim}
\makeatletter
\AtBeginDocument{\let\LS@rot\@undefined} 
\makeatother

\begin{document}

\title{Quantum statistics in an extended collider coupled to a qubit}

\author{Rishav Chaudhuri}
\affiliation{Department of Physics and Astronomy, Purdue University, West Lafayette, Indiana 47907, USA}
\affiliation{Purdue Quantum Science and Engineering Institute, West Lafayette, Indiana 47907, USA}

\author{Sai Satyam Samal}
\affiliation{Department of Physics and Astronomy, Purdue University, West Lafayette, Indiana 47907, USA}

\date{\today}

\begin{abstract}

Mesoscopic colliders provide an effective platform for probing the mutual statistics of quantum particles. Recent experiments have successfully extracted the mutual statistics of fermions, and more exotic anyons using quantum point contacts (QPCs). Coupling a point-like collider, such as a quantum point contact, to a two-level impurity or qubit can induce statistical transmutation of fermions, causing them to display boson-like bunching tendencies. Here, we extend the analysis to an extended collider. We investigate the scattering of two incoming fermionic and bosonic wave packets in the presence of postselection on the impurity state, and systematically analyze the possible benchmarks used to characterize bunching and infer the underlying mutual statistics. We show that only a specific benchmark faithfully captures the mutual statistics of the colliding particles, while alternative choices can produce spurious statistical signatures. Hence, the correct benchmark for probing the quantum statistics depends on the intricate details of the mesoscopic collider.

\end{abstract}

\maketitle

\section{Introduction}

Mutual statistics constitute a fundamental property of quantum particles. It encodes information about how a wave function is modified under adiabatic exchange of two particles~\cite{Khare2005,leinaas_theory_1977}. Based on this, in three or more dimensions, particles can be identified as either bosons (symmetric wave function under exchange) or fermions (anti-symmetric wave function under exchange)~\cite{Fetter,Sakurai_Napolitano_2020}. However, in two dimensions, we have a plethora of particles called anyons (abelian and non-abelian)~\cite{leinaas_theory_1977,PhysRevLett.49.957,PhysRevLett.53.722,10.1093/acprof:oso/9780199227259.001.0001,Khare2005,RevModPhys.80.1083,rao2016introduction} whose statistics are intermediate between bosons and fermions. It turned out that fractional quantum Hall systems~\cite{PhysRevLett.52.1583,PhysRevLett.48.1559,10.1093/acprof:oso/9780199227259.001.0001,girvin1999quantum} host such exotic particles~\cite{PhysRevLett.49.957,Khare2005}. Understanding and probing the mutual statistics of such exotic particles have been one of the most active areas of research for their potential applications in topological quantum computation~\cite{PhysRevLett.94.166802,RevModPhys.80.1083}.

An extensive amount of work has been done in both theory and experiment to probe the mutual statistics in interference and collider-based setups~\cite{PhysRevB.55.2331,PhysRevLett.91.196803,PhysRevB.74.045319,PhysRevB.76.085333,Halperin_2011,PhysRevLett.109.106802,PhysRevB.88.235415,neder_interference_2007,PhysRevB.108.L241302,PhysRevB.107.104406,rp5m-r8fr,puster2026extractinganyonicexchangephase,PhysRevLett.116.156802,Lee2022NonAbelianAnyonCollider,PhysRevLett.131.186601,Lee2023PartitioningDilutedAnyons,PhysRevB.111.L201407,zhang2025effectivelinearresponsenonequilibrium,PhysRevLett.134.096303,td98-5ltj,nakamura_direct_2020,ronen_aharonovbohm_2021,PhysRevX.13.041012,Kundu_2023,kim_aharonovbohm_2024,werkmeister_strongly_2024,doi:10.1126/science.aaz5601,PhysRevX.13.011031}. In this article, we work with collider-based setups, where there are no interference loops, and statistics can be captured by studying the cross-current correlations~\cite{Blanter_2000,PhysRevLett.116.156802,doi:10.1126/science.aaz5601,td98-5ltj}. We assume that incoming particles can only interact with the other particles via statistical interaction, this is because colliders can be designed where Coulomb interactions between the particles can be neglected~\cite{zhang2023measuring}. By colliding particles, one probes the statistical interaction and the scattering result encodes the information on mutual statistics~\cite{PhysRevLett.109.106802,PhysRevLett.116.156802}. A simple manifestation of mutual statistics can be seen in a two-particle collision event. With two incoming particles, fermions tend to move to different detectors after the collision~\cite{Blanter_2000} in contrast to bosons, which tend to exit to the same detector.

Information on mutual statistics of colliding particles is characterized by either excessive bunching (boson-like) or anti-bunching (fermion-like) compared to the corresponding quantity for classical particles (waves) in point-like (extended) colliders~\cite{Blanter_2000,PhysRevLett.109.106802,PhysRevLett.116.156802,td98-5ltj}. However, the anti-bunching behavior of fermions in a point-like collider can be affected by coupling the point-like collider to an impurity, for example, a qubit and may exhibit statistical transmutation~\cite{PhysRevB.99.045430}. Such an impurity can distort the statistical information of colliding particles hence may lead to incorrect inference. Further, it was demonstrated in Ref.~\cite{td98-5ltj}, that microscopic geometric details of the scatterer (an extended collider) can also misguide to an incorrect inference about the statistics of the colliding particles for an inappropriately chosen benchmark. It is therefore natural to investigate the effect of a qubit impurity in an extended collider~\cite{PhysRevB.56.9692,PhysRevB.75.195332,PhysRevB.80.035319,td98-5ltj}, and in particular determine the appropriate benchmark for extracting the underlying quantum statistics.

By focusing on fermions and bosons, we investigate the effect of qubit-state postselection on scattering through an extended collider coupled to a qubit. To reliably extract the underlying mutual statistics, we systematically analyze the possible choices of benchmarks and identify the one that faithfully captures the intrinsic statistics of the colliding particles. We proceed in two steps. First, we consider a specific coupling between the collider and qubit, in which the qubit modifies only the phases of the reflection and transmission amplitudes. We then demonstrate that our proposed benchmark of quantum statistics distinguishes bosons from fermions for a general coupling between the extended collider and the qubit. Recent experimental studies have demonstrated the relevance of non-point-like geometry of the collider via measuring excess shot noise or resonances in the transmission curves~\cite{PhysRevB.80.035319,PhysRevX.13.011031,PhysRevLett.134.076302,td98-5ltj,garg2025enhancedshotnoisegraphene}. It is therefore important to understand the interplay between extended collider geometries and additional quantum degrees of freedom, such as coupling to a qubit. Our analysis highlights that the extraction of quantum statistics in such setups requires careful choice of benchmark to avoid spurious statistical signatures.

\begin{figure*}[htb!]
    \centering
    \includegraphics[width=1.0\linewidth]{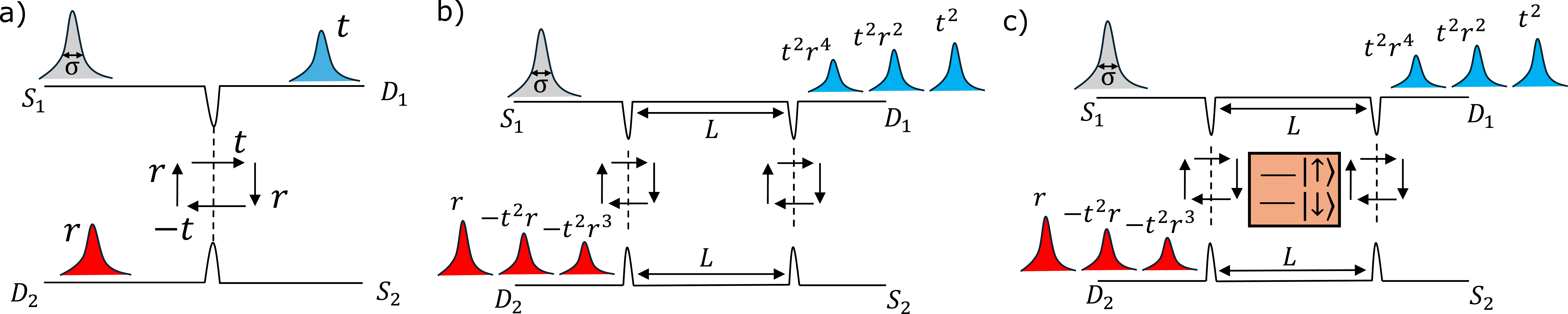}
    \caption{Mesoscopic colliders with two sources $S_{1}$, $S_{2}$ and two detectors $D_{1}$, $D_{2}$. A part of the incoming wave packet (grey-colored) is transmitted to the detector $D_{1}$, transmitted wave packet (blue-colored) and other part of the incoming wave packet is reflected to the detector $D_{2}$, reflected wave packet (red-colored). (a) Point-like collider with transmission and reflection amplitude given by Eq.~\eqref{eq:s}. The transmitted (reflected) wave packet involves a single wavelet on the lead with detector $D_{1}$ ($D_{2}$). (b) Extended collider with effective transmission and reflection amplitude given by Eq.~\eqref{eq:seff_main}. In this case the reflected and the transmitted wave packet involves an infinite number of wavelets, each separated by $2L$ due to the non-point-like geometry of the collider. (c) Extended collider coupled to a two-level impurity. The reflection and transmission amplitude in this case, depend on the state of the qubit ($|\uparrow\rangle$ and $|\downarrow\rangle$) as given in Eq.~\eqref{eq:seffqubit_main}.}
    \label{fig:colliders}
\end{figure*}

\section{Model of the collider}
We begin by reviewing the physics of the point-like collider with two sources, $S_{1}$ and $S_{2}$, and two detectors $D_{1}$ and $D_{2}$~\cite{Blanter_2000} as shown in the Fig.~\ref{fig:colliders}a. We begin by assuming the following form of the $S-$matrix  (also assuming time-reversal symmetry),
\begin{align}\label{eq:s}
S & =\left(\begin{array}{cc}
t & r\\
r & -t
\end{array}\right)\,,
\end{align}
where $t$ is the transmission amplitude and $r$ is the reflection amplitude are chosen to real. Consider a particular event where the particle goes from the source $S_{1}$ to the detector $D_{1}$, denoted by $(1\to1)$ and we call the associated probability the transmission probability. Using the $S-$matrix, Eq.~\eqref{eq:s}, the transmission probability for an incoming particle with momentum $k$ is given as $P(1\to1) = |t|^{2}$, which is independent of the momentum. Similarly, the probability of reflection (the event in which the particle goes from $S_{1}$ to $D_{2}$) is given as $P(1\to2) = |r|^{2} = 1-|t|^{2}$.

One can generalize the point-like collider to a case with two tunneling points separated by $L$, Fig.~\ref{fig:colliders}b and we call this an ``extended collider"~\cite{PhysRevB.56.9692,PhysRevB.75.195332,PhysRevB.80.035319,td98-5ltj}. Note that it resembles the Fabry-Perot interference setup~\cite{PhysRevB.55.2331,nakamura_direct_2020,PhysRevX.13.041012,x4w5-h3bb,girdhar2026fabryperotinterferometrystochasticanyonic} but here we are modeling it as an extended quantum point contact or extended collider that can have multiple tunneling points where we cannot control the area between the two QPCs~\cite{PhysRevB.80.035319}. Our extended collider model (see Fig.~\ref{fig:colliders}b and Fig.~\ref{fig:colliders}c) bears closer resemblance to the setup considered in Ref.~\cite{PhysRevB.80.035319} than to that investigated in Ref.~\cite{td98-5ltj}. Let us start by considering a single particle emitted from the source $S_{1}$ with momentum $k$, then the amplitude for the particle to reach the detector $D_{1}$ is given by a geometric series, $A(1\to 1) = \sum_{n=0}^{\infty}t^{2}r^{2n}e^{i(2n+1)kL} = \frac{t^{2}e^{ikL}}{1-r^{2}e^{i2kL}}$. For simplicity we assume that scattering matrix is identical for both the QPCs, Fig.~\ref{fig:colliders}b and Fig.~\ref{fig:colliders}c. Similarly, one can obtain the amplitude of reflection and write down the full scattering matrix, denoted by $S_{\text{eff}}$ (see Appendix~\ref{app:resonant_tunneling} for the details of the derivation), 
\begin{equation}
S_{\text{eff}}=\left[\begin{array}{cc}
\mathcal{T}_{k} & \mathcal{R}_{k} \\
\mathcal{R}_{k} & \mathcal{T}_{k}
\end{array}\right]=\left[\begin{array}{cc}
\frac{t^{2}e^{ikL}}{1-r^{2}e^{i2kL}} & \frac{r(1-e^{i2kL})}{1-r^{2}e^{i2kL}} \\
\frac{r(1-e^{i2kL})}{1-r^{2}e^{i2kL}} & \frac{t^{2}e^{ikL}}{1-r^{2}e^{i2kL}}
\end{array}\right]\, ,\label{eq:seff_main}
\end{equation}
Note that the effective matrix ($S_{\text{eff}}$) is also time-reversal invariant but explicitly depends on the momentum of the incoming particle in contrast to a point-like collider, Eq.~\eqref{eq:s}. Now, in this case, we have $P(1\to1) = |\mathcal{T}_{k}|^{2}$ and $P(1\to2) = |\mathcal{R}_{k}|^{2}$. As a result of explicit momentum dependence, the extended collider demonstrates resonant tunneling~\cite{gordanQM,Nazarov_Blanter_2009} as a function of the orbital phase $\varphi = kL$. These resonances can be found by setting $\mathcal{P}(1\to1) = 1$ which is satisfied when the orbital phase ($\varphi$) is an integer multiple of $\pi$ i.e., $\varphi = kL = n\pi$ where $n$ is an integer, see Appendix~\ref{app:resonant_tunneling} for more details.

Next, we would like to couple our extended collider to a two-level impurity, namely a qubit~\cite{PhysRevB.99.045430}. This models a two-level impurity in the system that can accidentally couple to the extended collider. Let us denote the two states of the qubit by $|\uparrow\rangle$ and $|\downarrow\rangle$. As a result of coupling to the external degree of freedom (qubit) the new scattering matrix of the system now depends on the state of the qubit. We model the coupling of the qubit by modifying the $S_{\text{eff}}$, Eq.~\eqref{eq:seff_main}. This is done by introducing extra phases which depend on the state of the qubit as follows~\cite{Nazarov_Blanter_2009},
\begin{align}\label{eq:seffqubit_main}
S_{\text{eff}}^{(m)}&=\left[\begin{array}{cc}\mathcal{T}_{k}e^{i\eta_{m}} & \mathcal{R}_{k}e^{i\theta_{m}}\\
\mathcal{R}_{k}e^{i(2\eta_{m}-\theta_{m})} & \mathcal{T}_{k}e^{i\eta_{m}}
\end{array}\right] \,,
\end{align}
where the $\mathcal{T}_{k}, \mathcal{R}_{k}$ are the transmission and reflection amplitudes as given in Eq.~\eqref{eq:seff_main} and $\theta_{m}, \eta_{m}$ are the coupling parameters with $m \in \{\uparrow, \downarrow\}$ and depends on the state of the qubit $\{|\uparrow\rangle, |\downarrow\rangle\}$. In this scheme the qubit effectively couples to the extended collider. Another scheme is to first couple the qubit to either of the two scatterers and then get the effective $S-$matrix. For clarity in the presentation, we consider the former case where the qubit effectively couples to the extended collider due to simplicity in the calculations. We shall comment on the case of more general coupling scheme later in the manuscript.

\section{Single-particle physics}
As mentioned previously, resonant tunneling is a characteristic feature of an extended collider. A particle with a definite momentum $k$ would be fully transmitted whenever the orbital phase $\varphi$ is an integer multiple of $\pi$ i.e., $\varphi=kL=n\pi$. However, in reality, an emitted particle from any of the sources is localized in space and has a finite width~\cite{PhysRevLett.132.156501,PhysRevLett.132.216601}. Therefore, it is more realistic to consider an incoming wave packet with finite width than a particle with a definite momentum. A wave packet is a linear superposition of different momentum modes and is characterized by a normalized momentum distribution, $\phi(k)$ such that $\int_{-\infty}^{\infty}dk|\phi(k)|^{2}=1$. In the analysis we present here, we are going to work with a Gaussian wave packet $\phi(k) = e^{-k^{2}\sigma^{2}/2}\sqrt{\sigma}/\pi^{1/4}$ with standard deviation $\sigma^{-1}/\sqrt{2}$. In real space, obtained after Fourier transform we obtain a Gaussian wave packet with standard deviation $\sigma/\sqrt{2}$. The transmission and reflection probabilities for a wave packet in the extended collider are given by, $\mathcal{P}(1\to1)= \int_{-\infty}^{\infty} dk |\mathcal{T}_{k}|^{2}|\phi(k)|^{2}$ (transmission) and $\mathcal{P}(1\to2)= \int_{-\infty}^{\infty} dk |\mathcal{R}_{k}|^{2}|\phi(k)|^{2}$ (reflection), where $\mathcal{T}_{k}$ and $\mathcal{R}_{k}$ are given in Eq.~\eqref{eq:seff_main}. Since a wave packet is a linear superposition of different momentum modes, in general we do not have resonances as in the case of a single particle with definite momentum $k$.

The above discussion gets significantly modified in the presence of coupling to an external degree of freedom (a qubit in our case). With a qubit, we have extra knobs that we can tune and hence it can give rise to some special and non-trivial scenarios. As a result of coupling to the qubit, the effective $S-$matrix is given by Eq.~\eqref{eq:seffqubit_main}, depends explicitly on the state of the qubit ($|\uparrow\rangle$ or $|\downarrow\rangle$). Following ref.~\cite{PhysRevB.99.045430}, we construct the density matrix of the system (after the scattering has taken place), denoted by $\rho_{\text{sys}}$, and is given by,
\begin{align}
\rho_{\text{sys}}=\sum_{m,m^{'}}\gamma_{m}\gamma_{m'}^{*}S_{\text{eff}}^{(m)}{|{i\rangle|m\rangle\langle m'|\langle i|S_{\text{eff}}^{(m')\dagger}}}\label{eq:denisty_system}\,,
\end{align}
where $m,\, m' \in \{\uparrow,\,\downarrow\}$, and $|i\rangle$ denotes the incoming single-particle state, see Appendix~\ref{app:scattering} for more details. We also assume that initially the qubit is in the normalized state $|\psi_{0}\rangle =\gamma_{\uparrow}|\uparrow\rangle+\gamma_{\downarrow}|\downarrow\rangle$ and hence $|\gamma_{\uparrow}|^{2} + |\gamma_{\downarrow}|^{2} = 1$. Note that here we are postselecting~\cite{PhysRev.134.B1410,PhysRevLett.95.200405} the qubit to be in the state $|\psi\rangle$, see Appendix~\ref{app:post_selection} for a pedagogical example on the postselection prescription. We construct the reduced density matrix of the qubit from $\rho_{\text{sys}}$ by tracing out the particle degrees of freedom. We denote it by $\rho_{\text{qubit}}$, and is given by,
\begin{align}\label{eq:rhoqubit}
\rho_{\text{qubit}}&=\left[\begin{array}{cc}|\gamma_{\uparrow}|^{2} & \gamma_{\uparrow}\gamma_{\downarrow}^{*}\varepsilon^{*}\\
\gamma_{\uparrow}^{*}\gamma_{\downarrow}\varepsilon & |\gamma_{\downarrow}|^{2}
\end{array}\right] \,,
\end{align}
where $\varepsilon = \langle i | \left(S_{\text{eff}}^{\uparrow}\right)^{\dagger}S_{\text{eff}}^{\downarrow} | i \rangle$. For the scatterer coupled with a qubit, there are also special scenarios in which the scattering processes are independent of the state of the qubit (\emph{disentanglement}). This disentanglement occurs whenever the reduced density matrix of the qubit, $\rho_{\text{qubit}}$ has $0$ and $1$ as its eigenvalues. This turns out to be the case whenever $\varepsilon = \langle i | \left(S_{\text{eff}}^{\uparrow}\right)^{\dagger}S_{\text{eff}}^{\downarrow} | i \rangle$ is a pure phase~\cite{PhysRevB.99.045430}.

Now, with this new effective $S-$matrix, Eq.~\eqref{eq:seffqubit_main} we look at the event in which a single wave packet is emitted from the source $S_{1}$ and goes to the detector $D_{1}$ where the qubit is now postselected to be in the state $|\psi\rangle$, see Appendix~\ref{app:post_selection} for pedagogical example of post selection prescription and also Ref.~\cite{PhysRevB.99.045430}. Since the scattering matrix, Eq.~\eqref{eq:seffqubit_main} depends on the state of the qubit, therefore the transmission probability will also be a function of the postselected qubit state $|\psi\rangle$. The single-particle probabilities with the qubit postselected or projected on to the state $\ket{\psi}$ are given as follows (see Appendix~\ref{app:single_particle_with_qubit} for the details),
\begin{widetext}
    \begin{align}
    P(1\to1;|\psi\rangle) &= \frac{\big\{1+2\cos(\eta_{\uparrow}-\eta_{\downarrow}+\phi_{0})|\bar{\gamma}_{\uparrow}\bar{\gamma}_{\downarrow}|\big\}\mathcal{P}(1\to1)}{1+2\cos(\eta_{\uparrow}-\eta_{\downarrow}+\phi_{0})|\bar{\gamma}_{\uparrow}\bar{\gamma}_{\downarrow}|\mathcal{P}(1\to1) + 2\cos(\eta_{\uparrow}-\eta_{\downarrow}+\phi_{1}+\phi_{0})|\bar{\gamma}_{\uparrow}\bar{\gamma}_{\downarrow}|\mathcal{P}(1\to2)} \,, \label{eq:1to1qubit_main}\\
    P(1\to2;|\psi\rangle) &=\frac{\big\{1+2\cos(\eta_{\uparrow}-\eta_{\downarrow}+\phi_{1}+\phi_{0})|\bar{\gamma}_{\uparrow}\bar{\gamma}_{\downarrow}|\big\}\mathcal{P}(1\to2)}{1+2\cos(\eta_{\uparrow}-\eta_{\downarrow}+\phi_{0})|\bar{\gamma}_{\uparrow}\bar{\gamma}_{\downarrow}|\mathcal{P}(1\to1) + 2\cos(\eta_{\uparrow}-\eta_{\downarrow}+\phi_{1}+\phi_{0})|\bar{\gamma}_{\uparrow}\bar{\gamma}_{\downarrow}|\mathcal{P}(1\to2)}\,,\label{eq:1to2qubit_main}
\end{align}
\end{widetext}
\noindent where $\phi_{1}=\eta_{\uparrow}-\eta_{\downarrow}-\theta_{\uparrow}+\theta_{\downarrow}$, $\tilde{\gamma}_{m}=\gamma_m\langle\psi|m\rangle = \bar{\gamma}_{m}\sqrt{\mathcal{C}}$, where $m\in\{\uparrow,\downarrow\}$, $\mathcal{C}= |\tilde{\gamma}_{\uparrow}|^{2}+|\tilde{\gamma}_{\downarrow}|^{2}\in \mathbb{R}^{+}$, and $\phi_{0} = \text{arg}(\tilde{\gamma}_{\uparrow}) - \text{arg}(\tilde{\gamma}_{\downarrow})$ are the parameters that depend completely on the impurity (qubit) degrees of freedom. We have defined the variable $\bar{\gamma}_{m} = \tilde{\gamma}_{m}/\sqrt{\mathcal{C}}$, such that we now have $|\bar{\gamma}_{\uparrow}|^{2} + |\bar{\gamma}_{\downarrow}|^{2} = 1 $. The denominator is a normalization factor which comes from postselecting the qubit to a particular state $|\psi\rangle$ and therefore, we have $P(1\to1;|\psi\rangle)+P(1\to2;|\psi\rangle)=1$, see Appendix~\ref{app:post_selection} for an example on postselection normalization. Further, note that the denominator may vanish for some choice of parameters, and therefore the domain of the parameters excludes all points where the probability, Eq.~\eqref{eq:1to1qubit_main} and Eq.~\eqref{eq:1to2qubit_main} diverges.

A closer look at the expression for the single-particle probabilities, Eq.~\eqref{eq:1to1qubit_main}-\eqref{eq:1to2qubit_main} shows that if we set  $\phi_{1}=\eta_{\uparrow}-\eta_{\downarrow}-\theta_{\uparrow}+\theta_{\downarrow}=0$ then, we obtain $P(1\to1;|\psi\rangle) = \mathcal{P}(1\to1)$ and similarly $P(1\to2;|\psi\rangle) = \mathcal{P}(1\to2)$ and hence the single-particle probabilities are independent of the state of the qubit $|\psi\rangle$. This implies we have disentanglement in our system i.e., the transmission/reflection probabilities are independent of the state of the qubit that we project onto (or postselect). The coupling between the qubit and the extended collider offers additional tuning knobs that can affect the single-particle physics in a non-trivial way. One such fine-tuned case is identified with vanishing of the reflection probability i.e., $P(1\to2;|\psi\rangle)=0$ whenever $1+2\cos(\eta_{\uparrow}-\eta_{\downarrow}+\phi_{1}+\phi_{0})|\bar{\gamma}_{\uparrow}\bar{\gamma}_{\downarrow}|=0$. Note that, in general $\mathcal{P}(1\to2)\neq 0$ for incoming wave packets in the absence of qubit. However, in the presence of the qubit, we can have $P(1\to2;|\psi\rangle)=0$, irrespective of the junction $S-$matrix, Eq.~\eqref{eq:s} and the shape of the incoming wave packet, $\phi(k)$. Here, all the momentum modes are blocked and cannot go from the source $S_{1}$ to $D_{2}$, which is in contrast to the case where the qubit is absent. However, $1+2\cos(\eta_{\uparrow}-\eta_{\downarrow}+\phi_{1}+\phi_{0})|\bar{\gamma}_{\uparrow}\bar{\gamma}_{\downarrow}|=0$ has real solutions when we have, $|\bar{\gamma}_{\uparrow}|\cdot|\bar{\gamma}_{\downarrow}|=\frac{1}{2}$. Hence, $\eta_{\uparrow}-\eta_{\downarrow}+\phi_{1}+\phi_{0}=(2n+1)\pi$ are the points of resonance i.e., $P(1\to2;|\psi\rangle)=0$ in an extended collider coupled to qubit. 

Note that in our setup we only deal with elastic processes, and no energy transfer takes place that can introduce qubit transitions. Further, the vanishing of the single-particle probability in this case is specific to our choice of coupling between the qubit and the extended collider, where  the qubit only modifies the overall phase of the scattering amplitudes. Later in the manuscript we generalize our analysis to the case where qubit coupling can modify the magnitude, as well as the  phase of the scattering amplitudes. Such an exact cancellation requires fine tuned parameters and are therefore not universal feature.

\section{Two-particle physics }

We now move on to understand the two-particle physics. Consider a situation where two particles (or equivalently two finite width wave packets) are emitted from sources $S_{1}$ (say from $x_{1}^{(0)}$ at time $t_{1}^{(0)}$) and $S_{2}$ (say from $x_{2}^{(0)}$ at time $t_{2}^{(0)}$). We will first focus on understanding the physics of an event where each of the detectors receives one particle. Let us say that the probability of the event where each detector receives one particle in the absence of any coupling to the qubit is denoted by $\mathcal{P}(11)_{\text{B/F}}$ and let $P(11;|\psi\rangle)_{\text{B/F}}$ be the corresponding probability in the presence of the qubit (projected onto, or postselected in the state $|\psi\rangle$), and the subscripts correspond to bosons (B) or fermions (F). The two-particle probability $P(11;|\psi\rangle)_{\text{B/F}}$ comprises of two sub-processes (1) direct process, where the particle from $S_{1(2)}$ goes to $D_{1(2)}$, and (2) exchange process, where the particle from $S_{1(2)}$ goes to $D_{2(1)}$. In the absence of the qubit, $\mathcal{P}(11)_{\text{B/F}} = \mathcal{P}(1\to 1)^{2} + \mathcal{P}(1\to 2)^{2} \pm 2\chi(x_{1}^{(0)},t_{1}^{(0)};x_{2}^{(0)},t_{2}^{(0)}) = \mathcal{P}(11)_{\text{CW}}\pm 2\chi(x_{1}^{(0)},t_{1}^{(0)};x_{2}^{(0)},t_{2}^{(0)})$. The term $\pm \chi(x_{1}^{(0)},t_{1}^{(0)};x_{2}^{(0)},t_{2}^{(0)})$ includes all the two-particle interference effects (interference between direct and exchange processes) and is sensitive to the mutual statistics~\cite{Blanter_2000,PhysRevLett.109.106802,td98-5ltj} of the particles, see Appendix~\ref{app:scattering} for a detailed derivation. The probability $\mathcal{P}(11)_{\text{CW}}$ is the two particle probability for classical waves where each detector receives one particle.
\vspace{0.85mm}

\begin{figure}[t!]
\centering
\includegraphics[width=1.0\columnwidth]{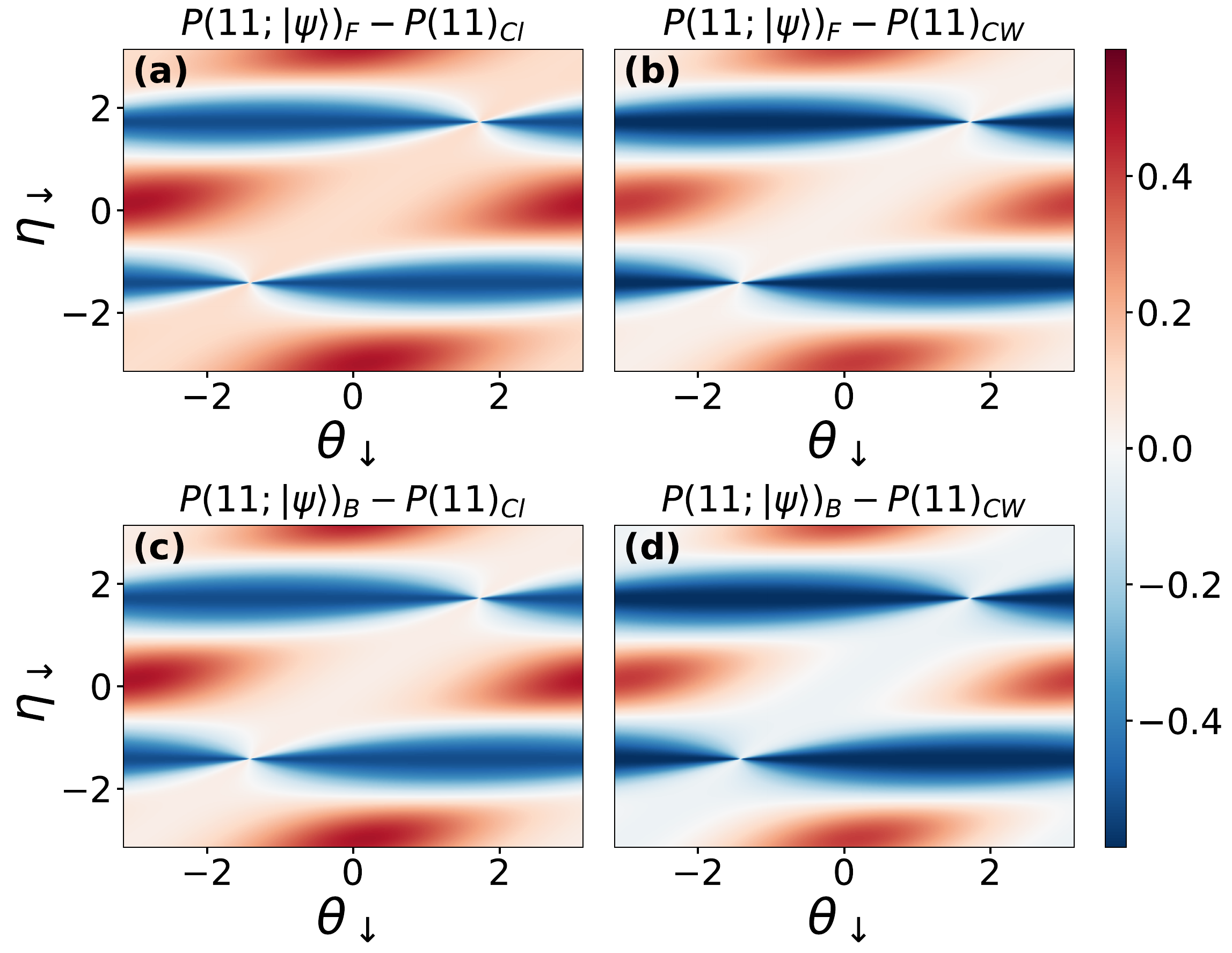}
\caption{Comparison of the probability of fermions [(a)-(b)] and bosons [(c)-(d)] with benchmark $\mathcal{B}_{1}$ and $\mathcal{B}_2$ in an extended collider coupled with a qubit where the scattering matrix as given in Eq.~\eqref{eq:seffqubit_main}. Panel (a) and panel (b) demonstrates $P(11;|\psi\rangle)_{\text{F}}-\mathcal{B}_{1}$ and $P(11;|\psi\rangle)_{\text{F}}-\mathcal{B}_{2}$ as a function of $\theta_{\downarrow}$ and $\eta_{\downarrow}$ for a wave packet obeying a normal distribution in momentum space with standard deviation $1/\sqrt{2}\sigma$.  Upon Fourier transformation, this gives a Gaussian wave packet in real space with standard deviation $\sigma/\sqrt{2}$, and hence the spatial width in real space is $\sigma$. Here, the initial state of the qubit is $\ket{\psi_{0}} = (\ket{\uparrow} + e^{-i0.3}\ket{\downarrow})/\sqrt{2}$, and hence $\gamma_{\uparrow} = 1/\sqrt{2},\,\gamma_{\downarrow} = e^{-i0.3}/\sqrt{2}$. Similarly, the final state of the qubit is $\ket{\psi} = (\ket{\uparrow} + \ket{\downarrow})/\sqrt{2}$, and therefore we find $\tilde{\gamma}_{\uparrow} = \gamma_{\uparrow}\cdot \langle \psi|\uparrow\rangle = 1/2$, and $\tilde{\gamma}_{\downarrow} = \gamma_{\downarrow}\cdot \langle \psi|\downarrow\rangle = e^{-i0.3}/2$. Hence, the normalization factor $\mathcal{C} = 1/2$. The following parameters have been kept fixed, $\eta_{\uparrow}=\theta_{\uparrow}=0,\, r=0.5,\, \sigma^{-1} \cdot L=1$. The above plot clearly demonstrates statistical transmutation in the parameter space ($\eta_{\downarrow}$,$\theta_{\downarrow}$) using the benchmarks $\mathcal{B}_{1}$ and $\mathcal{B}_{2}$ respectively.} \label{fig:transmutation_prob}
\end{figure}

As in the single-particle case, we can evaluate the two-particle probability in the presence of the qubit, and the expression for the probability is given as follows,
\begin{widetext}
\begin{align}
    P(11;|\psi\rangle)_{F} &=\mathcal{S}(11;|\psi\rangle)\mathcal{P}(11)_{\text{F}} \ \text{with} \ \mathcal{S}(11;|\psi\rangle)=\frac{1+2|\bar{\gamma}_{\uparrow}\bar{\gamma}_{\downarrow}|\cos[\phi_{(1,1)}]}{1+2|\bar{\gamma}_{\uparrow}\bar{\gamma}_{\downarrow}|\cos\phi_{(1,1)}\big\{ \mathcal{P}(11)_{\text{F}} + 2\mathcal{P}(20)_{\text{F}}\cos\phi_{1}\big\}}\,,\label{eq:p11qubit_main}
\end{align}
\end{widetext}
\noindent where $\phi_{(1,1)}= 2(\eta_{\uparrow}-\eta_{\downarrow})+\phi_{0}$, and the denominator is a result of postselection normalization, see Appendix~\ref{app:post_selection} for a pedagogical derivation of the postselection normalization scheme and  Appendix~\ref{app:two_particle_qubit} for detailed derivation of the normalization factor. As we have pointed out in the single particle case, the domain of the parameters in the expression for probability, Eq.~\eqref{eq:p11qubit_main} excludes all the points where the denominator goes to zero and making the probability undefined. The probability depends on the state of the qubit that we are projecting onto i.e.,  $|\psi\rangle$. As in the case of single particle, there also exist fine-tuned parameters that can lead to \emph{disentanglement} and the probability becomes independent of the state of the qubit, $|\psi\rangle$. If we set $\phi_{1}=\eta_{\uparrow}-\eta_{\downarrow}-\theta_{\uparrow}+\theta_{\downarrow}=0$ we obtain $P(11;|\psi\rangle)=\mathcal{P}(11)_{\text{F/B}}$, $P(20;|\psi\rangle)=\mathcal{P}(20)_{\text{F/B}}$ and $P(02;|\psi\rangle)=\mathcal{P}(02)_{\text{F/B}}$ and hence, the probabilities are independent of the state of the qubit $|\psi\rangle$. Similar to the single-particle case, the disentanglement condition here has been obtained by computing the reduced density matrix of the qubit and demanding its eigenvalues be 0 and 1, see Appendix~\ref{app:two_particle_qubit}. This corresponds to computing the quantity $\mathcal{E}= \langle i|S_{\text{eff}}^{\uparrow}S_{\text{eff}}^{\downarrow}|i\rangle$ and demanding it to be a pure phase, note that here, $|i\rangle$ denotes a two-particle incoming state of fermions or bosons.

In order to extract the mutual statistics of colliding particles, we would like to compare the anti-bunching probability of fermions [$P(11;|\psi\rangle)_{\text{F}}$] with a benchmark. Statistics is determined by noting whether we have an excess of bunching (boson-like behavior) or deficit of bunching (fermions-like behavior) with respect to a given benchmark. In our case of an extended collider coupled to a two-level impurity (qubit), there are four possible choices of benchmarks. The first benchmark $\mathcal{B}_{1}=\mathcal{P}(11)_{\text{Cl}}$ is the probability of the corresponding event with classical, distinguishable particles, which gives a false statistical signature i.e., the quantity $P(11;|\psi\rangle)_{\text{F}}-\mathcal{B}_{1}$ is not sign definite~\cite{PhysRevB.99.045430}, see Fig.~\ref{fig:transmutation_prob}a. The second benchmark, $\mathcal{B}_{2}$ is the corresponding probability for classical waves i.e., $\mathcal{B}_{2}=\mathcal{P}(11)_{\text{CW}}$~\cite{td98-5ltj}. It turns out that the benchmark, $\mathcal{B}_{2}$ is also an incorrect choice and the quantity $P(11;|\psi\rangle)_{\text{F}}-\mathcal{B}_{2}$ is also not sign definite, see Fig.~\ref{fig:transmutation_prob}b. Next, we propose two more nontrivial benchmarks $\mathcal{B}_{3}= \mathcal{S}(11;|\psi\rangle)_{\text{Cl}}\mathcal{P}(11)_{\text{Cl}}$ and $\mathcal{B}_{4}= \mathcal{S}(11;|\psi\rangle)_{\text{CW}}\mathcal{P}(11)_{\text{CW}}$ that can be defined in our setup of an extended collider coupled to a two-level impurity. The multiplicative factor $\mathcal{S}(11;|\psi\rangle)_{\text{Cl}}$ and $\mathcal{S}(11;|\psi\rangle)_{\text{CW}}$ is the same as shown in Eq.~\eqref{eq:p11qubit_main} with $\mathcal{P}(11)_{\text{F}}$ and $\mathcal{P}(20)_{\text{F}}$ replaced with the corresponding quantity for classical particles and classical waves i.e.,  $\mathcal{P}(11)_{\text{Cl}}$ and $\mathcal{P}(20)_{\text{Cl}}$, and $\mathcal{P}(11)_{\text{CW}}$ and $\mathcal{P}(20)_{\text{CW}}$.

\begin{figure}[b!]
    \centering
    \includegraphics[width=1.0\linewidth]{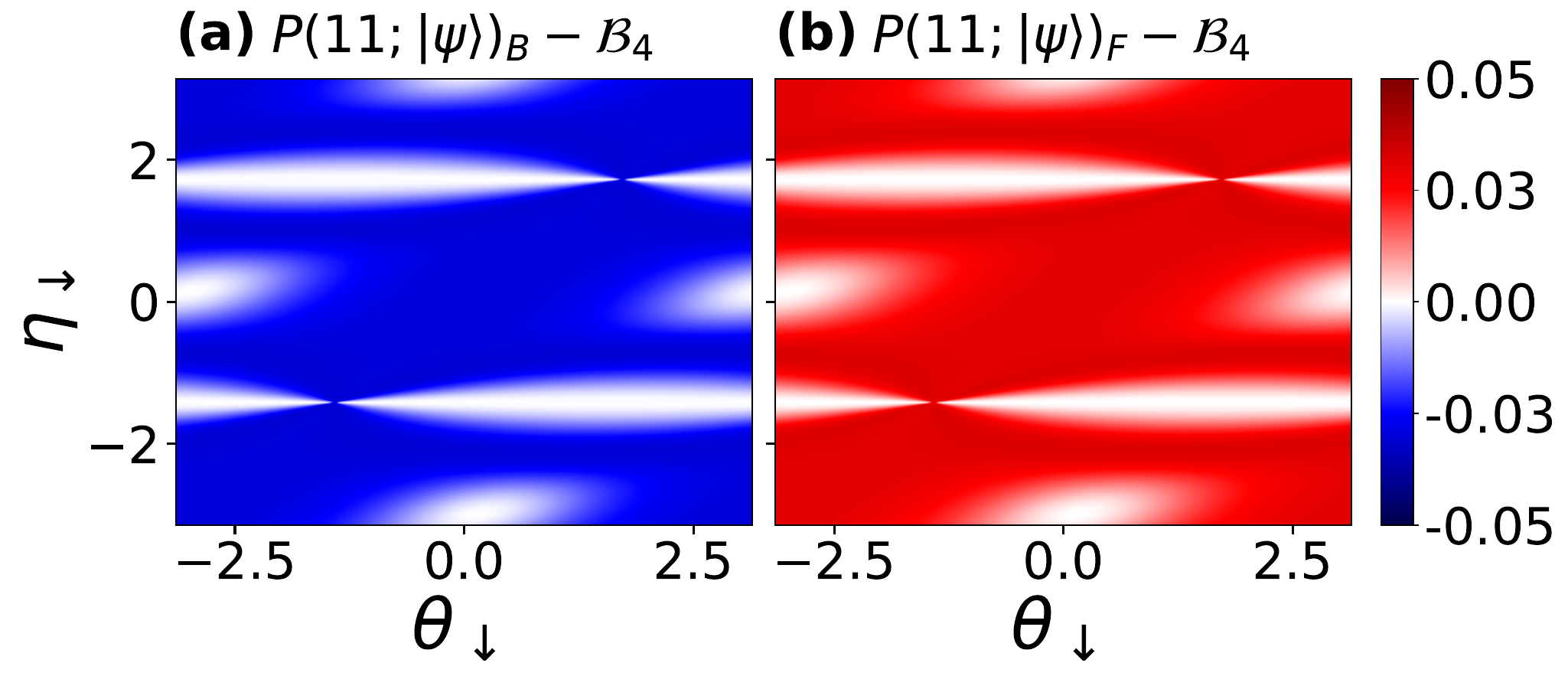}
    \caption{Benchmark $\mathcal{B}_{4}$ distinguishing bosons and fermions based on the mutual statistics as a function of parameters $\eta_{\downarrow}$ and $\theta_{\downarrow}$ for a wave packet obeying a normal distribution in momentum space with standard deviation $1/\sqrt{2}\sigma$. Upon Fourier transformation, this gives a Gaussian wave packet in real space with standard deviation $\sigma/\sqrt{2}$, and hence the spatial width in real space is $\sigma$. In this case, the qubit couples to the collider via phase coupling, cf. Eq.~\eqref{eq:seffqubit_main}. Here, the initial state of the qubit is $\ket{\psi_{0}} = (\ket{\uparrow} + e^{-i0.3}\ket{\downarrow})/\sqrt{2}$, and hence $\gamma_{\uparrow} = 1/\sqrt{2},\,\gamma_{\downarrow} = e^{-i0.3}/\sqrt{2}$. Similarly, the final state of the qubit is $\ket{\psi} = (\ket{\uparrow} + \ket{\downarrow})/\sqrt{2}$, and therefore we find $\tilde{\gamma}_{\uparrow} = \gamma_{\uparrow}\cdot \langle \psi|\uparrow\rangle = 1/2$, and $\tilde{\gamma}_{\downarrow} = \gamma_{\downarrow}\cdot \langle \psi|\downarrow\rangle = e^{-i0.3}/2$. Hence, the normalization factor $\mathcal{C} = 1/2$. The following parameters have been kept fixed, $\eta_{\uparrow}=\theta_{\uparrow}=0,\, r=0.5,\, \sigma^{-1} \cdot L=1$. (a) The quantity $P(11;|\psi\rangle)_{\text{B}}-\mathcal{B}_{4}$ is always negative and hence sign definite. (b) The corresponding quantity for fermions $P(11;|\psi\rangle)_{\text{F}}-\mathcal{B}_{4}$ is positive and hence also sign definite.}
    \label{fig:bosons_fermions_benchmark}
\end{figure}

At first, it might seem unreasonable to define the benchmarks, $\mathcal{B}_{3}$ and $\mathcal{B}_{4}$ that contain contributions from classical degrees and quantum degrees of freedom. However, the benchmarks $\mathcal{B}_{3}$ and $\mathcal{B}_{4}$ can be obtained from the probability of fermions $P(11;|\psi\rangle)_{F}$ (or equivalently of bosons) by varying the width of the wave packets and by introducing delay in the emission time of incoming wave packets. Let us start by considering narrow-width ($\sigma \ll L$) incoming wave packets of fermions (or equivalently bosons) from sources $S_{1}$ and $S_{2}$ at times $t_{1}^{(0)}$ and $t_{2}^{(0)}$, such that $v|t_{1}^{(0)}-t_{2}^{(0)}|\gg \sigma$, where $L$ is the size of the extended collider and $v$ is the velocity of the wave packets, cf. Fig.~\ref{fig:colliders}c. Note that the emission rate and the width of the wave packets are experimentally controllable parameters as demonstrated in Ref.~\cite{doi:10.1126/science.1141243,jullien_quantum_2014}. As a result of well-localized wave packets and a delay in the emission times, there will not be any interference between any of the sub-processes (or histories) and hence the probability for fermions $P(11;|\psi\rangle)_{F}$ would reduce to benchmark $\mathcal{B}_{3}= \mathcal{S}(11;|\psi\rangle)_{\text{Cl}}\mathcal{P}(11)_{\text{Cl}}$. From here, we find that the quantity $P(11;|\psi\rangle)_{\text{F}}-\mathcal{B}_{3} \propto \mathcal{P}(11)_{\text{F}} - \mathcal{P}(11)_{\text{Cl}}$ and is not sign definite in an extended collider as shown in Ref.~\cite{td98-5ltj}.

\begin{table*}[t!]
    \centering
    \renewcommand\arraystretch{1.5}
    \resizebox{\textwidth}{!}{%
    \begin{tabular}{|c|c|c|c|c|}
        \hline
         & Classical particle(s) & Classical wave(s) & Quantum particle(s) & Quantum wave(s) \\ \hline 
         \makecell{Different \\ benchmarks \\ of bunching } & $\mathcal{B}_{1} = \mathcal{P}(11)_{\text{Cl}}$ &  $\mathcal{B}_{2} = \mathcal{P}(11)_{\text{CW}}$ & $\mathcal{B}_{3}=\mathcal{S}(11;|\psi\rangle)_{\text{Cl}}\mathcal{P}(11)_{\text{Cl}}$  & $\mathcal{B}_{4}=\mathcal{S}(11;|\psi\rangle)_{\text{CW}}\mathcal{P}(11)_{\text{CW}}$ \\ \hline
         \makecell{Comment on \\ obtaining the \\ benchmarks of \\ bunching.}  & \makecell{Benchmark $\mathcal{B}_{1}$ is equal to\\ the two-particle probability \\ $P(11)_{\text{Cl}}$ for classical particles.} & \makecell{Benchmark $\mathcal{B}_{2}$ is equal\\ to the probability $P(11)_{\text{CW}}$ \\for classical waves.} & \makecell{Benchmark $\mathcal{B}_{3}$ can be obtained\\ from the probability $\mathcal{P}(11;|\psi\rangle)_{\text{F/B}}$ \\in the limit of narrow\\ wave packets i.e. $\sigma\ll L$} & \makecell{Benchmark $\mathcal{B}_{4}$ can be obtained\\ from the probability $\mathcal{P}(11;|\psi\rangle)_{\text{F/B}}$ \\with well separated incoming\\ particles i.e. $v|t_{1}^{(0)}-t_{2}^{(0)}|\gg \sigma$.}\\
      \hline
      \makecell{Comment on the \\ different benchmarks\\ of bunching} & \makecell{$P(11;|\psi\rangle)_{\text{F}}-\mathcal{B}_{1}\leq 0$ for \\certain set of parameters.\\(False statistical signature)} & \makecell{$P(11;|\psi\rangle)_{\text{F}}-\mathcal{B}_{2}\leq 0$ for \\certain set of parameters.\\(False statistical signature)} & \makecell{$P(11;|\psi\rangle)_{\text{F}}-\mathcal{B}_{3}\leq 0$ for \\certain set of parameters.\\(False statistical signature)} & \makecell{$P(11;|\psi\rangle)_{\text{F}}-\mathcal{B}_{4}\geq 0$ for \\any set of parameters.\\(True statistical signature)} \\
      \hline
    \end{tabular}
    }
    \caption{Summary of all the possible benchmarks in the system of an extended collider coupled to a qubit, where the coupling modifies the phase of the scattering matrix elements, Eq.~\eqref{eq:seffqubit_main}. The benchmarks $\mathcal{B}_{1}$ and $\mathcal{B}_{2}$ are purely classical and hence doesn't involve qubit. However, the benchmarks $\mathcal{B}_{3}$ and $\mathcal{B}_{4}$ are extracted using quantum particles and in the presence of the qubit. The benchmark $\mathcal{B}_{3}$ is obtained in the limit of very narrow wave packets i.e. $\sigma \ll L$ such that there are no interference terms at all (hence the subscript ``Cl" in $\mathcal{B}_{3}$). For obtaining the benchmark $\mathcal{B}_{4}$, we assume that the colliding particles are emitted with a very large delay i.e. $v|t_{1}^{(0)}-t_{2}^{(0)}|\gg \sigma$ such that the mutual interference term $\chi(x_{1}^{(0)},t_{1}^{(0)};x_{2}^{(0)},t_{2}^{(0)})$ can be neglected. Note that the benchmark $\mathcal{B}_{3}$ and $\mathcal{B}_{4}$, and similarly the postselected probability $P(11;\ket{\psi})$ is undefined for certain set of parameters, therefore the domain of the choice of parameters excludes all such points where the benchmark and the probability are not well defined.}  
    \label{tab:summary}
\end{table*}

Similarly, the benchmark $\mathcal{B}_{4}=\mathcal{S}(11;|\psi\rangle)_{\text{CW}}\mathcal{P}(11)_{\text{CW}}$ can be obtained by considering incoming wave packets that are emitted from sources $S_{1}$ and $S_{2}$ with a very large time delay, i.e., $v|t_{1}^{(0)}-t_{2}^{(0)}|\gg \sigma$. In other words, $P(11;|\psi\rangle)_{\text{B/F}} = \mathcal{S}(11;|\psi\rangle)\mathcal{P}(11)_{\text{B/F}} \to \mathcal{S}(11;|\psi\rangle)_{\text{CW}}\mathcal{P}(11)_{\text{CW}} = \mathcal{B}_{4}$ whenever we have $v|t_{1}^{(0)}-t_{2}^{(0)}|\gg \sigma$~\cite{td98-5ltj} and also see Appendix~\ref{app:scattering}. This is because $\mathcal{P}(11)_{\text{B/F}} = \mathcal{P}(11)_{\text{CW}}\pm 2\chi(x_{1}^{(0)},t_{1}^{(0)};x_{2}^{(0)},t_{2}^{(0)})$. In the limit $v|t_{1}^{(0)}-t_{2}^{(0)}|\gg \sigma$ we have $\chi(x_{1}^{(0)},t_{1}^{(0)};x_{2}^{(0)},t_{2}^{(0)})\to 0$, (see Appendix~\ref{app:scattering}) and we obtain the desired benchmark, $\mathcal{B}_{4}$. Intuitively, whenever two particles (wave packets) are emitted at very different times, then one particle cannot see, i.e., statistically interact with, the other particle (or wave packet) and hence, we obtain a classical-wave result, which includes all the self-interference contributions. From here, we evaluate the quantity $P(11;|\psi\rangle)_{\text{F}}-\mathcal{B}_{4}$ and find,
\begin{align}\label{eq:correct_benchmark}
    P(11;|\psi\rangle)_{\text{F}}-\mathcal{B}_{4} & =\mathcal{K}_{\text{F}}\cdot \big[\mathcal{P}(11)_{\text{F}}-\mathcal{P}(11)_{\text{CW}}\big]\,,
\end{align}
and $\mathcal{K}_{\text{F}}$ is given as follows,
\begin{widetext}
    \begin{align}\label{eq:full_mathcalk}
\mathcal{K}_{\text{F}} &=\frac{\Big\{1+2|\bar{\gamma}_{\uparrow}\bar{\gamma}_{\downarrow}|\cos\phi_{(1,1)}\cos\phi_{1}\Big\}\Big\{1+2|\bar{\gamma}_{\uparrow}\bar{\gamma}_{\downarrow}|\cos\phi_{(1,1)}\Big\}}{\Big[1+2|\bar{\gamma}_{\uparrow}\bar{\gamma}_{\downarrow}|\cos\phi_{(1,1)}\big\{\mathcal{P}(11)_{\text{F}}+2\mathcal{P}(20)_{\text{F}}\cos\phi_{1}\big\}\Big]\Big[1+2|\bar{\gamma}_{\uparrow}\bar{\gamma}_{\downarrow}|\cos\phi_{(1,1)}\big\{\mathcal{P}(11)_{\text{CW}}+2\mathcal{P}(20)_{\text{CW}}\cos\phi_{1}\big\}\Big]}\,.
\end{align}
\end{widetext}
\noindent Hence, we find that the quantity $P(11;|\psi\rangle)_{\text{F}}-\mathcal{B}_{4}$ has different signs for bosons and fermions. Therefore, the statistical information is correctly captured. Further, the quantity $\mathcal{K}_{\text{F}}$ is positive (see Appendix~\ref{app:benchmark_qubit}, for details of the derivation), and $\mathcal{P}(11)_{\text{F}}-\mathcal{P}(11)_{\text{CW}}$ is sign definite for fermions, and is of opposite sign for bosons as proven in Ref.~\cite{td98-5ltj}. Therefore, the benchmark $\mathcal{B}_{4} = \mathcal{S}(11;|\psi\rangle)_{\text{CW}}\mathcal{P}(11)_{\text{CW}}$ determines the true mutual statistics of particles in an extended collider that is coupled to a two-level impurity. We have summarized our discussion in the Table~\ref{tab:summary}.

\begin{figure}[b!]
    \centering
    \includegraphics[width=1.0\linewidth]{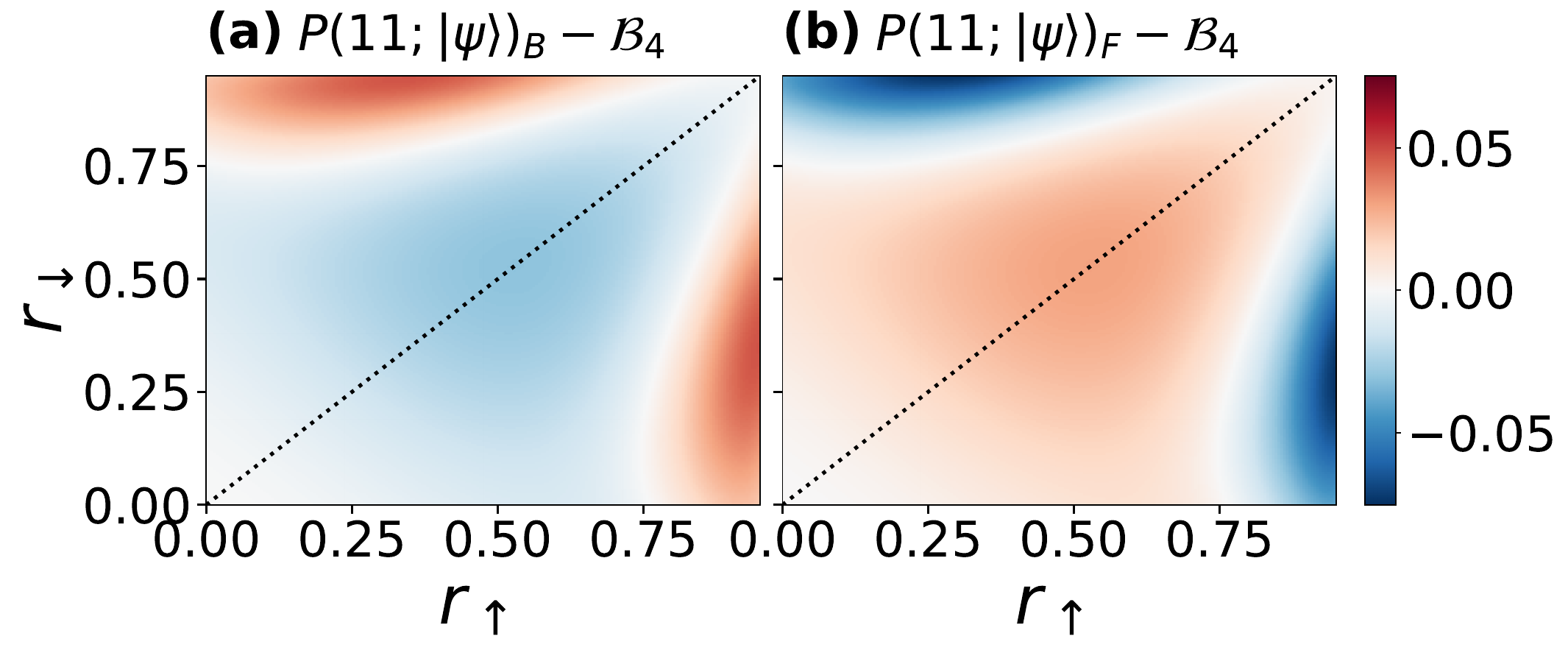}
    \caption{Extracting the mutual statistics of bosons and fermions using the benchmark $\mathcal{B}_{4}$ for a general coupling between the qubit impurity and the scattering matrix, cf. Eq.~\eqref{eq:seffqubit_main_general} with the choice $\mathcal{T}_{k}^{(m)} = (1-r_{m}^2)e^{ikL}/(1-r_{m}^2e^{2ikL})  = \mathcal{T'}_{k}^{(m)}$, and $\mathcal{R}_{k}^{(m)} = r_{m}(1-e^{2ikL})/(1-r_{m}^2e^{2ikL})  = \mathcal{R'}_{k}^{(m)}$. We are using a wave packet obeying a normal distribution in momentum space with standard deviation $1/\sqrt{2}\sigma$.  Upon Fourier transformation, this gives a Gaussian wave packet in real space with standard deviation $\sigma/\sqrt{2}$, and hence the spatial width in real space is $\sigma$ and we choose $\sigma^{-1} \cdot L=1$. The qubit is initially in the state $|\psi_{0}\rangle = (\ket{\uparrow} + e^{-i0.3}\ket{\downarrow})/\sqrt{2}$, and its postselected quantum state is $\ket{\psi} = (\ket{\uparrow} + \ket{\downarrow})/\sqrt{2}$. The quantity $P(11;|\psi\rangle)_{\text{F/B}} - \mathcal{B}_4$ is not sign definite in general, however the corresponding quantity for bosons and fermions (subscript ``B'' and ``F'') differ by a relative sign, which originates from the distinct mutual statistics of bosons and fermion. Hence, the benchmark $\mathcal{B}_{4}$ successfully distinguishes bosons and fermions based on their mutual statistics.  The diagonal line corresponds to the case where there is no coupling between the qubit and the extended collider and hence the quantity $P(11;|\psi\rangle)_{\text{F/B}} - \mathcal{B}_4$ is sign definite.}
    \label{fig:P11_minus_B4}
\end{figure}

It is also important to realize that in all of the above analysis, we have relied on the specific coupling of the impurity qubit with the extended collider, cf. Eq.~\eqref{eq:seffqubit_main}. It is very natural to raise the concern of whether the formalism holds for a more general and non-trivial coupling between the extended collider and impurity qubit. A more general coupling of the qubit with the extended collider would modify the scattering matrix as follows,
\begin{align}\label{eq:seffqubit_main_general}
S_{\text{eff gen}}^{(m)}&=\left[\begin{array}{cc}\mathcal{T}_{k}^{(m)} & \mathcal{R'}_{k}^{(m)}\\
\mathcal{R}_{k}^{(m)} & \mathcal{T'}_{k}^{(m)}
\end{array}\right] \,,
\end{align}
where the scattering matrix elements now depend on the qubit state $m\in \{ \uparrow, \downarrow \}$. Our entire formalism discussed so far can be applied to obtain the two particle probability for any event in this case as well. 

Since, we are interested in the obtaining the mutual statistics of bosons and fermions, we shall focus on the event where we have two incoming particles, one from source $S_{1}$ and another from $S_{2}$, and detectors $D_{1}$ and $D_{2}$ receives one particle each. The corresponding probability is given as follows,
\begin{align}
P(&11;|\psi\rangle)_{\text{F/B}} = \frac{1}{\mathcal{N}_{\text{F/B}}} \mathcal{P}(11;|\psi\rangle)_{\text{F/B}}\,,\\
&=\frac{1}{\mathcal{N}_{\text{F/B}}}\iint^{\infty}_{-\infty}v^{2}dt_{1}dt_{2}|\tilde{\gamma}_{\uparrow}A^{\text{F/B}}_{\uparrow}(11)+\tilde{\gamma}_{\downarrow}A^{\text{F/B}}_{\downarrow}(11)|^{2}\,,\notag
\end{align}
where the qubit is initially in the state $\ket{\psi_{0}} = \sum_{m}\gamma_{m}\ket{m}$ and its postselected quantum state is $\ket{\psi}$, such that we have $\tilde{\gamma}_{m} = \gamma_{m} \langle \psi |  m\rangle $, and the factor $\mathcal{N}_\text{F/B}$ is the normalization factor resulting from postselection for fermions and bosons respectively. In obtaining the probability, we integrate over the entire range of the detection times $t_{1}$ and $t_{2}$. The amplitude for the two particle event is $A^{\text{F/B}}_{\uparrow/\downarrow}(11) =A^{\text{D}}_{\uparrow/\downarrow}(11) \mp  A^{\text{E}}_{\uparrow/\downarrow}(11)$, where the amplitude $A^{\text{D}}_{\uparrow/\downarrow}(11)$ corresponds to the direct event, i.e., particle from $S_{1}$ (and $S_{2}$) goes to detector $D_{1}$ (and $D_{2}$), and similarly amplitude $A^{\text{E}}_{\uparrow/\downarrow}(11)$ corresponds to the exchange event, i.e., particle from $S_{1}$ (and $S_{2}$) goes to detector $D_{2}$ (and $D_{1}$), see Appendix~\ref{app:gen_scat_matrix} for full expression for the amplitudes.

The statistical difference between bosons and fermions leads to a relative sign in the exchange amplitudes, i.e. $A^{\text{F/B}}_{\uparrow/\downarrow}(11) =A^{\text{D}}_{\uparrow/\downarrow}(11) \mp A^{\text{E}}_{\uparrow/\downarrow}(11)$. Next, we exploit this feature and write down the probability schematically as follows,
\begin{align}
    \mathcal{P}(11;|\psi\rangle)_{\text{F/B}} = \mathcal{W}_{11} \mp \mathcal{\chi}_{11}\,,
\end{align}
where the positive and negative sign showing here is a result of mutual statistics of bosons and fermions. Similarly, the expression for probabilities for the event in which either detector $D_{1}$ or $D_{2}$ receives both the particles can schematically be written as $\mathcal{P}(20;|\psi\rangle)_{\text{F/B}} = \mathcal{W}_{20} \mp \mathcal{\chi}_{20}$, and $\mathcal{P}(02;|\psi\rangle)_{\text{F/B}} = \mathcal{W}_{02} \mp \mathcal{\chi}_{02}$. From here we identify the normalization factor $\mathcal{N}_{\text{F/B}} = \mathcal{P}(11;|\psi\rangle)_{\text{F/B}} + \mathcal{P}(20;|\psi\rangle)_{\text{F/B}} + \mathcal{P}(02;|\psi\rangle)_{\text{F/B}}$.

Next, note that the terms $\chi_{11}$, $\chi_{02}$ and $\chi_{20}$ are interference terms composed of products of direct and exchange amplitudes, i.e. combination of terms of the form $A_{m}^{\text{D}}A_{m'}^{\text{E}}$ where $m,m'\in \{ \uparrow,\downarrow \}$. In this case, for obtaining the benchmark $\mathcal{B}_{4}$, we introduce a very large delay in the emission times, i.e., $v|t_{1}^{(0)} - t_{2}^{(0)}|\gg \sigma$ for the particles emitted from the sources $S_{1}$ and $S_{2}$ at times $t_{1}^{(0)}$ and $t_{2}^{(0)}$~\cite{doi:10.1126/science.1141243,jullien_quantum_2014}. As a result of  very large delay in the emission times, the interference terms $\chi_{11},\chi_{20},\chi_{02}\to 0$ do not contribute to the probability $P(11;\ket{\psi})_{\text{B/F}}$ and we obtain the benchmark $\mathcal{B}_{4} = \mathcal{W}_{11}/(\mathcal{W}_{11}+\mathcal{W}_{20}+\mathcal{W}_{02})$. By using the benchmark $\mathcal{B}_{4}$, and comparing the probability of bosons and fermions we find the following,
\begin{align}
    [P(11;|\psi\rangle)_{\text{B}} - \mathcal{B}_4]\times[P(11;|\psi\rangle)_{\text{F}} - \mathcal{B}_4] \leq 0 \,,
\end{align}
and hence, the benchmark always identifies the statistical difference between bosons and fermions, see Appendix~\ref{app:gen_scat_matrix} for details. 

Hence, we have established that there is a relative sign between the two quantities $P(11;|\psi\rangle)_{\text{B}} - \mathcal{B}_4$ and $P(11;|\psi\rangle)_{\text{F}} - \mathcal{B}_4$, more precisely, they are not sign definite by themselves. However, as a result of their statistics, whenever the corresponding quantity is positive for fermions, it is negative for bosons and vice-versa. This is explicitly demonstrated by considering a $S_{\text{eff gen}}^{(m)}$, where $\mathcal{T}_{k}^{(m)} = (1-r_{m}^2)e^{ikL}/(1-r_{m}^2e^{2ikL})  = \mathcal{T'}_{k}^{(m)}$, and $\mathcal{R}_{k}^{(m)} = r_{m}(1-e^{2ikL})/(1-r_{m}^2e^{2ikL})  = \mathcal{R'}_{k}^{(m)}$, and hence in this case the magnitude of the reflection amplitude at each of the QPCs depends on the state of the qubit, i.e., $r_{\uparrow} \neq r_{\downarrow}$. With this choice of effective scattering matrix, we find that  $P(11;|\psi\rangle)_{\text{F}} - \mathcal{B}_4$ is not in general sign definite but the corresponding quantity for bosons have a relative sign and this difference is the statistical signature, Fig.~\ref{fig:P11_minus_B4}(a)-(b).

\section{Conclusion}

In summary, we present a study of the collision of bosons and fermions in an extended collider coupled to a two-level impurity and comment on the correct benchmark for extracting the mutual statistics. We begin by exploring the single-particle physics in an extended collider coupled to a qubit, cf. Eq.~\eqref{eq:seffqubit_main}. We demonstrate that the probability of different scattering events depends on the postselected qubit state $\ket{\psi}$, where the initial state of the qubit is $\ket{\psi_{0}}$. By postselecting specific states of the qubit, we obtain either full transmission or reflection of the incoming wave packets, cf. Eq.~\eqref{eq:1to1qubit_main}-\eqref{eq:1to2qubit_main}. Following this, we study the two-particle physics and explicitly show the dependence of the probabilities of different scattering events on the postselected qubit state, Eq.~\eqref{eq:p11qubit_main}.

Further, we present a thorough analysis of extracting the mutual statistics of the colliding particles in the setup.  We obtain apparent statistical transmutation with respect to benchmarks $\mathcal{B}_{1} = \mathcal{P}(11)_{\text{Cl}}$ and $\mathcal{B}_{2} = \mathcal{P}(11)_{\text{CW}}$, see Fig.~\ref{fig:transmutation_prob}, similar to previous study on point-like colliders~\cite{PhysRevB.99.045430,td98-5ltj}. In order to extract the mutual statistics of the colliding particles, we propose two new benchmarks $\mathcal{B}_{3}= \mathcal{S}(11;|\psi\rangle)_{\text{Cl}}\mathcal{P}(11)_{\text{Cl}}$ and $\mathcal{B}_{4}= \mathcal{S}(11;|\psi\rangle)_{\text{CW}}\mathcal{P}(11)_{\text{CW}}$, where $\mathcal{S}(11;|\psi\rangle)_{\text{Cl/CW}}$ is given in Eq.~\eqref{eq:p11qubit_main} with two-particle probabilities of fermions replaced with the corresponding probabilities of classical particles and waves. The new benchmarks $\mathcal{B}_{3}$ and $\mathcal{B}_{4}$ can be obtained from the two-particle probability $P(11;|\psi\rangle)_{\text{B/F}}$, see Tab.~\ref{tab:summary}. It turns out that, between $\mathcal{B}_{3}$ and $\mathcal{B}_{4}$, the correct benchmark for extracting the mutual statistics is $\mathcal{B}_{4}$. It yields a sign-definite result for bosons and fermions, Eq.~\eqref{eq:correct_benchmark}, thereby determining the statistics unambiguously, see Fig.~\ref{fig:bosons_fermions_benchmark}(a)-(b). Note all this analysis is for the case where the extended collider and qubit are coupled in a specific way via phase coupling, see Eq.~\eqref{eq:seffqubit_main}.

Further we generalize the choice of coupling between the qubit and the extended collider by considering a new effective scattering matrix, Eq.~\eqref{eq:seffqubit_main_general} that depends on the state of the qubit $\ket{\uparrow}$ and $\ket{\downarrow}$. We demonstrate that in this general case, the proposed benchmark of statistics $\mathcal{B}_{4}$ distinguishes bosons from fermions, see Fig.~\ref{fig:P11_minus_B4}(a)-(b). However, in the general case the quantity $P(11;|\psi\rangle)_{\text{B/F}} - \mathcal{B}_{4}$ is not sign definite, but the effect of statistical difference between bosons and fermions introduces a relative sign between them. Hence, whenever the quantity $P(11;|\psi\rangle)_{\text{B}} - \mathcal{B}_{4}$ is positive, the corresponding quantity for fermions $P(11;|\psi\rangle)_{\text{F}} - \mathcal{B}_{4}$ is negative and vice-versa, cf. Fig.~\ref{fig:P11_minus_B4}(a)-(b).

Our analysis  establishes that the true benchmark of exchange statistics depends highly on the system's configuration. There is no universal benchmark that can be employed for determining the mutual statistics exactly. However, we demonstrated that by choosing correlators carefully one can get rid of all the unnecessary single particle interference effects, and the influence of the impurity, to extract correct mutual statistics of the colliding particles. Our findings can help researchers to propose and implement the correct scheme of measurements of physical observables in quantum systems. Furthermore, recent developments in measurement induced quantum steering and monitored quantum dynamics provide a broader context for our analysis. These studies demonstrate how coupling to auxiliary degrees of freedom and measurement back action can modify quantum states and particle dynamics, and different linear and non-linear observables~\cite{PhysRevResearch.2.033347,gurarie2025randomlymeasuredquantumparticle}. Here, we address the corresponding question for observables that probe the mutual statistics of bosons and fermions.

\section{Acknowledgment}
It is a pleasure to thank Jukka I. V\"{a}yrynen, Smitha Vishveshwara, and Yuval Gefen for many useful discussions. S.S.S. is grateful to Yuval Gefen for drawing attention to Ref.~\cite{PhysRevB.99.045430} and for numerous valuable discussions on extended mesoscopic colliders. S.S.S. acknowledges support through the Grant No.~2022391 from the United States--Israel Binational Science Foundation, Jerusalem, Israel awarded to Jukka I. V\"{a}yrynen and Yuval Gefen. S.S.S. also acknowledges support from Bilsland Graduate Fellowship, and R.C. gratefully acknowledges support through Summer Research Grant awarded by Department of Physics and Astronomy, Purdue University. S.S.S. also thanks the APS Division of Condensed Matter Physics for a travel grant to attend the 2026 APS Global Physics Summit, where this work was presented.

\begin{appendix}
\begin{widetext}

\setcounter{equation}{0}
\setcounter{figure}{0}
\setcounter{table}{0}
\setcounter{section}{0}

\makeatletter
\renewcommand{\theequation}{\Alph{section}\arabic{equation}}
\renewcommand{\thetable}{\Alph{section}\arabic{table}}
\renewcommand{\thefigure}{\Alph{section}\arabic{figure}}
\@addtoreset{equation}{section}
\@addtoreset{table}{section}
\@addtoreset{figure}{section}
\makeatother

\section{Resonant tunneling}\label{app:resonant_tunneling}
In this section, we are going to derive the effective scattering matrix for a setup with two quantum point contacts (QPC). The scattering matrix $S-$matrix (unitary), for a single QPC is given as follows,
\begin{align}
S_{1} =\left(\begin{array}{cc}
t & r\\
r & -t
\end{array}\right)=S_{2} \hspace{0.5cm} \text{and} \hspace{0.5cm} 
S_{1}^{\dagger}S_{1} =\left(\begin{array}{cc}
1 & 0\\
0 & 1
\end{array}\right)=S_{2}^{\dagger}S_{2}\,,
\end{align}
where $t$ refers to the transmission amplitude i.e., the probability amplitude for 
the particle to go from source $S_1$ to detector $D_1$ and similarly,
amplitude for the particle to go from source
$S_1$ to detector $D_2$ is $r$. With this we can derive the effective
$S-$matrix of the setup which includes two scatterers (or QPCs). This is derived as follows,
\begin{align}
\mathcal{T}_k & = t\cdot t\cdot e^{i\varphi}+t\cdot r\cdot r\cdot t\cdot e^{i3\varphi}+t\cdot r\cdot r\cdot r\cdot r\cdot t\cdot e^{i5\varphi}+\cdots=\frac{t^{2}e^{i\varphi}}{1-r^{2}e^{i2\varphi}}\,,
\end{align}
and similarly, we derive the reflection amplitude as follows,
\begin{align}
\mathcal{R}_k & =r-t\cdot r\cdot t\cdot e^{i2\varphi}-t\cdot r\cdot r\cdot r\cdot t\cdot e^{i4\varphi}-\cdots=\frac{r-re^{i2\varphi}}{1-r^{2}e^{i2\varphi}}\,,
\end{align}
where that the phase $\varphi$ corresponds to the orbital phase that the particle of momentum $k$ acquires as it propagates a distance $L$ between the two QPCs, i.e. $\varphi=kL$. Hence, the effective $S-$matrix is given as,
\begin{equation}
S_{\text{eff}}=\left[\begin{array}{cc}
\frac{t^{2}e^{i\varphi}}{1-r^{2}e^{i2\varphi}} & \frac{r(1-e^{i2\varphi})}{1-r^{2}e^{i2\varphi}} \\
\frac{r(1-e^{i2\varphi})}{1-r^{2}e^{i2\varphi}} & \frac{t^{2}e^{i\varphi}}{1-r^{2}e^{i2\varphi}}
\end{array}\right]\,.\label{eq:seff}
\end{equation}
For simplicity in the calculations, let us also assume that all the scattering amplitudes for the individual QPC (or scatterer) are real. Hence, we obtain the following for the single-particle probabilities of the
reflection and transmission,
\begin{align}
P(1\to1)=|\mathcal{T}_k|^{2} =\frac{(1-r^{2})^{2}}{1+r^{4}-2r^{2}\cos2\varphi} \hspace{0.5cm} \text{and} \hspace{0.5cm} P(1\to2)=|\mathcal{R}_k|^{2} =\frac{2r^{2}(1-\cos2\varphi)}{1+r^{4}-2r^{2}\cos2\varphi}\,,
\end{align}
As a result of unitarity, we have,
\begin{equation}
|\mathcal{T}_k|^{2}+|\mathcal{R}_k|^{2}=1\,.
\end{equation}
As a sanity check, we must also have the effective $S-$matrix to be unitary in order to conserve probability, $S_{\text{eff}}^{\dagger}S_{\text{eff}}=1$. Additionally, we must have vanishing off-diagonal terms for $S_{\text{eff}}^{\dagger}S_{\text{eff}}$. This is verified as follows,
\begin{align}
\big[S_{\text{eff}}^{\dagger}S_{\text{eff}}\big]_{12} =\Big[\frac{t^{2}e^{i\varphi}}{1-r^{2}e^{i2\varphi}}\Big]\Big{[}\frac{r(1-e^{i2\varphi})}{1-r^{2}e^{i2\varphi}}\Big{]}^{*}+\Big[\frac{t^{2}e^{i\varphi}}{1-r^{2}e^{i2\varphi}}\Big]^{*}\Big{[}\frac{r(1-e^{i2\varphi})}{1-r^{2}e^{i2\varphi}}\Big{]}= 0\,.
\end{align}
Hence, we find that the effective $S-$matrix is also unitary, i.e.,
\begin{equation}
S_{\text{eff}}^{\dagger}S_{\text{eff}}=1\,.
\end{equation}
Now, in a setup like this where we have multipath interference effects, this can lead to resonances in the system. This can be witnessed by choosing a particular value of the phase, $\varphi= kL = n\pi$,
and this implies,
\begin{align}
|\mathcal{R}_k|^{2} & =\frac{2r^{2}(1-\cos2\varphi)}{1+r^{4}-2r^{2}\cos2\varphi}=0 \hspace{0.5cm} \text{and} \hspace{0.5cm} |\mathcal{T}_k|^{2} =1\,.
\end{align}
Hence, irrespective of the value of $r$ and $t$, we are going to have a
complete transmission in the system. This is known as resonant tunneling.

\section{Scattering formalism}\label{app:scattering}

In this section we are going to establish the scattering formalism that we use to obtain probabilities of the different scattering events. Most of the results in this section have already been proven in the supplementary material of the reference~\cite{td98-5ltj}. However, for the sake of completeness, we give some important formulas and ideas here as well. Upper edge in the collider constitute the right movers, and the lower edge constitute the left movers. We have two sources $S_{1}$ in the upper edge and $S_{2}$ in the lower edge, similarly, we have two detectors $D_{1}$ and $D_{2}$ in the upper and the lower edge, respectively. The operator $\Tilde{\phi}_{s_1}^{\dagger}$, creates a wave packet at $(x_{1}^{(0)},t_{1}^{(0)})$ originating from $S_{1}$ and is given as, 
\begin{align}
    \Tilde{\phi}^{\dagger}_{s_{1}}(x_{1}^{(0)},t_{1}^{(0)}) = \int_{-\infty}^{\infty} dk e^{ik(-x_{1}^{(0)}+vt_{1}^{(0)})} \phi^{*}(k) c_{k,s_{1}}^{\dagger} \,,
\end{align}
where $\phi(k)$ is the wave form in momentum space, normalized such that  $\int_{-\infty}^{\infty}dk |\phi(k)|^{2}=1$. The creation operator at position $(x_{1},t_{1})$ in the detector lead ($D_{1}$) is given as,
\begin{align}
    \phi_{d_{1}}^{\dagger}(x_{1},t_{1}) = \int_{-\infty}^{\infty} \frac{dk}{\sqrt{2\pi}} e^{ik(-x_{1}+vt_{1})} c_{k,d_{1}}^{\dagger} \,.
\end{align}
The creation operators at the sources $S_{1}$, $S_{2}$ and the detectors $D_{1}$, $D_{2}$ are related by the effective scattering matrix, Eq.~\eqref{eq:seffqubit_main} as follows,
\begin{align}\label{eq:seffqubit_main_appenB}
    \left(\begin{array}{c}  c_{k,d_{1}}\\
c_{k,d_{2}}
\end{array}\right) = \left(\begin{array}{cc}\mathcal{T}_{k}e^{i\eta_{m}} & \mathcal{R}_{k}e^{i\theta_{m}}\\
\mathcal{R}_{k}e^{i(2\eta_{m}-\theta_{m})} & \mathcal{T}_{k}e^{i\eta_{m}}
\end{array}\right) \left(\begin{array}{c}  c_{k,s_{1}}\\
c_{k,s_{2}}
\end{array}\right) = \left(\begin{array}{cc} t_{k,m} & r'_{k,m}\\
r_{k,m} & t'_{k,m} \end{array}\right) \left(\begin{array}{c}  c_{k,s_{1}}\\
c_{k,s_{2}}
\end{array}\right)\,.
\end{align}
From here we find that the single particle scattering amplitude for the particle (or wave packet) to start from $S_{1}$ with coordinates $(x_{1}^{(0)},t_{1}^{(0)})$ and reaching the detector $D_{1}$ at $(x_{1},t_{1})$ is given as follows,
\begin{align}\label{eq:singleamp_appenb}
    \mathcal{I}_{1}^{(m)}(x_{1},t_{1};x_{1}^{(0)},t_{1}^{(0)}) = A_{m}(1\rightarrow 1) =& \langle\Omega|\phi_{d_{1}}(x_{1},t_{1})\tilde{\phi}^{\dagger}_{s_{1}}(x_{1}^{0},t_{1}^{0})|\Omega\rangle = \int_{-\infty}^{\infty} \frac{dk}{\sqrt{2\pi}} \, e^{ik(x_{1}-x_{1}^{(0)}-vt_{1}+vt_{1}^{(0)})}\phi(k)^{*} \, t_{k,m}\,,
\end{align}
where $|\Omega\rangle$ is the vacuum state or equivalently the filled Fermi-sea. Note that we have introduced a notation $A_{m}(i\to j)$ referring to the tunneling event where the wave packet starts from the source $S_{i}$ and reaches the detector $D_{j}$, with $i,\,j \in \{1,\,2\}$. Similarly, we can write down all other single particle amplitudes as follows,
\begin{align}\label{eq:singleamps_appenb}
A_{m}(1\to1) = \mathcal{I}_{1}^{(m)}(x_{1},t_{1};x_{1}^{(0)},t_{1}^{(0)}) &= \int_{-\infty}^{\infty} \frac{dk}{\sqrt{2\pi}} \, e^{ik(x_{1}-x_{1}^{(0)}-vt_{1}+vt_{1}^{(0)})}\phi(k)^{*} \, t_{k,m}\,, \\
A_{m}(2\to2) = \mathcal{I}_{2}^{(m)}(x_{2},t_{2};x_{2}^{(0)},t_{2}^{(0)}) &=  \int_{-\infty}^{\infty} \frac{dk_{2}}{\sqrt{2\pi}}e^{ik_{2}(-x_{2}-vt_{2}+x_{2}^{(0)}+vt_{2}^{(0)})}\phi^{*}(k_{2}) \, t'_{k,m} \,, \\
A_{m}(1\to2)  =  \mathcal{J}_{1}^{(m)}(x_{2},t_{2};x_{1}^{(0)},t_{1}^{(0)}) &= \int_{-\infty}^{\infty} \frac{dk_{1}}{\sqrt{2\pi}}e^{ik_{1}(-x_{2}-vt_{2}-x_{1}^{(0)}+vt_{1}^{(0)})}\phi^{*}(k_{1}) \, r_{k,m} \,, \\
A_{m}(2\to1)  =  \mathcal{J}_{2}^{(m)}(x_{1},t_{1};x_{2}^{(0)},t_{2}^{(0)}) &= \int_{-\infty}^{\infty} \frac{dk_{2}}{\sqrt{2\pi}}e^{ik_{2}(x_{1}-vt_{1}+x_{2}^{(0)}+vt_{2}^{(0)})}\phi^{*}(k_{2}) \, r'_{k,m} \,,
\end{align}
where we assume that the particle (wave packet) starts from the source $S_{1}$ at $(x_{1}^{(0)},t_{1}^{(0)})$, $S_{2}$ at $(x_{2}^{(0)},t_{2}^{(0)})$, and reach the detector $D_{1}$ at $(x_{1},t_{1})$ and $D_{2}$ at $(x_{2},t_{2})$. From here, the probability of scattering is given as follows,
\begin{align}
    P_{m} (1 \to 1) &= \int_{-\infty}^{\infty} vdt_{1} \big{|}\mathcal{I}_{1}^{(m)}(x_{1},t_{1};x_{1}^{(0)},t_{1}^{(0)})\big{|}^{2} = \int_{-\infty}^{\infty} dk |t_{k,m}|^{2} |\phi(k)|^{2} \,,\\
    P_{m} (1 \to 2) &= \int_{-\infty}^{\infty} vdt_{2} \big{|}\mathcal{J}_{1}^{(m)}(x_{2},t_{2};x_{1}^{(0)},t_{1}^{(0)})\big{|}^{2} = \int_{-\infty}^{\infty} dk |r_{k,m}|^{2} |\phi(k)|^{2}\,,\\
    P_{m} (2 \to 1) &= \int_{-\infty}^{\infty} vdt_{1} \big{|}\mathcal{J}_{2}^{(m)}(x_{1},t_{1};x_{2}^{(0)},t_{2}^{(0)})\big{|}^{2} = \int_{-\infty}^{\infty} dk |r'_{k,m}|^{2} |\phi(k)|^{2} \,,\\
    P_{m} (2 \to 2) &= \int_{-\infty}^{\infty} vdt_{2} \big{|}\mathcal{I}_{2}^{(m)}(x_{2},t_{2};x_{2}^{(0)},t_{2}^{(0)})\big{|}^{2} =\int_{-\infty}^{\infty} dk |t'_{k,m}|^{2} |\phi(k)|^{2}\,,
\end{align}
where the integration is over the detection time $t_{1}$, $t_{2}$ at the two detectors $D_{1}$ and $D_{2}$ respectively. Also, note that integral of the times $t_{1}$, $t_{2}$ gives an $\delta-$function and hence we get a single integral of the momentum variable $k$. Next, it is important to realize that in our case, the qubit coupling to the extended collider is via phase modification of the scattering amplitude, Eq.~\eqref{eq:seffqubit_main_appenB} and as a result, we get the probability to the case where there is no qubit coupled to the extended collider,
\begin{align}\label{eq:scattering_no_qubit}
    P_{m}(1\to 1) = \mathcal{P}(1\to 1) &= \int_{-\infty}^{\infty} dk |\mathcal{T}_{k}|^{2} |\phi(k)|^{2} \,, \qquad \, P_{m}(1\to 2) = \mathcal{P}(1\to 2) = \int_{-\infty}^{\infty} dk |\mathcal{R}_{k}|^{2} |\phi(k)|^{2} \,,\\
    P_{m}(2\to 2) = \mathcal{P}(2\to 2) &= \int_{-\infty}^{\infty} dk |\mathcal{T}_{k}|^{2} |\phi(k)|^{2} \,, \qquad \, P_{m}(2\to 1) = \mathcal{P}(2\to 1) = \int_{-\infty}^{\infty} dk |\mathcal{R}_{k}|^{2} |\phi(k)|^{2} \,,
\end{align}
and it vital to understand that this corresponds to the case of no postselection. The state of the qubit is unchanged as the scattering takes place, and detection of the particles at the detectors $D_{1}$ and $D_{2}$. 

\subsection{Two-particle scattering}
 Let us generalize the above analysis for the case where we have two incoming particles from the two sources $S_{1}$ and $S_{2}$. Let us start by looking at the event where we receive one particle at each detector $D_{1}$ and $D_{2}$. Scattering amplitude of this event, denoted by $A_{m}(11)$ is given as follows,
 \begin{align}
     A_{m}(11) = \langle\Omega|\phi_{d_{2}}(x_{2},t_{2})\phi_{d_{1}}(x_{1},t_{1})\tilde{\phi}_{s_{1}}^{\dagger}(x_{1}^{(0)},t_{1}^{(0)})\tilde{\phi}_{s_{2}}^{\dagger}(x_{2}^{(0)},t_{2}^{(0)})|\Omega\rangle\,.
 \end{align}
where $m=\{\uparrow,\downarrow\}$, is the state of the qubit.
From here the probability is extracted as follows by integrating over the times $t_{1}$ and $t_{2}$,
\begin{align}\label{eq:p11wavepackets}
    \mathcal{P}(11)^{(m)}_{\text{B}/\text{F}} =& \iint_{-\infty}^{\infty} v^{2}dt_{1}dt_{2}|\mathcal{I}_{1}^{(m)}(x_{1},t_{1};x_{1}^{(0)},t_{1}^{(0)})\mathcal{I}_{2}^{(m)}(x_{2},t_{2};x_{2}^{(0)},t_{2}^{(0)}) \pm  \mathcal{J}_{2}^{(m)}(x_{1},t_{1};x_{2}^{(0)},t_{2}^{(0)})\mathcal{J}_{1}^{(m)}(x_{2},t_{2};x_{1}^{(0)},t_{1}^{(0)})|^{2}\,.
\end{align}
where we have used out single particle results from Eq.~\eqref{eq:singleamps_appenb}. Next, we move forward by simplifying the result and we obtain,
\begin{align}
    \mathcal{P}(11)_{\text{B}/\text{F}}^{(m)} =& \Big( \int_{-\infty}^{\infty} dk |\mathcal{T}_{k}|^{2}|\phi(k)|^{2} \Big)^{2} + \Big( \int_{-\infty}^{\infty} dk |\mathcal{R}_{k}|^{2}|\phi(k)|^{2} \Big)^{2}  \pm 2\chi^{(m)}(x_{1}^{(0)},t_{1}^{(0)};x_{2}^{(0)},t_{2}^{(0)})\,,\\
    =& \mathcal{P}(1\to1)\mathcal{P}(2\to2)+\mathcal{P}(1\to2)\mathcal{P}(2\to1) \pm 2\chi^{(m)}(x_{1}^{(0)},t_{1}^{(0)};x_{2}^{(0)},t_{2}^{(0)})\,,\\
    =& \mathcal{P}(11)_{\text{CW}}  \pm 2 \chi^{(m)}(x_{1}^{(0)},t_{1}^{(0)};x_{2}^{(0)},t_{2}^{(0)}) \,,
\end{align}
where we have,
\begin{equation}\label{eq:chi_sign}
\chi^{(m)}(x_{1}^{(0)},t_{1}^{(0)};x_{2}^{(0)},t_{2}^{(0)}) = -\Bigg{|}\int_{-\infty}^{\infty} dk|\phi(k)|^{2}{r}_{k,m}^{*}{t'}_{k,m}e^{ik\big[x_{0}^{(2)}+vt_{0}^{(2)}+x_{0}^{(1)}-vt_{0}^{(1)}\big]}\Bigg{|}^{2}\,.
\end{equation}
Hence, we find that if the qubit state is unaffected during the scattering process followed by detection, then the probability is unaffected by the state of the qubit for our choice of coupling of the qubit with the collider, Eq.~\eqref{eq:seffqubit_main_appenB}. The probability of the event where both the particles end up in the same detector is obtained in a similar way. We finish this section by giving the formula for the probability of scattering for classical waves,
\begin{align}
    \mathcal{P}(11)_{\text{CW}} =& \Big( \int_{-\infty}^{\infty} dk |\mathcal{T}_{k}|^{2}|\phi(k)|^{2} \Big)^{2} + \Big( \int_{-\infty}^{\infty} dk |\mathcal{R}_{k}|^{2}|\phi(k)|^{2} \Big)^{2}  =  \mathcal{P}(1\to1)\mathcal{P}(2\to2)+\mathcal{P}(1\to2)\mathcal{P}(2\to1)\,,
\end{align}
This can be derived from the corresponding quantity for fermions or bosons with well separated emission times, i.e. $v|t_{1}^{(0)}-t_{2}^{(0)}|\gg \sigma$. This is obtained using the fact that we have,
\begin{align}
\chi(x_{1}^{(0)},t_{1}^{(0)};x_{2}^{(0)},t_{2}^{(0)}) \to 0 \qquad \left(\text{in the limit } v|t_{1}^{(0)}-t_{2}^{(0)}|\gg \sigma \right)\,,
\end{align}
and hence we find,
\begin{align}
    P(11)_{\text{B/F}} \to \mathcal{P}(11)_{\text{CW}}  \qquad \left(\text{in the limit } v|t_{1}^{(0)}-t_{2}^{(0)}|\gg \sigma \right)\,.
\end{align}
A thorough derivation already exists in the supplementary material of the reference~\cite{td98-5ltj}, therefore we do not redo the calculations here.

\section{Normalization in postselected quantum state}\label{app:post_selection}

Consider a system of qubit in the quantum state $|Q\rangle$ and an
ancillary qubit in the state $|A\rangle$. The quantum states are
given as follows,
\begin{align}
|Q\rangle & =q_{1}|\uparrow\rangle+q_{2}|\downarrow\rangle\,,\\
|A\rangle & =a_{1}|\uparrow\rangle+a_{2}|\downarrow\rangle\,,
\end{align}
with the assumption that the states are normalized,
\begin{equation}
\langle Q|Q\rangle=|q_{1}|^{2}+|q_{2}|^{2}=1=|a_{1}|^{2}+|a_{2}|^{2}=\langle A|A\rangle\,.
\end{equation}
From here, we can construct the density matrix as follows,
\begin{equation}
\rho_{i}=|QA\rangle\langle QA|\,,
\end{equation}
where,
\begin{align}
|QA\rangle=|Q\rangle\otimes|A\rangle & =a_{1}q_{1}|\uparrow\rangle\otimes|\uparrow\rangle+a_{2}q_{1}|\uparrow\rangle\otimes|\downarrow\rangle+a_{1}q_{2}|\downarrow\rangle\otimes|\uparrow\rangle+a_{2}q_{2}|\downarrow\rangle\otimes|\downarrow\rangle\,,\\
 & =a_{1}q_{1}|\uparrow\uparrow\rangle+a_{2}q_{1}|\uparrow\downarrow\rangle+a_{1}q_{2}|\downarrow\uparrow\rangle+a_{2}q_{2}|\downarrow\downarrow\rangle\,.
\end{align}
Now, let us perform a measurement of the qubit provided we postselect the
ancilla qubit in the state,
\begin{equation}
|A_{f}\rangle=|\uparrow\rangle\,.
\end{equation}
Let us say that the probability of finding the qubit in the state
$|\uparrow\rangle$ is $P(|\uparrow\rangle;|A_{f}\rangle)$. This
is given as follows,
\begin{equation}
\mathcal{P}(|\uparrow\rangle;|A_{f}\rangle)=P(|\uparrow\rangle;|\uparrow\rangle)=\langle\uparrow\uparrow|\rho_{i}|\uparrow\uparrow\rangle=|a_{1}|^{2}|q_{1}|^{2}\,,
\end{equation}
and we obtain the probability of the qubit to in the state $|\downarrow\rangle$,
\begin{equation}
\mathcal{P}(|\downarrow\rangle;|A_{f}\rangle)=P(|\downarrow\rangle;|\uparrow\rangle)=\langle\downarrow\uparrow|\rho_{i}|\downarrow\uparrow\rangle=|a_{1}|^{2}|q_{2}|^{2}\,.
\end{equation}
Naively adding these probabilities, we find them to be not normalized,
\begin{equation}
\mathcal{P}(|\uparrow\rangle;|A_{f}\rangle)+\mathcal{P}(|\downarrow\rangle;|A_{f}\rangle)=|a_{1}|^{2}\neq1\,.
\end{equation}
Hence, the correct probability with postselection is obtained by
normalization as follows,
\begin{align}
P(|\uparrow\rangle;|A_{f}\rangle) & =\frac{\mathcal{P}(|\uparrow\rangle;|A_{f}\rangle)}{\mathcal{P}(|\uparrow\rangle;|A_{f}\rangle)+\mathcal{P}(|\downarrow\rangle;|A_{f}\rangle)}=|q_{1}|^{2}\,,\\
P(|\downarrow\rangle;|A_{f}\rangle) & =\frac{\mathcal{P}(|\downarrow\rangle;|A_{f}\rangle)}{\mathcal{P}(|\uparrow\rangle;|A_{f}\rangle)+\mathcal{P}(|\downarrow\rangle;|A_{f}\rangle)}=|q_{2}|^{2}\,.
\end{align}
As a result, this  exercise tells us that post selection leads to an unphysical
probability, i.e., the total probability may not add up to $1$, and
therefore it is necessary to normalize in the end of postselection.

\section{Derivation of single particle probabilities and the disentangling condition}\label{app:single_particle_with_qubit}
In this section, we work out the details of a single-particle scattering in the presence of a qubit. The effective scattering matrix is as follows:
\begin{align}\label{eq:seffqubit}
S_{\text{eff}}^{(m)}
&=\left\{\begin{array}{cc}\mathcal{T}_{k}e^{i\eta_{m}} & \mathcal{R}_{k}e^{i\theta_{m}}\\
\mathcal{R}_{k}e^{i(2\eta_{m}-\theta_{m})} & \mathcal{T}_{k}e^{i\eta_{m}}
\end{array}\right\} =\left\{\begin{array}{cc}|\mathcal{T}_{k}|e^{i({\theta_{2}}+\eta_{m})} & |\mathcal{R}_{k}|e^{i({\theta_{1}}+\theta_{m})}\\
|\mathcal{R}_{k}|e^{i({\theta_{1}}+2\eta_{m}-\theta_{m})} & |\mathcal{T}_{k}|e^{i({\theta_{2}}+\eta_{m})}
\end{array}\right\} \,,
\end{align}
where $m=\{\uparrow,\downarrow\}$ denote the two possible states of the qubit $\{|\uparrow\rangle, |\downarrow\rangle\}$ and determines the scattering matrix. The parameters $\eta_m$, $\theta_m$ depends on the state of the qubit, whereas $\mathcal{T}_{k}$, $\mathcal{R}_{k}$ are intrinsic to the collider. Note that for an extended collider, the reflection and transmission amplitudes depend on the momentum of the particles via a geometric phase $\varphi= kL$, where $k$ is the momentum of the particle and $L$ is the size of the extended collider. For a single particle emitted from the source $S_{1}$, the probability of being scattered to the detector $D_{1}$ in the presence of a qubit is given as follows using Eq.~\eqref{eq:singleamp_appenb},
\begin{align}
P(1\to1;|\psi\rangle)\propto\mathcal{P}(1\to1;|\psi\rangle)= \int_{-\infty}^{\infty} vdt_1 \langle f |\rho_{\text{sys}}| f \rangle  = \int_{-\infty}^{\infty} vdt_1\Big|\sum_{m}\gamma_{m}\mathcal{I}^{(m)}_{1}(x_{1},t_{1};x_{1}^{(0)},t_{1}^{(0)})\langle\psi|\psi_{m}\rangle\Big|^{2}\,,
\end{align}
where the density matrix $\rho_{\text{sys}}$ is given in the main text Eq.~\eqref{eq:denisty_system}, $\gamma_m\langle\psi|m\rangle = \tilde{\gamma}_m$ and  $(x_{1}^{(0)},t_{1}^{(0)})$ are the coordinates of the injection point of the incoming particles from the source $S_{1}$ and the detector $D_{1}$ is located at $(x_{1},t_{1})$ as follows,
\begin{align}
    |i\rangle &= \Tilde{\phi}^{\dagger}_{s_{1}}(x_{1}^{(0)},t_{1}^{(0)})|\Omega\rangle\,,\\
    |f\rangle &= \phi^{\dagger}_{d_{1}}(x_{1},t_{1})|\Omega\rangle\,.
\end{align}
From here, we can see that there will be four contributions to the probability, given as follows,
\begin{align}
    \mathcal{P}(1\to1;|\psi\rangle) & = \int_{-\infty}^{\infty} vdt_1 \langle f,\,\psi |\rho_{\text{sys}}| f,\,\psi \rangle \,,\\
    &=\int_{-\infty}^{\infty} vdt_1\Big|\sum_{m}\gamma_{m}\mathcal{I}^{(m)}_{1}(x_{1},t_{1};x_{1}^{(0)},t_{1}^{(0)})\langle\psi|\psi_{m}\rangle\Big|^{2}\,,\\ &= |\tilde{\gamma}_{\uparrow}|^{2}\mathcal{P}(1\to1) + |\tilde{\gamma}_{\downarrow}|^{2}\mathcal{P}(1\to1) + \mathcal{P}(1\to1)|\tilde{\gamma}_{\uparrow}\tilde{\gamma}_{\downarrow}|\big{[}e^{i\eta_{\uparrow}}e^{-i\eta_{\downarrow}}e^{i\phi_{0}}+ e^{-i\eta_{\uparrow}}e^{i\eta_{\downarrow}}e^{-i\phi_{0}}\big{]}\,.
\end{align}
This implies that we have,
\begin{align}
    &P(1\to1;|\psi\rangle) \propto \big\{\mathcal{C}+2\cos(\eta_{\uparrow}-\eta_{\downarrow}+\phi_{0})|\tilde{\gamma}_{\uparrow}\tilde{\gamma}_{\downarrow}|\big\}\mathcal{P}(1\to1)\,,
\end{align}
where we have used the fact that the qubit state i.e., $|\tilde{\gamma}_{\uparrow}|^{2} + |\tilde{\gamma}_{\downarrow}|^{2}=\mathcal{C}$ and the expression for $\mathcal{P}(1 \to 1)$ is given in Eq.~\eqref{eq:scattering_no_qubit} . Similarly, we are going to have the following for the probability of the event where the particle from $S_{1}$ goes to the detector $D_{2}$,
\begin{align}
    &P(1\to2;|\psi\rangle) \propto  \big\{\mathcal{C}+2\cos(\eta_{\uparrow}-\eta_{\downarrow}+\phi_{1}+\phi_{0})|\tilde{\gamma}_{\uparrow}\tilde{\gamma}_{\downarrow}|\big\}\mathcal{P}(1\to2)\,, \qquad \phi_1=\eta_{\uparrow}-\eta_{\downarrow}-\theta_{\uparrow}+\theta_{\downarrow}\,.
\end{align}
As a result of coupling to the impurity (qubit), the obtained probability is not normalized to unity and therefore one needs to post-normalize with the following factor, $N$, where:
\begin{align}
    &N = \mathcal{C}+2\cos(\eta_{\uparrow}-\eta_{\downarrow}+\phi_{0})|\tilde{\gamma}_{\uparrow}\tilde{\gamma}_{\downarrow}|\mathcal{P}(1\to1)+ 2\cos(\eta_{\uparrow}-\eta_{\downarrow}+\phi_{1}+\phi_{0})|\tilde{\gamma}_{\uparrow}\tilde{\gamma}_{\downarrow}|\mathcal{P}(1\to2)\,.
\end{align}
Hence, the single particle probabilities in the presence of the qubit is given as,
\begin{align}
    P(1\to1;|\psi\rangle) &= \frac{\big\{\mathcal{C}+2\cos(\eta_{\uparrow}-\eta_{\downarrow}+\phi_{0})|\tilde{\gamma}_{\uparrow}\tilde{\gamma}_{\downarrow}|\big\}\mathcal{P}(1\to1)}{\mathcal{C}+2\cos(\eta_{\uparrow}-\eta_{\downarrow}+\phi_{0})|\tilde{\gamma}_{\uparrow}\tilde{\gamma}_{\downarrow}|\mathcal{P}(1\to1) + 2\cos(\eta_{\uparrow}-\eta_{\downarrow}+\phi_{1}+\phi_{0})|\tilde{\gamma}_{\uparrow}\tilde{\gamma}_{\downarrow}|\mathcal{P}(1\to2)} \,, \label{eq:1to1qubit}\\
    &=\frac{\big\{1+2\cos(\eta_{\uparrow}-\eta_{\downarrow}+\phi_{0})|\bar{\gamma}_{\uparrow}\bar{\gamma}_{\downarrow}|\big\}\mathcal{P}(1\to1)}{1+2\cos(\eta_{\uparrow}-\eta_{\downarrow}+\phi_{0})|\bar{\gamma}_{\uparrow}\bar{\gamma}_{\downarrow}|\mathcal{P}(1\to1) + 2\cos(\eta_{\uparrow}-\eta_{\downarrow}+\phi_{1}+\phi_{0})|\bar{\gamma}_{\uparrow}\bar{\gamma}_{\downarrow}|\mathcal{P}(1\to2)} \,, \\
    P(1\to2;|\psi\rangle) &=\frac{\big\{\mathcal{C}+2\cos(\eta_{\uparrow}-\eta_{\downarrow}+\phi_{1}+\phi_{0})|\tilde{\gamma}_{\uparrow}\tilde{\gamma}_{\downarrow}|\big\}\mathcal{P}(1\to2)}{\mathcal{C}+2\cos(\eta_{\uparrow}-\eta_{\downarrow}+\phi_{0})|\tilde{\gamma}_{\uparrow}\tilde{\gamma}_{\downarrow}|\mathcal{P}(1\to1) + 2\cos(\eta_{\uparrow}-\eta_{\downarrow}+\phi_{1}+\phi_{0})|\tilde{\gamma}_{\uparrow}\tilde{\gamma}_{\downarrow}|\mathcal{P}(1\to2)}\,, \\
    &=\frac{\big\{1+2\cos(\eta_{\uparrow}-\eta_{\downarrow}+\phi_{1}+\phi_{0})|\bar{\gamma}_{\uparrow}\bar{\gamma}_{\downarrow}|\big\}\mathcal{P}(1\to2)}{1+2\cos(\eta_{\uparrow}-\eta_{\downarrow}+\phi_{0})|\bar{\gamma}_{\uparrow}\bar{\gamma}_{\downarrow}|\mathcal{P}(1\to1) + 2\cos(\eta_{\uparrow}-\eta_{\downarrow}+\phi_{1}+\phi_{0})|\bar{\gamma}_{\uparrow}\bar{\gamma}_{\downarrow}|\mathcal{P}(1\to2)}\,,\label{eq:1to2qubit}
\end{align}
where for simplicity we have defined the following new variable such that we can get rid of the constant term $\mathcal{C}$ in the numerator and denominator,
\begin{align}
    \bar{\gamma}_{\uparrow} = \frac{\tilde{\gamma}_{\uparrow}}{\sqrt{\mathcal{C}}}\,,\qquad \bar{\gamma}_{\downarrow} = \frac{\tilde{\gamma}_{\downarrow}}{\sqrt{\mathcal{C}}} \quad  \implies \quad |\bar{\gamma}_{\uparrow}|^{2} + |\bar{\gamma}_{\downarrow}|^{2} = 1 \,.
\end{align}
In the expression for the probability, if we set $\phi_{1}=0,$ we obtain,
\begin{align}
P(1\to2;|\psi\rangle)&=\mathcal{P}(1\to2)= \int_{-\infty}^{\infty} dk |\mathcal{R}_{k}|^{2}|\phi(k)|^{2}
\label{A19}\,,\\
P(1\to1;|\psi\rangle)&=\mathcal{P}(1\to1)= \int_{-\infty}^{\infty} dk |\mathcal{T}_{k}|^{2}|\phi(k)|^{2}\,,
\label{A20}
\end{align}
and hence, the probability is equal to the case where there is no coupling to the qubit. This is the disentangling limit, where the qubit doesn't affect the outcome of the scattering. If we choose the wave packet shape as $|\phi(k)|^{2}=\delta(k-k')$, then we get,
\begin{align}
\mathcal{P}(1\to1)&=|\mathcal{T}_{k}|^{2}\,,
\\
\mathcal{P}(1\to2)&=|\mathcal{R}_{k}|^{2}\,.
\end{align}
Since the probability is momentum dependent, one also obtains resonant tunneling whenever $kL=n\pi$ where $n$ is an integer. In other words, whenever the geometric phase satisfies $\phi = kL = n\pi$, we obtain $\mathcal{P}(1\to1)=1$. Next, it is interesting  to analyze under what conditions the reflection probability becomes zero, i.e.
Eq.~\eqref{eq:1to2qubit} vanishes in the presence of qubit. In particular, we would like to find the conditions under which the numerator of Eq.~\eqref{eq:1to2qubit} i.e., $1+2|\bar{\gamma}_{\uparrow}\bar{\gamma}_{\downarrow}|\cos(\eta_{\uparrow}-\eta_{\downarrow}+\phi_{1}+\phi_{0})$ vanishes. If we demand, $1+2|\bar{\gamma}_{\uparrow}\bar{\gamma}_{\downarrow}|\cos(\eta_{\uparrow}-\eta_{\downarrow}+\phi_{1}+\phi_{0})=0$, we find,
\begin{equation}
\cos(\eta_{\uparrow}-\eta_{\downarrow}+\phi_{1}+\phi_{0})=\frac{-1}{2|\bar{\gamma}_{\uparrow}\bar{\gamma}_{\downarrow}|}\,.
\end{equation}
For any real solutions to exist, we must have:
\begin{equation}
-1\leq\frac{-1}{2|\bar{\gamma}_{\uparrow}\bar{\gamma}_{\downarrow}|}\leq+1\,,
\end{equation}
which implies,
\begin{equation}
|\bar{\gamma}_{\uparrow}\bar{\gamma}_{\downarrow}|\geq\frac{1}{2}\,.
\end{equation}
Next, note that we have $|\bar{\gamma}_{\uparrow}|^{2}+|\bar{\gamma}_{\downarrow}|^{2}=1$, this enforces,
\begin{equation}\label{eq:prod_coeff_post_selection}
|\bar{\gamma}_{\uparrow}\bar{\gamma}_{\downarrow}|\leq\frac{1}{2}\,,
\end{equation}
therefore, real solutions exist iff $|\bar{\gamma}_{\uparrow}|=|\bar{\gamma}_{\downarrow}|=\frac{1}{\sqrt{2}}$.
Under these conditions, we can have
$\cos(\eta_{\uparrow}-\eta_{\downarrow}+\phi_{1}+\phi_{0})=-1$. Now, without the loss of generality, let us assume $\eta_{\uparrow}=0=\theta_{\uparrow}$ and hence we obtain,
\begin{equation}
\cos(\theta_{\downarrow}-2\eta_{\downarrow}+\phi_{0})=-1\,,
\end{equation}
or
\begin{equation}
\phi_{0}=(2n+1)\pi-\theta_{\downarrow}+2\eta_{\downarrow}\,.
\end{equation}
Hence, we get extra resonances solely due to the presence of the qubit
degrees of freedom which are irrespective of the geometric phase $\phi=kL$
and the shape of the incoming wave packets. 

\subsection{Disentanglement condition for a single particle incoming state}
To get the disentanglement condition for a single particle case, we
calculate the quantity $\epsilon=\langle i|S_{\uparrow}^{\dagger}S_{\downarrow}|i\rangle$. Recall that the effective scattering matrix of a particle with momentum $k$ is given as:
\begin{align}\label{eq:seffqubit}
S_{\text{eff}}^{(m)}&=\left\{\begin{array}{cc}\mathcal{T}_{k}e^{i\eta_{m}} & \mathcal{R}_{k}e^{i\theta_{m}}\\
\mathcal{R}_{k}e^{i(2\eta_{m}-\theta_{m})} & \mathcal{T}_{k}e^{i\eta_{m}}
\end{array}\right\} =\left\{\begin{array}{cc}t_{k,m} & r'_{k,m}\\
r_{k,m} & t'_{k,m}
\end{array}\right\} \,,
\end{align}
where the elements of the scattering matrix can be written as $\mathcal{T}_{k} = |\mathcal{T}_{k}|e^{i{\theta_{2}}}$ and $\mathcal{R}_{k} = |\mathcal{R}_{k}|e^{i{\theta_{1}}}$, where ${\theta}_{1}=\text{arg}\,\mathcal{R}_{k} $
${\theta_{2}}=\text{arg}\, \mathcal{T}_{k}$ with the geometric phase $\phi=kL$ and $r$ is the reflection amplitude for the individual quantum point contact. The explicit expression for $\epsilon=\langle i|S_{\uparrow}^{\dagger}S_{\downarrow}|i\rangle$ is,
\begin{align}
\epsilon&=\int_{-\infty}^{\infty} dkdk'\phi^{*}(k')\phi(k)e^{-ik(-x_{1}^{(0)}+vt_{1}^{(0)})}e^{ik'(-x_{1}^{(0)}+vt_{1}^{(0)})}\langle0|[t_{k,\uparrow}^{*}c_{d_{1},k}+{r}_{k,\uparrow}^{*}c_{d_{2},k}][t_{k',\downarrow}c^{\dagger}_{d_{1},k'}+{r}_{k',\downarrow}c^{\dagger}_{d_{2},k'}]|0\rangle\label{102}\,.
\end{align}
There are two non-zero terms contributing to $\epsilon$, given as follows,
\begin{align}
T_{1} &= \int_{-\infty}^{\infty} dkdk't_{k,\uparrow}^{*}t_{k',\downarrow}\phi^{*}(k')\phi(k)\langle0|c_{d_{1},k}c^{\dagger}_{d_{1},k'}|0\rangle e^{-ik(-x_{1}^{(0)}+vt_{1}^{(0)})}e^{ik'(-x_{1}^{(0)}+vt_{1}^{(0)})}\label{103} \,,\\
&=\int_{-\infty}^{\infty} dkdk't_{k,\uparrow}^{*}t_{k',\downarrow}\delta(k-k')\phi^{*}(k')\phi(k)e^{-ik(-x_{1}^{(0)}+vt_{1}^{(0)})}e^{ik'(-x_{1}^{(0)}+vt_{1}^{(0)})}\,,\\
&=e^{-i(\eta_{\uparrow}-\eta_{\downarrow})}\intop dk|\mathcal{T}_k|^{2}|\phi(k)|^{2}\,.\label{104}
\end{align}
and
\begin{align}
T_{2} &=\int_{-\infty}^{\infty} dkdk' {r}_{k,\uparrow}^{*}{r}_{k',\downarrow}\phi^{*}(k')\phi(k)\langle 0|c_{d_{2},k}c^{\dagger}_{d_{2},k'}|0\rangle e^{-ik(-x_{1}^{(0)}+vt_{1}^{(0)})}e^{ik'(-x_{1}^{(0)}+vt_{1}^{(0)})} \,,\\
&=\int_{-\infty}^{\infty} dkdk' {r}_{k,\uparrow}^{*}r_{k',\downarrow}\delta(k-k')\phi^{*}(k')\phi(k)e^{-ik(-x_{1}^{(0)}+vt_{1}^{(0)})}e^{ik'(-x_{1}^{(0)}+vt_{1}^{(0)})}\,, \\
&=e^{-i(2\eta_{\uparrow}-\theta_{\uparrow})}e^{i(2\eta_{\downarrow}-\theta_{\downarrow})}\int_{-\infty}^{\infty} dk|\mathcal{R}_k|^{2}|\phi(k)|^{2}\label{105}\,.
\end{align}
Hence, we find the quantity, $\epsilon$,
\begin{align}
    \epsilon &= T_1 + T_2=e^{-i(\eta_{\uparrow}-\eta_{\downarrow})}\Big\{\int_{-\infty}^{\infty} dk|\mathcal{T}_k|^{2}|\phi(k)|^{2}+\int_{-\infty}^{\infty} dk|\mathcal{R}_k|^{2}|\phi(k)|^{2}e^{-i[(\eta_{\uparrow}-\eta_{\downarrow})-(\theta_{\uparrow}-\theta_{\downarrow})]}\Big\}\,.
\end{align}
Now, in the particular case, where we have $\phi_{1}=(\eta_{\uparrow}-\eta_{\downarrow})-(\theta_{\uparrow}-\theta_{\downarrow})=0$, this implies $\eta_{\uparrow}-\eta_{\downarrow}=\theta_{\uparrow}-\theta_{\downarrow}$
and $\epsilon$ simplifies to:
\begin{align}
\epsilon&=e^{-i(\eta_{\uparrow}-\eta_{\downarrow})}\Big\{\int_{-\infty}^{\infty} dk|\mathcal{T}_k|^{2}|\phi(k)|^{2}+\int_{-\infty}^{\infty} dk|\mathcal{R}_k|^{2}|\phi(k)|^{2}\Big\}=e^{-i(\eta_{\uparrow}-\eta_{\downarrow})}.
\end{align}
and hence $\epsilon$ becomes a pure phase and the reduced density matrix
will have eigenvalues 0 and 1. Therefore, this is identified with the disentanglement
criterion. In the expression for the single-particle probabilities, Eq.~\eqref{eq:1to1qubit} and Eq.~\eqref{eq:1to2qubit} obtained for scatterer coupled to qubit, substituting $\phi_{1} = 0$ reduces to the case where there is no coupling with the qubit i.e., Eq.~\eqref{A19} and Eq.~\eqref{A20}.

\section{Derivation of two particle probabilities and the disentangling condition}\label{app:two_particle_qubit}

In this section, we are going to evaluate the two particle probability where we have a qubit coupled to extended collider. Let us start with scattering matrix of an extended collider coupled with an qubit,
\begin{align}\label{eq:seffqubit}
S_{\text{eff}}^{(m)}&=\left(\begin{array}{cc}|\mathcal{T}_{k}|e^{i({\theta_{2}}+\eta_{m})} & |\mathcal{R}_{k}|e^{i({\theta_{1}}+\theta_{m})}\\
|\mathcal{R}_{k}|e^{i({\theta_{1}}+2\eta_{m}-\theta_{m})} & |\mathcal{T}_{k}|e^{i({\theta_{2}}+\eta_{m})}
\end{array}\right)=\left(\begin{array}{cc}t_{k,m} & r'_{k,m}\\
r_{k,m} & t'_{k,m}
\end{array}\right)\,,
\end{align}
where $\theta_{1}$ and $\theta_{2}$ are the arguments of scattering amplitude $\mathcal{R}_{k}$ and $\mathcal{T}_{k}$. We start with an initial state where we have two incoming particles, one from each source, given as follows,
\begin{align}
    |i\rangle=\Tilde{\phi}^{\dagger}_{s_{1}}(x_{1}^{(0)},t_{1}^{(0)})\Tilde{\phi}^{\dagger}_{s_{2}}(x_{2}^{(0)},t_{2}^{(0)})|\Omega\rangle
\end{align}
and the final state has one particle in detector $D_{1}$ and the other in detector $D_{2}$, which is given as follows,
\begin{align}
    |f\rangle=\phi^{\dagger}_{d_{1}}(x_{1},t_{1}) \phi^{\dagger}_{d_{2}}(x_{2},t_{2})|\Omega\rangle
\end{align}
From here the two particle probability, $P(11;|\psi\rangle)_{\text{F}}$ where the qubit is projected onto the state $|\psi\rangle$ is given as follows, 
\begin{align}
    P(11;|\psi\rangle)_{\text{F}}\propto \mathcal{P}(11;|\psi\rangle)_{\text{F}} = \int_{-\infty}^{\infty} v^{2}dt_1dt_2 \big| \tilde{\gamma}_{\uparrow} A_{\uparrow}(11) + \tilde{\gamma}_{\downarrow} A_{\downarrow}(11) \big|^{2} \,,
\end{align}\label{153}
where,
\begin{align}
    A_{\uparrow}(11) &= \mathcal{I}_{1}^{(\uparrow)}(x_{1},t_{1};x_{1}^{(0)},t_{1}^{(0)})\mathcal{I}_{2}^{(\uparrow)}(x_{2},t_{2};x_{2}^{(0)},t_{2}^{(0)}) -\mathcal{J}_{1}^{(\uparrow)}(x_{2},t_{2};x_{1}^{(0)},t_{1}^{(0)})\mathcal{J}_{2}^{(\uparrow)}(x_{1},t_{1};x_{2}^{(0)},t_{2}^{(0)})\,,\\
    A_{\downarrow}(11) &= \mathcal{I}_{1}^{(\downarrow)}(x_{1},t_{1};x_{1}^{(0)},t_{1}^{(0)})\mathcal{I}_{2}^{(\downarrow)}(x_{2},t_{2};x_{2}^{(0)},t_{2}^{(0)}) - \mathcal{J}_{1}^{(\downarrow)}(x_{2},t_{2};x_{1}^{(0)},t_{1}^{(0)})\mathcal{J}_{2}^{(\downarrow)}(x_{1},t_{1};x_{2}^{(0)},t_{2}^{(0)})\,,
\end{align}
where we have used Eq.~\eqref{eq:singleamp_appenb} and Eq.~\eqref{eq:singleamps_appenb}. Evaluating the probability gives four terms, two of them are direct terms and two of them are interference terms. The direct terms, involves modulus square and integration over $t_1,\,t_2$ gives,
\begin{equation}
\int_{-\infty}^{\infty} v^2dt_1dt_2(|\tilde{\gamma}_{\uparrow}A_{\uparrow}(11)|^{2}+|\tilde{\gamma}_{\downarrow}A_{\downarrow}(11)|^{2}) = (|\tilde{\gamma}_{\uparrow}|^{2} + |\tilde{\gamma}_{\downarrow}|^{2})\cdot \mathcal{P}(11)_{\text{F}} =\mathcal{C}\cdot \mathcal{P}(11)_{\text{F}}\,, \ \  \label{154}
\end{equation}
where [see Appendix~\ref{app:scattering}]
\begin{align}
\mathcal{P}(11)_{\text{F}}&=\Big( \int_{-\infty}^{\infty} dk |\mathcal{T}_{k}|^{2}|\phi(k)|^{2} \Big)^{2} + \Big( \int_{-\infty}^{\infty} dk |\mathcal{R}_{k}|^{2}|\phi(k)|^{2} \Big)^{2}  - 2\chi(x_{1}^{(0)},t_{1}^{(0)};x_{2}^{(0)},t_{2}^{(0)}) \,,
\end{align}
is the two particle probability (for fermions) of receiving one particle in each detector in the absence of coupling of the extended collider with the qubit, and we have used the following,
\begin{align}
    \chi(x_{1}^{(0)},t_{1}^{(0)};x_{2}^{(0)},t_{2}^{(0)}) = -\Bigg{|}\int_{-\infty}^{\infty} dk|\phi(k)|^{2}\mathcal{R}_{k}^{*}\mathcal{T}_{k}e^{ik\big[x_{0}^{(2)}+vt_{0}^{(2)}+x_{0}^{(1)}-vt_{0}^{(1)}\big]}\Bigg{|}^{2}\,.
\end{align}
Now, let us concentrate on the interference term i.e., $\tilde{\gamma}_{\uparrow}A_{\uparrow}(11)\tilde{\gamma}_{{\downarrow}}^{*}A^{*}_{\downarrow}(11)$. For this we expand,
\begin{align}
\tilde{\gamma}_{\uparrow}A_{\uparrow}(11)
&=\tilde{\gamma}_{\uparrow}\mathcal{I}_{1}^{(\uparrow)}(x_{1},t_{1};x_{1}^{(0)},t_{1}^{(0)})\mathcal{I}_{2}^{(\uparrow)}(x_{2},t_{2};x_{2}^{(0)},t_{2}^{(0)})- \tilde{\gamma}_{\uparrow}\mathcal{J}_{1}^{(\uparrow)}(x_{2},t_{2};x_{1}^{(0)},t_{1}^{(0)})\mathcal{J}_{2}^{(\uparrow)}(x_{1},t_{1};x_{2}^{(0)},t_{2}^{(0)}) \,,\label{155}
\end{align}
and similarly,
\begin{align}
\tilde{\gamma}_{{\downarrow}}^{*}A^{*}_{\downarrow}(11)&=\tilde{\gamma}_{{\downarrow}}^{*}\mathcal{I}_{1}^{(\downarrow)*}(x_{1},t_{1};x_{1}^{(0)},t_{1}^{(0)})\mathcal{I}_{2}^{(\downarrow)*}(x_{2},t_{2};x_{2}^{(0)},t_{2}^{(0)})-  \tilde{\gamma}_{{\downarrow}}^{*}\mathcal{J}_{1}^{(\downarrow)*}(x_{2},t_{2};x_{1}^{(0)},t_{1}^{(0)})\mathcal{J}_{2}^{(\downarrow)*}(x_{1},t_{1};x_{2}^{(0)},t_{2}^{(0)})\,.\label{157}
\end{align}
We will calculate each term in the expression of $\tilde{\gamma}_{\uparrow}A_{\uparrow}(11)\tilde{\gamma}_{{\downarrow}}^{*}A^{*}_{\downarrow}$. Multiplying the first terms of Eq. $\eqref{155}$  and $\eqref{157}$ and then integrating
over $t_1,t_2$ give:
\begin{align}
|\tilde{\gamma}_{\uparrow}\tilde{\gamma}_{\downarrow}|e^{2i(\eta_{\uparrow}-\eta_{\downarrow})+i\phi_{0}}\mathcal{P}^{2}(1\to1)=\int_{-\infty}^{\infty} v^2
dt_1dt_2\tilde{\gamma}_{\uparrow}\tilde{\gamma}_{\downarrow}^{*}&\mathcal{I}_{1}^{(\uparrow)}(x_{1},t_{1};x_{1}^{(0)},t_{1}^{(0)})\mathcal{I}_{2}^{(\uparrow)}(x_{2},t_{2};x_{2}^{(0)},t_{2}^{(0)})
 \label{158}\\
\times & \mathcal{I}_{1}^{(\downarrow)*}(x_{1},t_{1};x_{1}^{(0)},t_{1}^{(0)})\mathcal{I}_{2}^{(\downarrow)*}(x_{2},t_{2};x_{2}^{(0)},t_{2}^{(0)})\,,\notag
\end{align}
Similarly, multiplying the second term of Eq. $\eqref{155}$ and Eq. $\eqref{157}$
and then integrating over $t_1,\,t_2$ gives,
\begin{align}
|\tilde{\gamma}_{\uparrow}\tilde{\gamma}_{\downarrow}|e^{2i(\eta_{\uparrow}-\eta_{\downarrow})+i\phi_{0}}\mathcal{P}^{2}(1\to2)=\int_{-\infty}^{\infty} v^2
dt_1dt_2\tilde{\gamma}_{\uparrow}\tilde{\gamma}_{\downarrow}^{*}&\mathcal{J}_{1}^{(\uparrow)}(x_{2},t_{2};x_{1}^{(0)},t_{1}^{(0)})\mathcal{J}_{2}^{(\uparrow)}(x_{1},t_{1};x_{2}^{(0)},t_{2}^{(0)})\label{159} \\
\times & \mathcal{J}_{1}^{(\downarrow)*}(x_{2},t_{2};x_{1}^{(0)},t_{1}^{(0)})\mathcal{J}_{2}^{(\downarrow)*}(x_{1},t_{1};x_{2}^{(0)},t_{2}^{(0)})\,. \notag
\end{align}
Complex conjugation of these terms also contribute to the probability and therefore we have,
\begin{align}
|\tilde{\gamma}_{\uparrow}\tilde{\gamma}_{\downarrow}|e^{2i(\eta_{\uparrow}-\eta_{\downarrow})+i\phi_{0}}\mathcal{P}^{2}(1\to1)+
|\tilde{\gamma}_{\uparrow}\tilde{\gamma}_{\downarrow}|e^{-2i(\eta_{\uparrow}-\eta_{\downarrow})-i\phi_{0}}\mathcal{P}^{2}(1\to1)&=2\mathcal{P}^{2}(1\to1)|\tilde{\gamma}_{\uparrow}\tilde{\gamma}_{\downarrow}|\cos\phi_{(1,1)}\,,\label{eq:directterms1}\\
|\tilde{\gamma}_{\uparrow}\tilde{\gamma}_{\downarrow}|e^{2i(\eta_{\uparrow}-\eta_{\downarrow})+i\phi_{0}}\mathcal{P}^{2}(1\to2)+|\tilde{\gamma}_{\uparrow}\tilde{\gamma}_{\downarrow}|e^{-2i(\eta_{\uparrow}-\eta_{\downarrow})-i\phi_{0}}\mathcal{P}^{2}(1\to2)&=2\mathcal{P}^{2}(1\to2)|\tilde{\gamma}_{\uparrow}\tilde{\gamma}_{\downarrow}|\cos\phi_{(1,1)}\,,\label{eq:directterms2}
\end{align}
where we have defined $\phi_{(1,1)} = 2(\eta_{\uparrow}-\eta_{\downarrow})+\phi_{0}$. Next, multiplying the first term of Eq. $\eqref{155}$ with the second term of Eq. $\eqref{157}$ give:
\begin{align}
-\int_{-\infty}^{\infty} v^{2}dt_1dt_2\tilde{\gamma}_{\uparrow}\tilde{\gamma}_{{\downarrow}}^{*}&\mathcal{I}_{1}^{(\uparrow)}(x_{1},t_{1};x_{1}^{(0)},t_{1}^{(0)})\mathcal{I}_{2}^{(\uparrow)}(x_{2},t_{2};x_{2}^{(0)},t_{2}^{(0)})
\mathcal{J}_{1}^{(\downarrow)*}(x_{2},t_{2};x_{1}^{(0)},t_{1}^{(0)})\mathcal{J}_{2}^{(\downarrow)*}(x_{1},t_{1};x_{2}^{(0)},t_{2}^{(0)}) \label{163}\\
=&-|\tilde{\gamma}_{\uparrow}\tilde{\gamma}_{\downarrow}|e^{2i(\eta_{\uparrow}-\eta_{\downarrow})+i\phi_{0}}\int_{-\infty}^{\infty} dk|\phi(k)|^{2}\mathcal{R}_{k}^{*}\mathcal{T}_{k}e^{ik\big[x_{0}^{(2)}+vt_{0}^{(2)}+x_{0}^{(1)}-vt_{0}^{(1)}\big]}\notag \\
&\hspace{3cm} \times \int_{-\infty}^{\infty} dk'|\phi(k')|^{2}\mathcal{R}_{k'}^{*}\mathcal{T}_{k'}e^{ik'\big[-x_{0}^{(2)}-vt_{0}^{(2)}-x_{0}^{(1)}+vt_{0}^{(1)}\big]} \,.\notag
\end{align}
Next, by using the fact that the $S_{\text{eff}}$ (effective scattering matrix of the scattering in the absence of the qubit) is unitary, we have $\mathcal{R}_{k}\mathcal{T}_{k}^{*} = -\mathcal{R}_{k}^{*}\mathcal{T}_{k}$ and this implies,
\begin{align}
-\int_{-\infty}^{\infty} v^{2}dt_1dt_2\tilde{\gamma}_{\uparrow}\tilde{\gamma}_{{\downarrow}}^{*} & \mathcal{I}_{1}^{(\uparrow)}(x_{1},t_{1};x_{1}^{(0)},t_{1}^{(0)})\mathcal{I}_{2}^{(\uparrow)}(x_{2},t_{2};x_{2}^{(0)},t_{2}^{(0)})
\mathcal{J}_{1}^{(\downarrow)*}(x_{2},t_{2};x_{1}^{(0)},t_{1}^{(0)})\mathcal{J}_{2}^{(\downarrow)*}(x_{1},t_{1};x_{2}^{(0)},t_{2}^{(0)}) \label{163}\\
= & |\tilde{\gamma}_{\uparrow}\tilde{\gamma}_{\downarrow}|e^{2i(\eta_{\uparrow}-\eta_{\downarrow})+i\phi_{0}}\Bigg{|}\int_{-\infty}^{\infty} dk|\phi(k)|^{2}\mathcal{R}_{k}^{*}\mathcal{T}_{k}e^{ik\big[x_{0}^{(2)}+vt_{0}^{(2)}+x_{0}^{(1)}-vt_{0}^{(1)}\big]}\Bigg{|}^{2} \,,\\
= & -e^{i\phi_{(1,1)}}|\tilde{\gamma}_{\uparrow}\tilde{\gamma}_{\downarrow}|\chi(x_{1}^{(0)},t_{1}^{(0)};x_{2}^{(0)},t_{2}^{(0)})\,,\notag
\end{align}
where we have defined, see Appendix~\ref{app:scattering},
\begin{equation}
\chi(x_{1}^{(0)},t_{1}^{(0)};x_{2}^{(0)},t_{2}^{(0)}) = -\Bigg{|}\int_{-\infty}^{\infty} dk|\phi(k)|^{2}\mathcal{R}_{k}^{*}\mathcal{T}_{k}e^{ik\big[x_{0}^{(2)}+vt_{0}^{(2)}+x_{0}^{(1)}-vt_{0}^{(1)}\big]}\Bigg{|}^{2}\,.
\end{equation}
Similarly, multiplying the second term of Eq. $\eqref{155}$ with the first term
of Eq. $\eqref{157}$ gives:
\begin{align}
-\int_{-\infty}^{\infty} v^{2}dt_1dt_2\tilde{\gamma}_{\uparrow}\tilde{\gamma}_{{\downarrow}}^{*}&\mathcal{J}_{1}^{(\uparrow)}(x_{2},t_{2};x_{1}^{(0)},t_{1}^{(0)})\mathcal{J}_{2}^{(\uparrow)}(x_{1},t_{1};x_{2}^{(0)},t_{2}^{(0)})\mathcal{I}_{1}^{(\downarrow)*}(x_{1},t_{1};x_{1}^{(0)},t_{1}^{(0)})\mathcal{I}_{2}^{(\downarrow)*}(x_{2},t_{2};x_{2}^{(0)},t_{2}^{(0)})\\&
=-e^{i\phi_{(1,1)}}|\tilde{\gamma}_{\uparrow}\tilde{\gamma}_{\downarrow}|\chi(x_{1}^{(0)},t_{1}^{(0)};x_{2}^{(0)},t_{2}^{(0)})\,.\label{166}
\end{align}
Next, we need to add the obtained expression with the complex conjugation partner to obtain,
\begin{align}\label{eq:interfterms}
-2|\tilde{\gamma}_{\uparrow}\tilde{\gamma}_{\downarrow}|\chi(x_{1}^{(0)},t_{1}^{(0)};x_{2}^{(0)},t_{2}^{(0)})\big{[} e^{i\phi_{(1,1)}} + e^{-i\phi_{(1,1)}} \big{]} = -4\cos\phi_{(1,1)}|\tilde{\gamma}_{\uparrow}\tilde{\gamma}_{\downarrow}|\chi(x_{1}^{(0)},t_{1}^{(0)};x_{2}^{(0)},t_{2}^{(0)}) \,.
\end{align}
Hence, compiling all the equations, Eq.~\eqref{eq:directterms1}, Eq.~\eqref{eq:directterms2}, and Eq.~\eqref{eq:interfterms}, we obtain the two particle probability in an extended collider which is coupled to a two-level impurity (qubit) upto normalization as follows,
\begin{align}
P(11;|\psi\rangle)_{\text{F}}& \propto \mathcal{C}\cdot\mathcal{P}(11)_{\text{F}}+2|\tilde{\gamma_{\uparrow}}\tilde{\gamma_{\downarrow}}|\cos[\phi_{(1,1)}]\{\mathcal{P}^{2}(1\to1)+\mathcal{P}^{2}(1\to2)-2\chi(x_{1}^{(0)},t_{1}^{(0)};x_{2}^{(0)},t_{2}^{(0)})\} \,,\nonumber \\&
=\{\mathcal{C}+2|\tilde{\gamma}_{\uparrow}\tilde{\gamma}_{\downarrow}|\cos[\phi_{(11)}]\}\mathcal{P}(11)_{\text{F}}\label{170}\,,
\end{align}
where $\mathcal{P}(11)_{\text{F}}$ corresponds to the two particle probability (for fermions) in the case extended collider is not coupled to the qubit (or disentangled limit) and we defined $|\tilde{\gamma}_{\uparrow}|^{2}+|\tilde{\gamma}_{\downarrow}|^{2}=\mathcal{C}$.

\subsection{Calculating the Normalization $\mathcal{N}$}
We found that the two particle probability for the event where each of the detector receives one particle in an extended collider coupled to a qubit is proportional to probability in the case where there is no coupling with the qubit. Next step is to evaluate the proportionality constant, $\mathcal{N}$. Therefore, it becomes necessary to evaluate the probability for the event where single detector receives both the particles. Let $P(20;|\psi\rangle)_{\text{F}}$ and $P(02;|\psi\rangle)_{\text{F}}$ be the probabilities of the event where both the particles (for fermions) either reach detector $D_{1}$ or $D_{2}$. In the following, we shall focus on the event where both the particles reach detector $D_{1}$. In this case, we take the final state as,
\begin{align}
|f\rangle=\frac{\phi^{\dagger}_{d_{1}}(x_{1},t_{2})\phi^{\dagger}_{d_{1}}(x_{1},t_{1})|\Omega\rangle}{\sqrt{2}}. 
\end{align}
Then the probability is given as,
\begin{align}
    P(20;|\psi\rangle)_{\text{F}}\propto \mathcal{P}(20;|\psi\rangle)_{\text{F}} = \int_{-\infty}^{\infty} v^{2}dt_1dt_2 \big| \tilde{\gamma}_{\uparrow} A_{\uparrow}(20) + \tilde{\gamma}_{\downarrow} A_{\downarrow}(20) \big|^{2} \,,\label{287}
\end{align}
where,
\begin{align}
    &A_{\uparrow}(20) = \frac{1}{\sqrt{2}}\big[\mathcal{I}_{1}^{(\uparrow)}(x_{1},t_{2};x_{1}^{(0)},t_{1}^{(0)})\mathcal{J}_{2}^{(\uparrow)}(x_{1},t_{1};x_{2}^{(0)},t_{2}^{(0)}) -  \mathcal{I}_{1}^{(\uparrow)}(x_{1},t_{1};x_{1}^{(0)},t_{1}^{(0)})\mathcal{J}_{2}^{(\uparrow)}(x_{1},t_{2};x_{2}^{(0)},t_{2}^{(0)})\big]\,,\\
    &A_{\downarrow}(20) = \frac{1}{\sqrt{2}}\big[\mathcal{I}_{1}^{(\downarrow)}(x_{1},t_{2};x_{1}^{(0)},t_{1}^{(0)})\mathcal{J}_{2}^{(\downarrow)}(x_{1},t_{1};x_{2}^{(0)},t_{2}^{(0)}) -  \mathcal{I}_{1}^{(\downarrow)}(x_{1},t_{1};x_{1}^{(0)},t_{1}^{(0)})\mathcal{J}_{2}^{(\downarrow)}(x_{1},t_{2};x_{2}^{(0)},t_{2}^{(0)})\big]\,.
\end{align}
Next, we expand the expression for $\mathcal{P}(20;|\psi\rangle)_{\text{F}}$ as follows,
\begin{align}
    \mathcal{P}(20;|\psi\rangle)_{\text{F}} &= \int_{-\infty}^{\infty} v^{2}dt_1dt_2 \big| \tilde{\gamma}_{\uparrow} A_{\uparrow}(20) + \tilde{\gamma}_{\downarrow} A_{\downarrow}(20) \big|^{2} \,,\\
    &= \int_{-\infty}^{\infty} v^{2}dt_1dt_2 \Big\{ \big| \tilde{\gamma}_{\uparrow} A_{\uparrow}(20) \big|^{2} + \big|  \tilde{\gamma}_{\downarrow} A_{\downarrow}(20) \big|^{2}+\tilde{\gamma}_{\uparrow} A_{\uparrow}(20)\tilde{\gamma}_{\downarrow}^{*} A_{\downarrow}^{*}(20)  + \tilde{\gamma}_{\uparrow}^{*} A_{\uparrow}^{*}(20)\tilde{\gamma}_{\downarrow} A_{\downarrow}(20) \Big\} \,.
\end{align}
The first two terms are easily simplified, and we get,
\begin{align}
    \int_{-\infty}^{\infty} v^{2}dt_1dt_2 \Big\{ \big| \tilde{\gamma}_{\uparrow} A_{\uparrow}(20) \big|^{2} + \big|  \tilde{\gamma}_{\downarrow} A_{\downarrow}(20) \big|^{2} \Big\}  =& \big{(} |\tilde{\gamma}_{\uparrow}|^{2} + |\tilde{\gamma}_{\downarrow}|^{2} \big{)}\cdot \big{[} \mathcal{P}(1\to1)\mathcal{P}(1\to2)+\chi(x_1^{(0)},t_1^{(0)};x_2^{(0)},t_2^{(0)}) \big{]} \,, \\ 
    =&\, \mathcal{C} \cdot \mathcal{P}(20)_{\text{F}} \,,
\end{align}
where we have defined $|\tilde{\gamma}_{\uparrow}|^{2} + |\tilde{\gamma}_{\downarrow}|^{2} = \mathcal{C}$ and $\mathcal{P}(20)_{\text{F}} = \mathcal{P}(1\to1)\mathcal{P}(1\to2)+\chi(x_1^{(0)},t_1^{(0)};x_2^{(0)},t_2^{(0)})$ is the two-particle probability with no coupling between the qubit and extended collider. The third and fourth terms are similarly calculated and we get,
\begin{align}
    \int_{-\infty}^{\infty} v^{2}dt_1dt_2 \Big\{ \tilde{\gamma}_{\uparrow} A_{\uparrow}(20)\tilde{\gamma}_{\downarrow}^{*} A_{\downarrow}^{*}(20)  + \tilde{\gamma}_{\uparrow}^{*} A_{\uparrow}^{*}(20)\tilde{\gamma}_{\downarrow} A_{\downarrow}(20) \Big\} = &|\tilde{\gamma}_{\uparrow}\tilde{\gamma}_{\downarrow}|\cdot \big{(}e^{i\phi_{(2,0)}} + e^{-i\phi_{(2,0)}} \big{)} \\
    &\cdot \big{[}\mathcal{P}(1\to1)\mathcal{P}(1\to2)+ \chi(x_1^{(0)},t_1^{(0)};x_2^{(0)},t_2^{(0)})\big{]}\,,\\
    =& 2 |\tilde{\gamma}_{\uparrow}\tilde{\gamma}_{\downarrow}| \cos \phi_{(2,0)} P(20)_{\text{F}} \,,
\end{align}
where we have defined,
\begin{align}
    \phi_{(2,0)} =\phi_0+\theta_{\uparrow}-\theta_{\downarrow}+\eta_{\uparrow}-\eta_{\downarrow}\,.
\end{align}
From here, we obtain the (non-normalized) probability as follows,
\begin{align}
    \mathcal{P}(20;|\psi\rangle)_{\text{F}} = \big{[}\mathcal{C} + 2 |\tilde{\gamma}_{\uparrow}\tilde{\gamma}_{\downarrow}| \cos \phi_{(2,0)} \big{]}\cdot \mathcal{P}(20)_{\text{F}}\,.
\end{align}
A similar set of calculation for the probability where the detector $D_{2}$ receives both the particles gives,
\begin{align}
    \mathcal{P}(02;|\psi\rangle)_{\text{F}} = \big{[}\mathcal{C} + 2 |\tilde{\gamma}_{\uparrow}\tilde{\gamma}_{\downarrow}| \cos \phi_{(0,2)} \big{]}\cdot \mathcal{P}(02)_{\text{F}}\,,
\end{align}
where we have defined,
\begin{align}
    \phi_{(0,2)} = 3\eta_{\uparrow}-3\eta_{\downarrow}-\theta_{\uparrow}+\theta_{\downarrow}+\phi_{0}\,.
\end{align}
In order to obtain the normalized probability, we defined the normalization constant as follows,
\begin{align}
\mathcal{N} &= \mathcal{P}(11;|\psi\rangle)_{\text{F}} + \mathcal{P}(20;|\psi\rangle)_{\text{F}} +\mathcal{P}(02;|\psi\rangle)_{\text{F}}\,,\\ 
&=\mathcal{C}+2|\tilde{\gamma}_{\uparrow}\tilde{\gamma}_{\downarrow}|\big\{ \cos\phi_{(1,1)}\mathcal{P}(11)_{\text{F}}+ \cos\phi_{(0,2)}\mathcal{P}(02)_{\text{F}} + 
\cos\phi_{(2,0)}\mathcal{P}(20)_{\text{F}}\big\} \,,\\
&= \mathcal{C}+2|\tilde{\gamma}_{\uparrow}\tilde{\gamma}_{\downarrow}|\cos\phi_{(1,1)}\big\{ \mathcal{P}(11)_{\text{F}} + 2\mathcal{P}(20)_{\text{F}}\cos\phi_{1}\big\} \,,
\end{align}
where we have used an additional identity,
\begin{align}
    \cos \phi_{(2,0)} + \cos \phi_{(0,2)} = 2\cos \phi_{1} \cos \phi_{(1,1)}\,.
\end{align}
Hence, our normalized probabilities are given as follows,
\begin{align}
    \label{eq:unnormalized}
    P(11;|\psi\rangle)_{\text{F}} &= \frac{1}{\mathcal{N}}\mathcal{P}(11;|\psi\rangle)_{\text{F}}\,; \hspace{0.5cm} P(20;|\psi\rangle)_{\text{F}} = \frac{1}{\mathcal{N}}\mathcal{P}(20;|\psi\rangle)_{\text{F}}\,; \hspace{0.5cm}
    P(02;|\psi\rangle)_{\text{F}} = \frac{1}{\mathcal{N}}\mathcal{P}(02;|\psi\rangle)_{\text{F}}\,,
\end{align}
and thereby, we have,
\begin{align}
    P(11;|\psi\rangle)_{\text{F}} + P(20;|\psi\rangle)_{\text{F}} + P(02;|\psi\rangle)_{\text{F}} = 1\,.
\end{align}

\subsection{Disentanglement Condition for a two-particle incoming state in the presence of a Geometric Phase}
In this section, we shall extract the  disentanglement condition for an extended collider considered here. The effective scattering matrix is given as,
\begin{align}\label{eq:seffqubit}
S_{\text{eff}}^{(m)}&=\left(\begin{array}{cc}\mathcal{T}_{k}e^{i\eta_{m}} & \mathcal{R}_{k}e^{i\theta_{m}}\\
\mathcal{R}_{k}e^{i(2\eta_{m}-\theta_{m})} & \mathcal{T}_{k}e^{i\eta_{m}}
\end{array}\right)=\left(\begin{array}{cc}t_{k,m} & r'_{k,m}\\
r_{k,m} & t'_{k,m}
\end{array}\right) \,.
\end{align}
For an extended collider the reflection and transmission amplitudes $\mathcal{R}_k$ and $\mathcal{T}_k$ are functions of $k$. Here also, we will try to calculate the quantity $\mathcal{E}=\langle i|S_{\uparrow}^{+}S_{\downarrow}|i\rangle$, the state $|i\rangle$ is an incoming two-particle state. Explicit expression for the quantity $\mathcal{E}$, with two incoming particles, one from each source is given as follows,
\begin{align}
\mathcal{E} =& \int_{-\infty}^{\infty} dkdk'dk_1dk_2e^{ik(-x_{1}^{(0)}+vt_1^{(0)})}e^{ik'(x_{2}^{(0)}+vt_2^{(0)})}e^{-ik_{2}(x_{2}^{(0)}+vt_2^{(0)})}e^{-ik_{1}(-x_{1}^{(0)}+vt_1^{(0)})}\phi^{*}(k)\phi^{*}(k')\phi(k_2)\phi(k_1) \nonumber
\notag\\
&\times\langle \Omega|[{r'}^{*}_{k_{2\uparrow}}c_{d_{1}k_{2}}+{t'}^{*}_{k_{2}\uparrow}c_{d_{2}k_{2}}][t^{*}_{k_1\uparrow}c_{d_{1}k_{1}}+{r}^{*}_{k_{1}\uparrow}c_{d_{2}k_{1}}][t_{k\downarrow}c^{\dagger}_{d_{1}k}+{r}_{k\downarrow}c^{\dagger}_{d_{2}k}][{r'}_{k'\downarrow}c^{\dagger}_{d_{1}k'}+{t'}_{k'\downarrow}c^{\dagger}_{d_{2}k'}]|\Omega\rangle\,.
\end{align}
From here, it is clear from here that there are six non-zero contribution given as follows,
\begin{align}
A&: t_{k\downarrow}{r'}_{k'\downarrow}{r'}^{*}_{k_{2\uparrow}}t^{*}_{k_{1}\uparrow}\langle \Omega|c_{d_{1}k_{2}}c_{d_{1}k_{1}}c^{\dagger}_{d_{1}k}c^{\dagger}_{d_{1}k'}|\Omega\rangle\,,\\
B&: {t'}^{*}_{k_{2}\uparrow}{r}^{*}_{k_1\uparrow}{r}_{k\downarrow}{t'}_{k'\downarrow}\langle \Omega|c_{d_{2}k_{2}}c_{d_{2}k_{1}}c^{\dagger}_{d_{2}k}c^{\dagger}_{d_{2}k'}|\Omega\rangle\,,\\
C&:{r'}^{*}_{k_{2}\uparrow}r^{*}_{k_1\uparrow}t_{k\downarrow}{t'}_{k'\downarrow}\langle \Omega|c_{d_{1}k_{2}}c_{d_{2}k_{1}}c^{\dagger}_{d_{1}k}c^{\dagger}_{d_{2}k'}|\Omega\rangle\,,\\
D&: {t'}^{*}_{k_{2}\uparrow}t^{*}_{k_1\uparrow}t_{k\downarrow}{t'}_{k'\downarrow}\langle \Omega|c_{d_{2}k_{2}}c_{d_{1}k_{1}}c^{\dagger}_{d_{1}k}c^{\dagger}_{d_{2}k'}|\Omega\rangle\,,\\
E&: {t'}^{*}_{k_{2}\uparrow}t^{*}_{k_1\uparrow}{r}_{k\downarrow}{r'}_{k'\downarrow}\langle \Omega|c_{d_{2}k_{2}}c_{d_{1}k_{1}}c^{\dagger}_{d_{2}k}c^{\dagger}_{d_{1}k'}|\Omega\rangle\,,\\
F&: {r'}^{*}_{k_{2}\uparrow}{r}^{*}_{k_1\uparrow}{r}_{k\downarrow}{r'}_{k'\downarrow}\langle \Omega|c_{d_{1}k_{2}}c_{d_{2}k_{1}}c^{\dagger}_{d_{2}k}c^{\dagger}_{d_{1}k'}|\Omega\rangle\,.
\end{align}
We will explicitly calculate term $A$ using Wick's theorem. There will be two contractions and give a pair of $\delta-$functions. Integrating over the momentum variables, we get,
\begin{align}
A  =& \int_{-\infty}^{\infty} dkdk'dk_1dk_2  e^{ik(-x_{1}^{(0)}+vt_1^{(0)})} e^{ik'(x_{2}^{(0)}+vt_2^{(0)})} e^{-ik_{2}(x_{2}^{(0)}+vt_2^{(0)})}e^{-ik_{1}(-x_{1}^{(0)}+vt_1^{(0)})} \phi^{*}(k)\phi^{*}(k')\phi(k_2)\phi(k_1) \nonumber \\
& \times t_{k\downarrow}{r'}_{k'\downarrow}{r'}^{*}_{k_{2\uparrow}}t^{*}_{k_{1}\uparrow}[\delta(k_1-k)\delta(k_2-k')-\delta(k_2-k)\delta(k_1-k')] \notag \,,\\
= & e^{i(\eta_{\downarrow}-\eta_{\uparrow}+\theta_{\downarrow}-\theta_{\uparrow})}[\mathcal{P}(1\to1)\mathcal{P}(1\to2) + \chi(x_{1}^{(0)},t_{1}^{(0)};x_{2}^{(0)},t_{2}^{(0)})]\,,
\end{align}
and a similar exercise gives all the other terms as follows,\begin{align}
    B &= e^{i[3(\eta_{\downarrow}-\eta_{\uparrow})-\theta_{\downarrow}+\theta_{\uparrow}]}[\mathcal{P}(1\to1)\mathcal{P}(1\to2) +\chi(x_{1}^{(0)},t_{1}^{(0)};x_{2}^{(0)},t_{2}^{(0)})]\,,\\
    C &= -e^{2i(\eta_{\downarrow}-\eta_{\uparrow})}\chi(x_{1}^{(0)},t_{1}^{(0)};x_{2}^{(0)},t_{2}^{(0)}) \,,\\
    D &= e^{2i(\eta_{\downarrow}-\eta_{\uparrow})}\mathcal{P}(1\to1)^{2}\,, \\
    E &= -e^{2i(\eta_{\downarrow}-\eta_{\uparrow})}\chi(x_{1}^{(0)},t_{1}^{(0)};x_{2}^{(0)},t_{2}^{(0)}) \,,\\
    F &= e^{2i(\eta_{\downarrow}-\eta_{\uparrow})}\mathcal{P}(1\to2)^{2}\,.
\end{align}
Adding up the contributions from all the terms, we get, 
\begin{align}
\mathcal{E} &= A+B+C+D+E+F\,, \\
&= e^{2i(\eta_{\downarrow}-\eta_{\uparrow})}[2\cos(\eta_{\downarrow}-\eta_{\uparrow}+\theta_{\uparrow}-\theta_{\downarrow})\{\mathcal{P}(1\to1)\mathcal{P}(1\to2)+\chi(x_{1}^{(0)},t_{1}^{(0)};x_{2}^{(0)},t_{2}^{(0)})\}-2\chi(x_{1}^{(0)},t_{1}^{(0)};x_{2}^{(0)},t_{2}^{(0)})\\
&\quad +\mathcal{P}(1\to1)^{2}+\mathcal{P}(1\to2)^{2}]\,.
\end{align}
Thus, the quantity $\mathcal{E}$ is not a pure phase if $\eta_{\downarrow}-\eta_{\uparrow}+\theta_{\uparrow}-\theta_{\downarrow}=-\phi_{1}\neq0$. These
are precisely the conditions under which the final state of the qubit
will be entangled to the scattering particles and for disentangling, we must have $\mathcal{E}$ to be a pure phase.

\section{Benchmark for extracting the mutual statistics}\label{app:benchmark_qubit}

In this section, we are going to explicitly show that the correct
benchmark for extracting the mutual statistics is $\mathcal{B}_{4}$.
Let us start with the expression for the two particle probability
of receiving one particle in each detector, $P(11;|\psi\rangle)_{F}$
with qubit in the state $|\psi\rangle$,
\begin{equation}
P(11;|\psi\rangle)_{\text{F}}=\mathcal{S}(11;|\psi\rangle)\mathcal{P}(11)_{\text{F}}\ \text{with}\ \mathcal{S}(11;|\psi\rangle)=\frac{\mathcal{C}+2|\tilde{\gamma}_{\uparrow}\tilde{\gamma}_{\downarrow}|\cos[\phi_{(11)}]}{\mathcal{C}+2|\tilde{\gamma}_{\uparrow}\tilde{\gamma}_{\downarrow}|\cos\phi_{(11)}\big\{\mathcal{P}(11)_{\text{F}}+2\mathcal{P}(20)_{\text{F}}\cos\phi_{1}\big\}}\,.
\end{equation}
Here the multiplicative factor $\mathcal{S}(11;|\psi\rangle)$ modifies
the two particle probability in the presence of a qubit and encodes
all the information of the qubit. Next, for simplicity we can define the following new variable such that we can get rid of the constant term $\mathcal{C}$ in the numerator and denominator,
\begin{align}
    \bar{\gamma}_{\uparrow} = \frac{\tilde{\gamma}_{\uparrow}}{\sqrt{\mathcal{C}}}\,,\qquad \bar{\gamma}_{\downarrow} = \frac{\tilde{\gamma}_{\downarrow}}{\sqrt{\mathcal{C}}} \quad   \implies \quad |\bar{\gamma}_{\uparrow}|^{2} + |\bar{\gamma}_{\downarrow}|^{2} = 1\,,
\end{align}
and as a result we are going to have the following,
\begin{align}
    \mathcal{S}(11;|\psi\rangle)=\frac{1+2|\bar{\gamma}_{\uparrow}\bar{\gamma}_{\downarrow}|\cos[\phi_{(11)}]}{1+2|\bar{\gamma}_{\uparrow}\bar{\gamma}_{\downarrow}|\cos\phi_{(11)}\big\{\mathcal{P}(11)_{\text{F}}+2\mathcal{P}(20)_{\text{F}}\cos\phi_{1}\big\}}\,.
\end{align}
In the expression, $\mathcal{P}(11)_{\text{F}}$
corresponds to the probability of receiving one particle in each detector
in the absence of any coupling with the qubit. The correct benchmark
for extracting the mutual statistics out of all the four benchmarks
($\mathcal{B}_{1}$, $\mathcal{B}_{2}$, $\mathcal{B}_{3}$, and $\mathcal{B}_{4}$)
is $\mathcal{B}_{4}$, and is given as follows,
\begin{equation}
\mathcal{B}_{4}=\frac{1+2|\bar{\gamma}_{\uparrow}\bar{\gamma}_{\downarrow}|\cos\phi_{(11)}}{1+2|\bar{\gamma}_{\uparrow}\bar{\gamma}_{\downarrow}|\cos\phi_{(11)}\big\{\mathcal{P}(11)_{\text{CW}}+2\mathcal{P}(20)_{\text{CW}}\cos\phi_{1}\big\}}\cdot\mathcal{P}(11)_{\text{CW}}\,.
\end{equation}
From the expression it is clear that the benchmark $\mathcal{B}_{4}$
is the corresponding probability with two sources for classical waves
in the presence of a qubit. This benchmark can be obtained from the
fermion probability in a limiting case where the emission times of
the incoming particles, $t_{1}^{(0)}\,,t_{2}^{(0)}$ are such that
$v|t_{1}^{(0)}-t_{2}^{(0)}|\gg \sigma$, where $\sigma$ is the width of
a wave packet and $v$ is the velocity. With well separated wave packets,
the contribution from the interference between the direct and the exchange
terms is small, but includes all the self-interference contributions
to the probability~\cite{td98-5ltj} and see the discussion in Appendix~\ref{app:scattering}. Therefore the benchmark $\mathcal{B}_{4}$ successfully
removes all the self-interference contribution and the result (interference
contribution between direct and exchange terms) is sensitive to the
mutual statistics, 
\begin{align}
P(11;|\psi\rangle)_{\text{F}}-\mathcal{B}_{4} & =\frac{1+2|\bar{\gamma}_{\uparrow}\bar{\gamma}_{\downarrow}|\cos\phi_{(11)}}{1+2|\bar{\gamma}_{\uparrow}\bar{\gamma}_{\downarrow}|\cos\phi_{(11)}\big\{\mathcal{P}(11)_{\text{F}}+2\mathcal{P}(20)_{\text{F}}\cos\phi_{1}\big\}}\cdot\mathcal{P}(11)_{\text{F}}\\
 & \quad-\frac{1+2|\bar{\gamma}_{\uparrow}\bar{\gamma}_{\downarrow}|\cos\phi_{(11)}}{1+2|\bar{\gamma}_{\uparrow}\bar{\gamma}_{\downarrow}|\cos\phi_{(11)}\big\{\mathcal{P}(11)_{\text{CW}}+2\mathcal{P}(20)_{\text{CW}}\cos\phi_{1}\big\}}\cdot\mathcal{P}(11)_{\text{CW}}\,,\\
 & =\frac{1+2|\bar{\gamma}_{\uparrow}\bar{\gamma}_{\downarrow}|\cos\phi_{(11)}}{\Big[1+2|\bar{\gamma}_{\uparrow}\bar{\gamma}_{\downarrow}|\cos\phi_{(11)}\big\{\mathcal{P}(11)_{\text{F}}+2\mathcal{P}(20)_{\text{F}}\cos\phi_{1}\big\}\Big]}\\
 & \quad\times\frac{1}{\Big[1+2|\bar{\gamma}_{\uparrow}\bar{\gamma}_{\downarrow}|\cos\phi_{(11)}\big\{\mathcal{P}(11)_{\text{CW}}+2\mathcal{P}(20)_{\text{CW}}\cos\phi_{1}\big\}\Big]}\\
 & \quad\times\Big[\mathcal{P}(11)_{\text{F}}+2|\bar{\gamma}_{\uparrow}\bar{\gamma}_{\downarrow}|\cos\phi_{(11)}\mathcal{P}(11)_{\text{F}}\big\{\mathcal{P}(11)_{\text{CW}}+2\mathcal{P}(20)_{\text{CW}}\cos\phi_{1}\big\}\\
 & \ \hspace{1cm}-\mathcal{P}(11)_{\text{CW}}-2|\bar{\gamma}_{\uparrow}\bar{\gamma}_{\downarrow}|\cos\phi_{(11)}\mathcal{P}(11)_{\text{CW}}\big\{\mathcal{P}(11)_{\text{F}}+2\mathcal{P}(20)_{\text{F}}\cos\phi_{1}\big\}\Big]\,.
\end{align}
Let us focus on the numerator part that is a coefficient of $\{1+2|\bar{\gamma}_{\uparrow}\bar{\gamma}_{\downarrow}|\cos\phi_{(11)}\}$,
\begin{align}
\text{Numerator} & =\mathcal{P}(11)_{\text{F}}-\mathcal{P}(11)_{\text{CW}}\\
 & \quad+2|\bar{\gamma}_{\uparrow}\bar{\gamma}_{\downarrow}|\cos\phi_{(11)}\Big[\cancel{\mathcal{P}(11)_{\text{F}}\mathcal{P}(11)_{\text{CW}}}+2\mathcal{P}(11)_{\text{F}}\mathcal{P}(20)_{\text{CW}}\cos\phi_{1}\\
 & \ \hspace{3cm}-\cancel{\mathcal{P}(11)_{\text{CW}}\mathcal{P}(11)_{\text{F}}}-2\mathcal{P}(11)_{\text{CW}}\mathcal{P}(20)_{\text{F}}\cos\phi_{1}\big\}\Big]\,,\\
 & =\mathcal{P}(11)_{\text{F}}-\mathcal{P}(11)_{\text{CW}}+4|\bar{\gamma}_{\uparrow}\bar{\gamma}_{\downarrow}|\cos\phi_{(11)}\cos\phi_{1}\big[\mathcal{P}(11)_{\text{F}}\mathcal{P}(20)_{\text{CW}}-\mathcal{P}(11)_{\text{CW}}\mathcal{P}(20)_{\text{F}}\big]\,,\\
 & =\mathcal{P}(11)_{\text{F}}-\mathcal{P}(11)_{\text{CW}}
+4|\bar{\gamma}_{\uparrow}\bar{\gamma}_{\downarrow}|\cos\phi_{(11)}\cos\phi_{1}\bigg[\mathcal{P}(11)_{\text{F}}\frac{1-\mathcal{P}(11)_{\text{CW}}}{2}-\mathcal{P}(11)_{\text{CW}}\frac{1-\mathcal{P}(11)_{\text{F}}}{2}\bigg]\,,\\
 & =\mathcal{P}(11)_{\text{F}}-\mathcal{P}(11)_{\text{CW}}+2|\bar{\gamma}_{\uparrow}\bar{\gamma}_{\downarrow}|\cos\phi_{(11)}\cos\phi_{1}\big[\mathcal{P}(11)_{\text{F}}-\mathcal{P}(11)_{\text{CW}}\big]\,,\\
 & =\big[\mathcal{P}(11)_{\text{F}}-\mathcal{P}(11)_{\text{CW}}\big]\cdot\Big\{1+2|\bar{\gamma}_{\uparrow}\bar{\gamma}_{\downarrow}|\cos\phi_{(11)}\cos\phi_{1}\Big\}\,.
\end{align}
Hence, we obtain,
\begin{align}
P(11;|\psi\rangle)_{\text{F}}-\mathcal{B}_{4} & =\frac{\big[\mathcal{P}(11)_{\text{F}}-\mathcal{P}(11)_{\text{CW}}\big]}{\Big[1+2|\bar{\gamma}_{\uparrow}\bar{\gamma}_{\downarrow}|\cos\phi_{(11)}\big\{\mathcal{P}(11)_{\text{F}}+2\mathcal{P}(20)_{\text{F}}\cos\phi_{1}\big\}\Big]}\\
 & \quad\times\frac{\Big\{1+2|\bar{\gamma}_{\uparrow}\bar{\gamma}_{\downarrow}|\cos\phi_{(11)}\cos\phi_{1}\Big\}\Big\{1+2|\bar{\gamma}_{\uparrow}\bar{\gamma}_{\downarrow}|\cos\phi_{(11)}\Big\}}{\Big[1+2|\bar{\gamma}_{\uparrow}\bar{\gamma}_{\downarrow}|\cos\phi_{(11)}\big\{\mathcal{P}(11)_{\text{CW}}+2\mathcal{P}(20)_{\text{CW}}\cos\phi_{1}\big\}\Big]}\,,\\
 & = \mathcal{K}_{\text{F}} \cdot \big[\mathcal{P}(11)_{\text{F}}-\mathcal{P}(11)_{\text{CW}}\big]\,,
\end{align}
and hence we arrive at Eq.~\eqref{eq:correct_benchmark} of the main text, where,
\begin{align}
    \mathcal{K}_{\text{F}} = \frac{\Big\{1+2|\bar{\gamma}_{\uparrow}\bar{\gamma}_{\downarrow}|\cos\phi_{(11)}\cos\phi_{1}\Big\}\Big\{1+2|\bar{\gamma}_{\uparrow}\bar{\gamma}_{\downarrow}|\cos\phi_{(11)}\Big\}}{\Big[1+2|\bar{\gamma}_{\uparrow}\bar{\gamma}_{\downarrow}|\cos\phi_{(11)}\big\{\mathcal{P}(11)_{\text{CW}}+2\mathcal{P}(20)_{\text{CW}}\cos\phi_{1}\big\}\Big]\Big[1+2|\bar{\gamma}_{\uparrow}\bar{\gamma}_{\downarrow}|\cos\phi_{(11)}\big\{\mathcal{P}(11)_{\text{F}}+2\mathcal{P}(20)_{\text{F}}\cos\phi_{1}\big\}\Big]}\,,
\end{align}
which is Eq.~\eqref{eq:full_mathcalk} in the main text. Next, by using Eq.~\eqref{eq:prod_coeff_post_selection}, we have,
\begin{align}
    |\bar{\gamma}_{\uparrow}\bar{\gamma}_{\downarrow}|\leq\frac{1}{2} \Rightarrow -1 \leq 2|\bar{\gamma}_{\uparrow}\bar{\gamma}_{\downarrow}|\cos\phi_{(11)}\cos\phi_{1} \leq 1\,,
\end{align}
because of the fact that the cosine function takes value between $-1$ and $1$. This implies, we have,
\begin{align}
    \Big\{1+2|\bar{\gamma}_{\uparrow}\bar{\gamma}_{\downarrow}|\cos\phi_{(11)}\cos\phi_{1}\Big\} \geq 0 \,,
\end{align}
and by similar arguments, the denominator is also always positive and hence, we find $\mathcal{K}_{\text{F}} \geq 0$ and positive. The quantity $P(11;|\psi\rangle)_{\text{F}}-\mathcal{B}_{4}$, we
demonstrated is proportional to $\mathcal{P}(11)_{\text{F}}-\mathcal{P}(11)_{\text{CW}}$
and therefore, removes all the self-interference contributions and
what remains is purely statistical signature. Note that all the multiplicative
factors to $\mathcal{P}(11)_{\text{F}}-\mathcal{P}(11)_{\text{CW}}$,
are nonnegative and therefore doesn't distort the statistical
signature. For bosons, we would get, 
\begin{align}
P(11;|\psi\rangle)_{\text{B}}-\mathcal{B}_{4} & =\frac{\big[\mathcal{P}(11)_{\text{B}}-\mathcal{P}(11)_{\text{CW}}\big]}{\Big[1+2|\bar{\gamma}_{\uparrow}\bar{\gamma}_{\downarrow}|\cos\phi_{(11)}\big\{\mathcal{P}(11)_{\text{B}}+2\mathcal{P}(20)_{\text{B}}\cos\phi_{1}\big\}\Big]}\\
 & \quad\times\frac{\Big\{1+2|\bar{\gamma}_{\uparrow}\bar{\gamma}_{\downarrow}|\cos\phi_{(11)}\cos\phi_{1}\Big\}\Big\{1+2|\bar{\gamma}_{\uparrow}\bar{\gamma}_{\downarrow}|\cos\phi_{(11)}\Big\}}{\Big[1+2|\bar{\gamma}_{\uparrow}\bar{\gamma}_{\downarrow}|\cos\phi_{(11)}\big\{\mathcal{P}(11)_{\text{CW}}+2\mathcal{P}(20)_{\text{CW}}\cos\phi_{1}\big\}\Big]} \,,\\
 & = \mathcal{K}_{\text{B}} \cdot \big[\mathcal{P}(11)_{\text{B}}-\mathcal{P}(11)_{\text{CW}}\big]\,,
\end{align}
where we have,
\begin{align}
    \mathcal{K}_{\text{B}} = \frac{\Big\{1+2|\bar{\gamma}_{\uparrow}\bar{\gamma}_{\downarrow}|\cos\phi_{(11)}\cos\phi_{1}\Big\}\Big\{1+2|\bar{\gamma}_{\uparrow}\bar{\gamma}_{\downarrow}|\cos\phi_{(11)}\Big\}}{\Big[1+2|\bar{\gamma}_{\uparrow}\bar{\gamma}_{\downarrow}|\cos\phi_{(11)}\big\{\mathcal{P}(11)_{\text{CW}}+2\mathcal{P}(20)_{\text{CW}}\cos\phi_{1}\big\}\Big]\Big[1+2|\bar{\gamma}_{\uparrow}\bar{\gamma}_{\downarrow}|\cos\phi_{(11)}\big\{\mathcal{P}(11)_{\text{B}}+2\mathcal{P}(20)_{\text{B}}\cos\phi_{1}\big\}\Big]}\,,
\end{align}
 Therefore, the quantities $P(11;|\psi\rangle)_{\text{B}}-\mathcal{B}_{4}$
and $P(11;|\psi\rangle)_{\text{F}}-\mathcal{B}_{4}$ come with a
relative sign and hence the benchmark $\mathcal{B}_{4}$ successfully
determines the mutual statistics.

\section{Extraction of quantum statistics with a general coupling between the scatterer and the qubit}\label{app:gen_scat_matrix}
Let us start with a general scattering matrix that is coupled to the
qubit,
\begin{equation}\label{eq:gen_scat}
S^{(m)}_{\text{eff gen}}=\left[\begin{array}{cc}
\mathcal{T}_{k}^{(m)} & \mathcal{R'}_{k}^{(m)}\\
\mathcal{R}_{k}^{(m)} & \mathcal{T'}_{k}^{(m)}
\end{array}\right]\,,
\end{equation}
where the scattering amplitude depends on the qubit state $m\in\{\uparrow,\downarrow\}$. Note that, now we are considering a case where the effect of the quantum impurity is beyond changing the phase of the tunneling amplitudes. In a general setting, the qubit in two different state can also give rise scattering amplitudes with different magnitudes. The unitarity
of the scattering matrix, $S^{(m)\dagger}_{\text{eff gen}}S^{(m)}_{\text{eff gen}} = S^{(m)}_{\text{eff gen}}S^{(m)\dagger}_{\text{eff gen}}=\mathbb{I}$, implies the following,
\begin{align}
|\mathcal{T}_{k}^{(m)}|^{2}+|\mathcal{R}_{k}^{(m)}|^{2} & =1=|\mathcal{T'}_{k}^{(m)}|^{2}+|\mathcal{R'}_{k}^{(m)}|^{2}\,,\\
\mathcal{T}_{k}^{(m)*}\mathcal{R'}_{k}^{(m)}+\mathcal{R}_{k}^{(m)*}\mathcal{T'}_{k}^{(m)} & =0=\mathcal{T}_{k}^{(m)}\mathcal{R'}_{k}^{(m)*}+\mathcal{R}_{k}^{(m)}\mathcal{T'}_{k}^{(m)*}\,.
\end{align}
By using the scattering theory formalism (see Sec.~\ref{app:scattering}), we have the following expression for the two-particle probability for the event with two incoming particles, one from each source, and each detector receives one particle,
\begin{equation}
P(11;|\psi\rangle)_{\text{F/B}}=\frac{1}{\mathcal{N}_{\text{F/B}}}\iint^{\infty}_{-\infty}v^{2}dt_{1}dt_{2}|\tilde{\gamma}_{\uparrow}A^{\text{F/B}}_{\uparrow}(11)+\tilde{\gamma}_{\downarrow}A^{\text{F/B}}_{\downarrow}(11)|^{2}\,,
\end{equation}
where $|\psi\rangle$ is the postselected qubit state and the $\mathcal{N}_{\text{F/B}}$ is the normalization factor after postselection. The two-particle
amplitudes are given as follows,
\begin{align}
A^{\text{F/B}}_{\uparrow}(11) & =A^{\text{D}}_{\uparrow}(11)\mp A^{\text{E}}_{\uparrow}(11)\,,\qquad
A^{\text{F/B}}_{\downarrow}(11)  =A^{\text{D}}_{\downarrow}(11)\mp A^{\text{E}}_{\downarrow}(11)\,,
\end{align}
where F/B refers to fermions and bosons respectively, and these are accompanied by $\mp$ sign between the amplitudes for direct and exchange processes and are a result of mutual statistics of fermions and bosons. Note that for simplicity in the notations, we have defined the following
direct (D) and exchange (E) amplitudes with the qubit state labeled by $m\in\{\uparrow,\downarrow\}$,
\begin{align}
A^{\text{D}}_{m}(11) & =\mathcal{I}^{(m)}_{1}(x_{1},t_{1};x^{(0)}_{1},t^{(0)}_{1})\mathcal{I}^{(m)}_{2}(x_{2},t_{2};x^{(0)}_{2},t^{(0)}_{2})\,,\\
A^{\text{E}}_{m}(11) & =\mathcal{J}^{(m)}_{1}(x_{2},t_{2};x^{(0)}_{1},t^{(0)}_{1})\mathcal{J}^{(m)}_{2}(x_{1},t_{1};x^{(0)}_{2},t^{(0)}_{2})\,,
\end{align}
where the full expression can be simply obtained by using the formalism developed in Sec.~\ref{app:scattering} and using the scattering matrix, Eq.~\eqref{eq:gen_scat}. Recall that a direct process corresponds to the case where the particle from $S_{1}$ (and $S_{2}$) goes to detector $D_{1}$ (and $D_{2}$), and particle from $S_{2}$ goes to $D_{2}$, similarly exchange process corresponds to the case where particle from $S_{1}$ (and $S_{2}$) goes to detector $D_{2}$ (and $D_{1}$). Now let us expand the expression of the probability,
\begin{align}
\label{eq:p11_general}
P(11;|\psi\rangle)_{\text{F/B}} & =\frac{1}{\mathcal{N}_{\text{F/B}}}\iint^{\infty}_{-\infty}v^{2}dt_{1}dt_{2}|\tilde{\gamma}_{\uparrow}A^{\text{F/B}}_{\uparrow}(11)+\tilde{\gamma}_{\downarrow}A^{\text{F/B}}_{\downarrow}(11)|^{2}\,,\\
  =\frac{1}{\mathcal{N}_{\text{F/B}}}&\iint^{\infty}_{-\infty}v^{2}dt_{1}dt_{2}|\tilde{\gamma}_{\uparrow}A^{\text{F/B}}_{\uparrow}(11)|^{2}+\frac{1}{\mathcal{N}_\text{F/B}}\iint^{\infty}_{-\infty}v^{2}dt_{1}dt_{2}|\tilde{\gamma}_{\downarrow}A^{\text{F/B}}_{\downarrow}(11)|^{2}\notag\\
 & \quad + \frac{2}{\mathcal{N}_\text{F/B}}\text{Re}\bigg[\iint^{\infty}_{-\infty}v^{2}dt_{1}dt_{2}\tilde{\gamma}^{*}_{\uparrow}\tilde{\gamma}_{\downarrow}A^{\text{F/B}*}_{\uparrow}(11)A^{\text{F/B}}_{\downarrow}(11)\bigg]\,,\\
  =\frac{1}{\mathcal{N}_{\text{F/B}}}&\iint^{\infty}_{-\infty}v^{2}dt_{1}dt_{2}\big[|\tilde{\gamma}_{\uparrow}|^{2}|A^{\text{D}}_{\uparrow}(11)|^{2}+|\tilde{\gamma}_{\uparrow}|^{2}|A^{\text{E}}_{\uparrow}(11)|^{2}\big]\mp\frac{2}{\mathcal{N}_{\text{F/B}}}|\tilde{\gamma}_{\uparrow}|^{2}\text{Re}\bigg[\iint^{\infty}_{-\infty}v^{2}dt_{1}dt_{2}A^{\text{D}*}_{\uparrow}(11)A^{\text{E}}_{\uparrow}(11)\bigg]\notag\\
  \quad+\frac{1}{\mathcal{N}_{\text{F/B}}}&\iint^{\infty}_{-\infty}v^{2}dt_{1}dt_{2}\big[|\tilde{\gamma}_{\downarrow}|^{2}|A^{\text{D}}_{\downarrow}(11)|^{2}+|\tilde{\gamma}_{\downarrow}|^{2}|A^{\text{E}}_{\downarrow}(11)|^{2}\big]\mp\frac{2}{\mathcal{N}_{\text{F/B}}}|\tilde{\gamma}_{\downarrow}|^{2}\text{Re}\bigg[\iint^{\infty}_{-\infty}v^{2}dt_{1}dt_{2}A^{\text{D}*}_{\downarrow}(11)A^{\text{E}}_{\downarrow}(11)\bigg]\notag\\
  \quad+\frac{2}{\mathcal{N}_{\text{F/B}}}&\text{Re}\bigg[\iint^{\infty}_{-\infty}v^{2}dt_{1}dt_{2}\tilde{\gamma}^{*}_{\uparrow}\tilde{\gamma}_{\downarrow}A^{\text{D}*}_{\uparrow}(11)A^{\text{D}}_{\downarrow}(11)\bigg]+\frac{2}{\mathcal{N}_{\text{F/B}}}\text{Re}\bigg[\iint^{\infty}_{-\infty}v^{2}dt_{1}dt_{2}\tilde{\gamma}^{*}_{\uparrow}\tilde{\gamma}_{\downarrow}A^{\text{E}*}_{\uparrow}(11)A^{\text{E}}_{\downarrow}(11)\bigg]\notag\\
  \quad\mp\frac{2}{\mathcal{N}_{\text{F/B}}}&\text{Re}\bigg[\iint^{\infty}_{-\infty}v^{2}dt_{1}dt_{2}\tilde{\gamma}^{*}_{\uparrow}\tilde{\gamma}_{\downarrow}A^{\text{E}*}_{\uparrow}(11)A^{\text{D}}_{\downarrow}(11)\bigg]\mp\frac{2}{\mathcal{N}_{\text{F/B}}}\text{Re}\bigg[\iint^{\infty}_{-\infty}v^{2}dt_{1}dt_{2}\tilde{\gamma}^{*}_{\uparrow}\tilde{\gamma}_{\downarrow}A^{\text{D}*}_{\uparrow}(11)A^{\text{E}}_{\downarrow}(11)\bigg]\,,\\
  = \frac{1}{\mathcal{N}_{\text{F/B}}}&\mathcal{P}(11;\ket{\psi})\,,
\end{align}
where $\mathcal{N}_{\text{F/B}}$ is the normalization factor resulting from the postselecting qubit to the state $\ket{\psi}$ for fermions and bosons respectively. From here we can clearly see that the terms that depend on the mutual
statistics of bosons and fermions are accompanied by positive
and negative signs respectively. For extracting the mutual statistics,
we would like to get rid of all the terms that are insensitive to
this sign difference. Let us start by introducing the following notation,
\begin{align}
    \mathcal{W}_{11} &= \text{Sum of all statistics \emph{independent} terms in the probability $\mathcal{P}(11;|\psi\rangle)_{\text{F/B}}$, Eq.~\eqref{eq:p11_general}}\,,\\
    \mathcal{\chi}_{11} &= \text{Sum of all statistics \emph{dependent} terms in the probability $\mathcal{P}(11;|\psi\rangle)_{\text{F/B}}$, Eq.~\eqref{eq:p11_general}}\,,
\end{align}
and more explicitly, we have,
\begin{align}
    \mathcal{W}_{11} = & \iint^{\infty}_{-\infty}v^{2}dt_{1}dt_{2}\big[|\tilde{\gamma}_{\uparrow}|^{2}|A^{\text{D}}_{\uparrow}(11)|^{2}+|\tilde{\gamma}_{\uparrow}|^{2}|A^{\text{E}}_{\uparrow}(11)|^{2}\big] + \iint^{\infty}_{-\infty}v^{2}dt_{1}dt_{2}\big[|\tilde{\gamma}_{\downarrow}|^{2}|A^{\text{D}}_{\downarrow}(11)|^{2}+|\tilde{\gamma}_{\downarrow}|^{2}|A^{\text{E}}_{\downarrow}(11)|^{2}\big] \\
    & + 2\text{Re}\bigg[\iint^{\infty}_{-\infty}v^{2}dt_{1}dt_{2}\tilde{\gamma}^{*}_{\uparrow}\tilde{\gamma}_{\downarrow}A^{\text{D}*}_{\uparrow}(11)A^{\text{D}}_{\downarrow}(11)\bigg]+2\text{Re}\bigg[\iint^{\infty}_{-\infty}v^{2}dt_{1}dt_{2}\tilde{\gamma}^{*}_{\uparrow}\tilde{\gamma}_{\downarrow}A^{\text{E}*}_{\uparrow}(11)A^{\text{E}}_{\downarrow}(11)\bigg]\,,\\
    \mathcal{\chi}_{11} = &  \,  2|\tilde{\gamma}_{\uparrow}|^{2}\text{Re}\bigg[\iint^{\infty}_{-\infty}v^{2}dt_{1}dt_{2}A^{\text{D}*}_{\uparrow}(11)A^{\text{E}}_{\uparrow}(11)\bigg] + 2|\tilde{\gamma}_{\downarrow}|^{2}\text{Re}\bigg[\iint^{\infty}_{-\infty}v^{2}dt_{1}dt_{2}A^{\text{D}*}_{\downarrow}(11)A^{\text{E}}_{\downarrow}(11)\bigg]  \\
    & +2\text{Re}\bigg[\iint^{\infty}_{-\infty}v^{2}dt_{1}dt_{2}\tilde{\gamma}^{*}_{\uparrow}\tilde{\gamma}_{\downarrow}A^{\text{E}*}_{\uparrow}(11)A^{\text{D}}_{\downarrow}(11)\bigg] + 2\text{Re}\bigg[\iint^{\infty}_{-\infty}v^{2}dt_{1}dt_{2}\tilde{\gamma}^{*}_{\uparrow}\tilde{\gamma}_{\downarrow}A^{\text{D}*}_{\uparrow}(11)A^{\text{E}}_{\downarrow}(11)\bigg]\,.
\end{align}
Let us focus on the statistics dependent term $\chi_{11}$. The first two terms in the expression are sign definite as shown in the Appendix~\ref{app:scattering}, and Eq.~\eqref{eq:chi_sign}, because they correspond to the interference terms when the qubit state is either $\ket{\uparrow}$ or $\ket{\downarrow}$. Let us recap the calculations, start by considering the qubit to be in the state $\ket{\uparrow}$. In this case we have the following probability of the two particle scattering event where each detector receives one particle each,
\begin{align}
    \mathcal{P}(11)_{\text{B/F}}^{(\uparrow)} &= \iint_{-\infty}^{\infty} v^{2}dt_{1}dt_{2} |A_{\uparrow}^{\text{D}}(11) \pm A_{\uparrow}^{\text{E}}(11)|^{2} \,,\\
    &= \iint_{-\infty}^{\infty} v^{2}dt_{1}dt_{2} |\mathcal{I}^{(\uparrow)}_{1}(x_{1},t_{1};x^{(0)}_{1},t^{(0)}_{1})\mathcal{I}^{(\uparrow)}_{2}(x_{2},t_{2};x^{(0)}_{2},t^{(0)}_{2}) \pm \mathcal{J}^{(\uparrow)}_{1}(x_{2},t_{2};x^{(0)}_{1},t^{(0)}_{1})\mathcal{J}^{(\uparrow)}_{2}(x_{1},t_{1};x^{(0)}_{2},t^{(0)}_{2})|^{2}\,,\notag\\
    &= \mathcal{P}(11)_{\text{CW}}^{(\uparrow)}  \pm 2\chi^{(\uparrow)}(x_{1}^{(0)},t_{1}^{(0)};x_{2}^{(0)},t_{2}^{(0)})\,,
\end{align}
by using the unitarity of the scattering matrix $S_{\text{eff gen}}^{(m)}$, Eq.~\eqref{eq:seffqubit_main_general} or Eq.~\eqref{eq:gen_scat},
\begin{equation}\label{eq:chi_sign}
\chi(x_{1}^{(0)},t_{1}^{(0)};x_{2}^{(0)},t_{2}^{(0)}) = -\Bigg{|}\int_{-\infty}^{\infty} dk|\phi(k)|^{2}\mathcal{R}_{k}^{(\uparrow)*}\mathcal{T'}_{k}^{(\uparrow)}e^{ik\big[x_{0}^{(2)}+vt_{0}^{(2)}+x_{0}^{(1)}-vt_{0}^{(1)}\big]}\Bigg{|}^{2}\,.
\end{equation}
Hence, we have the following, cf. Eq.~\eqref{eq:chi_sign},
\begin{align}\label{eq:chi_sign_gen}
        2|\tilde{\gamma}_{\uparrow}|^{2}\text{Re}\bigg[\iint^{\infty}_{-\infty}v^{2}dt_{1}dt_{2}A^{\text{D}*}_{\uparrow}(11)A^{\text{E}}_{\uparrow}(11)\bigg] + 2|\tilde{\gamma}_{\downarrow}|^{2}\text{Re}\bigg[\iint^{\infty}_{-\infty}v^{2}dt_{1}dt_{2}A^{\text{D}*}_{\downarrow}(11)A^{\text{E}}_{\downarrow}(11)\bigg] \leq 0\,.
\end{align}
However, the other two terms in the expression for $\chi_{11}$ are not in general sign definite because it involves product of amplitudes with two different scattering matrix, which in general can be very different. We also do a similar calculation for the probability where either detector $D_{1}$ or $D_{2}$ receives both the particles, that is $\mathcal{P}(20;|\psi\rangle)_{\text{F/B}}$ and $\mathcal{P}(02;|\psi\rangle)_{\text{F/B}}$ respectively. Similar to Eq.~\eqref{eq:p11_general}, we can write the expressions for the probabilities $P(02;|\psi\rangle)_{\text{F/B}}$ and $P(20;|\psi\rangle)_{\text{F/B}}$ respectively (not shown here). As a result, we have the following expression, 
\begin{align}
   \mathcal{P}(11;|\psi\rangle)_{\text{B/F}} = \mathcal{W}_{11} \pm \mathcal{\chi}_{11}\,,\qquad \mathcal{P}(20;|\psi\rangle)_{\text{B/F}} = \mathcal{W}_{20} \pm \mathcal{\chi}_{20}\,, \qquad \mathcal{P}(02;|\psi\rangle)_{\text{B/F}} = \mathcal{W}_{02} \pm \mathcal{\chi}_{02}\,,
\end{align}
where we would like to emphasize that the expression for $\mathcal{P}(20;|\psi\rangle)_{\text{B/F}}$ and $\mathcal{P}(02;|\psi\rangle)_{\text{B/F}}$ are not derived here, but they also follow similar structure. In the above equations, $\mathcal{W}_{02}$ and $\mathcal{W}_{20}$ denote the sum of all statistics independent terms, whereas $\mathcal{\chi}_{02}$ and $\mathcal{\chi}_{20}$ denote the sum of terms that are are statistics sensitive. Now we define the following,
\begin{align}
    \mathcal{N}_0 = \mathcal{W}_{11} + \mathcal{W}_{02} + \mathcal{W}_{20}\,, \qquad \mathcal{\chi}_{\text{tot}} = \mathcal{\chi}_{11} + \mathcal{\chi}_{02} + \mathcal{\chi}_{20}\,,
\end{align}
As a result of postselection of the qubit state (projecting on to) $\ket{\psi}$, the expression for the probabilities must be normalized properly. Hence, we have the following expression for the normalization, 
\begin{align}
    \mathcal{N}_{\text{F/B}} = \mathcal{P}(11;|\psi\rangle)_{\text{F/B}} + \mathcal{P}(20;|\psi\rangle)_{\text{F/B}} + \mathcal{P}(02;|\psi\rangle)_{\text{F/B}} = \mathcal{N}_0 \mp \mathcal{\chi}_{\text{tot}}\,,
\end{align}
here, $\mathcal{N}_{\text{B/F}}$ represents the postselected normalization for bosons and fermions respectively. Since, it is a sum of probabilities, we have $\mathcal{N}_{\text{B}},\mathcal{N}_{\text{F}} \geq 0$.  The correct benchmark of statistics $\mathcal{B}_4$ is obtained by requiring all the terms inside $\mathcal{\chi}_{\text{tot}}$ goes to zero. This is achieved by introducing large time delay between the emission of wave packets from the two sources. As a result, the correct benchmark $\mathcal{B}_4$, in this general setting is given by the following,
\begin{align}
    \mathcal{B}_4 = \lim_{v|t_{1}^{(0)} - t_{2}^{(0)}|\to \infty} P(11;\ket{\psi})_{\text{F/B}} =  \frac{\mathcal{W}_{11}}{\mathcal{N}_0}\,.
\end{align}
Now, the complete normalized two-particle probability $P(11;\ket{\psi})$ is given as follows,
\begin{align}
   &P(11;|\psi\rangle)_{\text{F/B}} = \frac{\mathcal{P}(11;|\psi\rangle)_{\text{F/B}}}{\mathcal{P}(11;|\psi\rangle)_{\text{F/B}}+\mathcal{P}(20;|\psi\rangle)_{\text{F/B}}+\mathcal{P}(02;|\psi\rangle)_{\text{F/B}}} = \frac{\mathcal{W}_{11}\mp\mathcal{\chi}_{11}}{\mathcal{N}_0\mp\mathcal{\chi}_{\text{tot}}}\,,
\end{align}
and hence, we have the following,
\begin{align}
    P(11;|\psi\rangle)_{\text{F/B}} - \mathcal{B}_4 &= \frac{\mathcal{W}_{11}\mp \mathcal{\chi}_{11}}{\mathcal{N}_0\mp\mathcal{\chi}_{\text{tot}}} - \frac{\mathcal{W}_{11}}{\mathcal{N}_0}= \mp \frac{\mathcal{N}_0\mathcal{\chi}_{11}-\mathcal{W}_{11}\chi_{\text{tot}}}{\mathcal{N}_0(\mathcal{N}_0\mp\chi_{\text{tot}})}\,.
\end{align}
Next, we exploit the sensitivity of the above result to mutual statistics to note the following,
\begin{align}
    [P(11;|\psi\rangle)_{\text{B}} - \mathcal{B}_4]\times[P(11;|\psi\rangle)_{\text{F}} - \mathcal{B}_4] = \frac{-(\mathcal{N}_0\mathcal{\chi}_{11}-\mathcal{W}_{11}\chi_{\text{tot}})^2}{\mathcal{N}_0^2\mathcal{N}_{\text{B}}\mathcal{N}_{\text{F}}} \leq 0\,.
\end{align}
Hence, we see that the fermions and bosons are always accompanied by opposite signs when compared against the correct benchmark $\mathcal{B}_4$. As a result, our prescription for extracting the mutual statistics of bosons and fermions using benchmark $\mathcal{B}_4$ is robust and it works well for a general scattering matrix that is coupled to a qubit, given by Eq.~\eqref{eq:gen_scat}. The important thing to note here is that it is not always guaranteed that the quantity $P(11;|\psi\rangle)_{\text{B}} - \mathcal{B}_4$ and $P(11;|\psi\rangle)_{\text{F}} - \mathcal{B}_4$ are in general not sign definite. The reason is the statistics dependent term $\chi_{11}$ is not in general sign definite, as we have argued in the above discussions around Eq.~\eqref{eq:chi_sign_gen}. However, the relative behavior of bosons and fermions is always captured using the benchmark $\mathcal{B}_{4}$. In our main text, the choice of the coupling between the scattering matrix and the qubit [cf. Eq.~\eqref{eq:seffqubit_main}] leads to a sign definite result for bosons and fermions, however, this is in general not true. In the following, we explicitly construct a scattering matrix that is coupled to the qubit and show that the benchmark $\mathcal{B}_{4}$ does not give a sign definite result of bosons and fermions. The quantity $P(11;|\psi\rangle)_{\text{B}} - \mathcal{B}_4$ and $P(11;|\psi\rangle)_{\text{F}} - \mathcal{B}_4$ comes with a relative sign, a difference arising from the statistical difference between bosons and fermions. Consider the following scattering matrix that depends on the state of the qubit $\ket{m}$, where $m\in \{ \uparrow, \downarrow \}$,
\begin{equation}
\label{eq:ex_scat}
S^{(m)}_{\text{eff gen}}=\frac{1}{1-r_{m}^2e^{2ikL}}\left[\begin{array}{cc}
(1-r_{m}^2)e^{ikL} & r_m(1-e^{2ikL})\\
r_m(1-e^{2ikL}) & (1-r_{m}^2)e^{ikL}
\end{array}\right]\,,
\end{equation}
where now the qubit state determines the strength of the reflection amplitude at the QPCs. In other words, the reflection amplitude $r_m$ now depends on the state of the qubit $m$. Therefore in general we have $r_{\uparrow}\neq r_{\downarrow}$ and only in the absence of coupling with qubit we have $r_{\uparrow} = r_{\downarrow}$. The two particle probability bosons and fermions when compared with the benchmark $\mathcal{B}_{4}$, we find that the quantity $P(11;\ket{\psi})_{\text{B}} - \mathcal{B}_{4}$ and $P(11;\ket{\psi})_{\text{F}} - \mathcal{B}_{4}$ are not sign definite individually but have a relative sign difference. More precisely, whenever $P(11;\ket{\psi})_{\text{F}} - \mathcal{B}_{4}$ is positive, $P(11;\ket{\psi})_{\text{B}} - \mathcal{B}_{4}$ is negative and vice-versa.

\end{widetext}

\end{appendix}

\bibliography{refs}

@misc{gurarie2025randomlymeasuredquantumparticle,
      title={Randomly measured quantum particle}, 
      author={V. Gurarie},
      year={2025},
      eprint={2504.05479},
      archivePrefix={arXiv},
      primaryClass={cond-mat.dis-nn},
      url={https://arxiv.org/abs/2504.05479}, 
}

@article{PhysRevResearch.2.033347,
  title = {Measurement-induced steering of quantum systems},
  author = {Roy, Sthitadhi and Chalker, J. T. and Gornyi, I. V. and Gefen, Yuval},
  journal = {Phys. Rev. Res.},
  volume = {2},
  issue = {3},
  pages = {033347},
  numpages = {20},
  year = {2020},
  month = {Sep},
  publisher = {American Physical Society},
  doi = {10.1103/PhysRevResearch.2.033347},
  url = {https://link.aps.org/doi/10.1103/PhysRevResearch.2.033347}
}

@article{PhysRevB.111.L201407,
  title = {Exchange-phase erasure in anyonic Hong-Ou-Mandel interferometry},
  author = {Varada, Sushanth and Sp\aa{}nsl\"att, Christian and Acciai, Matteo},
  journal = {Phys. Rev. B},
  volume = {111},
  issue = {20},
  pages = {L201407},
  numpages = {8},
  year = {2025},
  month = {May},
  publisher = {American Physical Society},
  doi = {10.1103/PhysRevB.111.L201407},
  url = {https://link.aps.org/doi/10.1103/PhysRevB.111.L201407}
}

@article{jullien_quantum_2014,
	title = {Quantum tomography of an electron},
	volume = {514},
	issn = {1476-4687},
	url = {https://doi.org/10.1038/nature13821},
	doi = {10.1038/nature13821},
	number = {7524},
	journal = {Nature},
	author = {Jullien, T. and Roulleau, P. and Roche, B. and Cavanna, A. and Jin, Y. and Glattli, D. C.},
	month = oct,
	year = {2014},
	pages = {603--607},
}

@article{
doi:10.1126/science.1141243,
author = {G. Fève  and A. Mahé  and J.-M. Berroir  and T. Kontos  and B. Plaçais  and D. C. Glattli  and A. Cavanna  and B. Etienne  and Y. Jin },
title = {An On-Demand Coherent Single-Electron Source},
journal = {Science},
volume = {316},
number = {5828},
pages = {1169-1172},
year = {2007},
doi = {10.1126/science.1141243},
URL = {https://www.science.org/doi/abs/10.1126/science.1141243},
}

@article{x4w5-h3bb,
  title = {Modified Interferometer to Measure Anyonic Braiding Statistics},
  author = {Kivelson, Steven A. and Murthy, Chaitanya},
  journal = {Phys. Rev. Lett.},
  volume = {135},
  issue = {12},
  pages = {126605},
  numpages = {6},
  year = {2025},
  month = {Sep},
  publisher = {American Physical Society},
  doi = {10.1103/x4w5-h3bb},
  url = {https://link.aps.org/doi/10.1103/x4w5-h3bb}
}

@misc{girdhar2026fabryperotinterferometrystochasticanyonic,
      title={Fabry-P\'erot interferometry with stochastic anyonic sources}, 
      author={Sarthak Girdhar and Edvin G. Idrisov and Thomas L. Schmidt},
      year={2026},
      eprint={2603.05052},
      archivePrefix={arXiv},
      primaryClass={cond-mat.mes-hall},
      url={https://arxiv.org/abs/2603.05052}, 
}

@article{Lee2022NonAbelianAnyonCollider,
  author  = {Lee, June-Young M. and Sim, H.-S.},
  title   = {Non-Abelian anyon collider},
  journal = {Nature Communications},
  volume  = {13},
  pages   = {6660},
  year    = {2022},
  doi     = {10.1038/s41467-022-34329-y},
  url     = {https://doi.org/10.1038/s41467-022-34329-y}
}

@article{Lee2023PartitioningDilutedAnyons,
  author  = {Lee, June-Young M. and Hong, Changki and Alkalay, Tomer and Schiller, Noam and Umansky, Vladimir and Heiblum, Moty and Oreg, Yuval and Sim, H.-S.},
  title   = {Partitioning of diluted anyons reveals their braiding statistics},
  journal = {Nature},
  volume  = {617},
  pages   = {277--281},
  year    = {2023},
  doi     = {10.1038/s41586-023-05883-2},
  url     = {https://doi.org/10.1038/s41586-023-05883-2}
}

@article{PhysRevLett.131.186601,
  title = {Anyon Statistics through Conductance Measurements of Time-Domain Interferometry},
  author = {Schiller, Noam and Shapira, Yotam and Stern, Ady and Oreg, Yuval},
  journal = {Phys. Rev. Lett.},
  volume = {131},
  issue = {18},
  pages = {186601},
  numpages = {7},
  year = {2023},
  month = {Nov},
  publisher = {American Physical Society},
  doi = {10.1103/PhysRevLett.131.186601},
  url = {https://link.aps.org/doi/10.1103/PhysRevLett.131.186601}
}

@article{PhysRevLett.134.096303,
  title = {Landscapes of an Out-of-Equilibrium Anyonic Sea},
  author = {Zhang, Gu and Gornyi, Igor and Gefen, Yuval},
  journal = {Phys. Rev. Lett.},
  volume = {134},
  issue = {9},
  pages = {096303},
  numpages = {8},
  year = {2025},
  month = {Mar},
  publisher = {American Physical Society},
  doi = {10.1103/PhysRevLett.134.096303},
  url = {https://link.aps.org/doi/10.1103/PhysRevLett.134.096303}
}

@misc{zhang2025effectivelinearresponsenonequilibrium,
      title={Effective linear response in non-equilibrium anyonic systems}, 
      author={Gu Zhang and Igor Gornyi and Yuval Gefen},
      year={2025},
      eprint={2510.03985},
      archivePrefix={arXiv},
      primaryClass={cond-mat.mes-hall},
      url={https://arxiv.org/abs/2510.03985}, 
}

@article{PhysRevLett.91.196803,
  title = {Revisiting the Hanbury Brown--Twiss Setup for Fractional Statistics},
  author = {Vishveshwara, Smitha},
  journal = {Phys. Rev. Lett.},
  volume = {91},
  issue = {19},
  pages = {196803},
  numpages = {4},
  year = {2003},
  month = {Nov},
  publisher = {American Physical Society},
  doi = {10.1103/PhysRevLett.91.196803},
  url = {https://link.aps.org/doi/10.1103/PhysRevLett.91.196803}
}

@misc{puster2026extractinganyonicexchangephase,
      title={Extracting the Anyonic Exchange Phase from Hanbury Brown-Twiss Correlations}, 
      author={Felix Puster and Matthias Thamm and Bernd Rosenow},
      year={2026},
      eprint={2603.13898},
      archivePrefix={arXiv},
      primaryClass={cond-mat.mes-hall},
      url={https://arxiv.org/abs/2603.13898}, 
}

@article{PhysRevLett.132.216601,
  title = {Finite Width of Anyons Changes Their Braiding Signature},
  author = {Iyer, K. and Ronetti, F. and Gr\'emaud, B. and Martin, T. and Rech, J. and Jonckheere, T.},
  journal = {Phys. Rev. Lett.},
  volume = {132},
  issue = {21},
  pages = {216601},
  numpages = {6},
  year = {2024},
  month = {May},
  publisher = {American Physical Society},
  doi = {10.1103/PhysRevLett.132.216601},
  url = {https://link.aps.org/doi/10.1103/PhysRevLett.132.216601}
}

@article{PhysRevLett.132.156501,
  title = {Effect of the Soliton Width on Nonequilibrium Exchange Phases of Anyons},
  author = {Thamm, Matthias and Rosenow, Bernd},
  journal = {Phys. Rev. Lett.},
  volume = {132},
  issue = {15},
  pages = {156501},
  numpages = {6},
  year = {2024},
  month = {Apr},
  publisher = {American Physical Society},
  doi = {10.1103/PhysRevLett.132.156501},
  url = {https://link.aps.org/doi/10.1103/PhysRevLett.132.156501}
}

@article{rp5m-r8fr,
  title = {Anyonic analogue of optical Mach-Zehnder interferometer},
  author = {Batra, Navketan and Wei, Zezhu and Vishveshwara, Smitha and Feldman, D. E.},
  journal = {Phys. Rev. B},
  volume = {112},
  issue = {12},
  pages = {125305},
  numpages = {24},
  year = {2025},
  month = {Sep},
  publisher = {American Physical Society},
  doi = {10.1103/rp5m-r8fr},
  url = {https://link.aps.org/doi/10.1103/rp5m-r8fr}
}

@article{PhysRevB.107.104406,
  title = {Thermal interferometry of anyons},
  author = {Wei, Zezhu and Batra, Navketan and Mitrovi\ifmmode \acute{c}\else \'{c}\fi{}, V. F. and Feldman, D. E.},
  journal = {Phys. Rev. B},
  volume = {107},
  issue = {10},
  pages = {104406},
  numpages = {32},
  year = {2023},
  month = {Mar},
  publisher = {American Physical Society},
  doi = {10.1103/PhysRevB.107.104406},
  url = {https://link.aps.org/doi/10.1103/PhysRevB.107.104406}
}

@article{PhysRevB.108.L241302,
  title = {Anyonic Mach-Zehnder interferometer on a single edge of a two-dimensional electron gas},
  author = {Batra, Navketan and Wei, Zezhu and Vishveshwara, Smitha and Feldman, D. E.},
  journal = {Phys. Rev. B},
  volume = {108},
  issue = {24},
  pages = {L241302},
  numpages = {6},
  year = {2023},
  month = {Dec},
  publisher = {American Physical Society},
  doi = {10.1103/PhysRevB.108.L241302},
  url = {https://link.aps.org/doi/10.1103/PhysRevB.108.L241302}
}

@article{PhysRevLett.95.200405,
  title = {Pre- and Post-Selection Paradoxes and Contextuality in Quantum Mechanics},
  author = {Leifer, M. S. and Spekkens, Robert W.},
  journal = {Phys. Rev. Lett.},
  volume = {95},
  issue = {20},
  pages = {200405},
  numpages = {4},
  year = {2005},
  month = {Nov},
  publisher = {American Physical Society},
  doi = {10.1103/PhysRevLett.95.200405},
  url = {https://link.aps.org/doi/10.1103/PhysRevLett.95.200405}
}

@article{PhysRev.134.B1410,
  title = {Time Symmetry in the Quantum Process of Measurement},
  author = {Aharonov, Yakir and Bergmann, Peter G. and Lebowitz, Joel L.},
  journal = {Phys. Rev.},
  volume = {134},
  issue = {6B},
  pages = {B1410--B1416},
  numpages = {0},
  year = {1964},
  month = {Jun},
  publisher = {American Physical Society},
  doi = {10.1103/PhysRev.134.B1410},
  url = {https://link.aps.org/doi/10.1103/PhysRev.134.B1410}
}

@article{PhysRevB.75.195332,
  title = {Transport of fractional Hall quasiparticles through an antidot},
  author = {Merlo, Matteo and Braggio, Alessandro and Magnoli, Nicodemo and Sassetti, Maura},
  journal = {Phys. Rev. B},
  volume = {75},
  issue = {19},
  pages = {195332},
  numpages = {10},
  year = {2007},
  month = {May},
  publisher = {American Physical Society},
  doi = {10.1103/PhysRevB.75.195332},
  url = {https://link.aps.org/doi/10.1103/PhysRevB.75.195332}
}

@article{PhysRevB.56.9692,
  title = {Aharonov-Bohm effect in the chiral Luttinger liquid},
  author = {Geller, Michael R. and Loss, Daniel},
  journal = {Phys. Rev. B},
  volume = {56},
  issue = {15},
  pages = {9692--9706},
  numpages = {0},
  year = {1997},
  month = {Oct},
  publisher = {American Physical Society},
  doi = {10.1103/PhysRevB.56.9692},
  url = {https://link.aps.org/doi/10.1103/PhysRevB.56.9692}
}

@article{PhysRevLett.52.1583,
  title = {Statistics of Quasiparticles and the Hierarchy of Fractional Quantized Hall States},
  author = {Halperin, B. I.},
  journal = {Phys. Rev. Lett.},
  volume = {52},
  issue = {18},
  pages = {1583--1586},
  numpages = {0},
  year = {1984},
  month = {Apr},
  publisher = {American Physical Society},
  doi = {10.1103/PhysRevLett.52.1583},
  url = {https://link.aps.org/doi/10.1103/PhysRevLett.52.1583}
}

@misc{rao2016introduction,
      title={Introduction to abelian and non-abelian anyons}, 
      author={Sumathi Rao},
      year={2016},
      eprint={1610.09260},
      archivePrefix={arXiv},
      primaryClass={cond-mat.mes-hall}
}

@misc{garg2025enhancedshotnoisegraphene,
      title={Enhanced shot noise in graphene quantum point contacts with electrostatic reconstruction}, 
      author={M. Garg and O. Maillet and N. L. Samuelson and T. Wang and J. Feng and L. A. Cohen and A. Zhang and K. Watanabe and T. Taniguchi and P. Roulleau and M. Sassetti and M. Zaletel and A. F. Young and D. Ferraro and P. Roche and F. D. Parmentier},
      year={2025},
      eprint={2503.17209},
      archivePrefix={arXiv},
      primaryClass={cond-mat.mes-hall},
      url={https://arxiv.org/abs/2503.17209}, 
}

@article{PhysRevLett.134.076302,
  title = {Spontaneous Localization at a Potential Saddle Point from Edge State Reconstruction in a Quantum Hall Point Contact},
  author = {Cohen, Liam A. and Samuelson, Noah L. and Wang, Taige and Klocke, Kai and Reeves, Cian C. and Taniguchi, Takashi and Watanabe, Kenji and Vijay, Sagar and Zaletel, Michael P. and Young, Andrea F.},
  journal = {Phys. Rev. Lett.},
  volume = {134},
  issue = {7},
  pages = {076302},
  numpages = {6},
  year = {2025},
  month = {Feb},
  publisher = {American Physical Society},
  doi = {10.1103/PhysRevLett.134.076302},
  url = {https://link.aps.org/doi/10.1103/PhysRevLett.134.076302}
}

@article{PhysRevB.80.035319,
  title = {Long tunneling contact as a probe of fractional quantum Hall neutral edge modes},
  author = {Overbosch, B. J. and Chamon, Claudio},
  journal = {Phys. Rev. B},
  volume = {80},
  issue = {3},
  pages = {035319},
  numpages = {5},
  year = {2009},
  month = {Jul},
  publisher = {American Physical Society},
  doi = {10.1103/PhysRevB.80.035319},
  url = {https://link.aps.org/doi/10.1103/PhysRevB.80.035319}
}

@article{zhang2023measuring,
	title = {Measuring statistics-induced entanglement entropy with a {Hong}–{Ou}–{Mandel} interferometer},
	volume = {15},
	issn = {2041-1723},
	url = {https://doi.org/10.1038/s41467-024-47335-z},
	doi = {10.1038/s41467-024-47335-z},
	number = {1},
	journal = {Nature Communications},
	author = {Zhang, Gu and Hong, Changki and Alkalay, Tomer and Umansky, Vladimir and Heiblum, Moty and Gornyi, Igor and Gefen, Yuval},
	month = apr,
	year = {2024},
	pages = {3428},
}

@book{10.1093/acprof:oso/9780199227259.001.0001,
    author = {Wen, Xiao-Gang},
    title = {Quantum Field Theory of Many-Body Systems: From the Origin of Sound to an Origin of Light and Electrons},
    publisher = {Oxford University Press},
    year = {2007},
    month = {09},
    isbn = {9780199227259},
    doi = {10.1093/acprof:oso/9780199227259.001.0001},
    url = {https://doi.org/10.1093/acprof:oso/9780199227259.001.0001},
}

@article{leinaas_theory_1977,
	title = {On the theory of identical particles},
	volume = {37},
	issn = {1826-9877},
	url = {https://doi.org/10.1007/BF02727953},
	doi = {10.1007/BF02727953},
	number = {1},
	journal = {Il Nuovo Cimento B (1971-1996)},
	author = {Leinaas, J. M. and Myrheim, J.},
	month = jan,
	year = {1977},
	pages = {1--23},
}

@article{werkmeister_strongly_2024,
	title = {Strongly coupled edge states in a graphene quantum {Hall} interferometer},
	volume = {15},
	issn = {2041-1723},
	url = {https://doi.org/10.1038/s41467-024-50695-1},
	doi = {10.1038/s41467-024-50695-1},
	number = {1},
	journal = {Nature Communications},
	author = {Werkmeister, Thomas and Ehrets, James R. and Ronen, Yuval and Wesson, Marie E. and Najafabadi, Danial and Wei, Zezhu and Watanabe, Kenji and Taniguchi, Takashi and Feldman, D. E. and Halperin, Bertrand I. and Yacoby, Amir and Kim, Philip},
	month = aug,
	year = {2024},
	pages = {6533},
}

@article{ronen_aharonovbohm_2021,
	title = {Aharonov–{Bohm} effect in graphene-based {Fabry}–{Pérot} quantum {Hall} interferometers},
	volume = {16},
	issn = {1748-3395},
	url = {https://doi.org/10.1038/s41565-021-00861-z},
	doi = {10.1038/s41565-021-00861-z},
	number = {5},
	journal = {Nature Nanotechnology},
	author = {Ronen, Yuval and Werkmeister, Thomas and Haie Najafabadi, Danial and Pierce, Andrew T. and Anderson, Laurel E. and Shin, Young Jae and Lee, Si Young and Lee, Young Hee and Johnson, Bobae and Watanabe, Kenji and Taniguchi, Takashi and Yacoby, Amir and Kim, Philip},
	month = may,
	year = {2021},
	pages = {563--569},
}

@article{kim_aharonovbohm_2024,
	title = {Aharonov–{Bohm} interference and statistical phase-jump evolution in fractional quantum {Hall} states in bilayer graphene},
	issn = {1748-3395},
	url = {https://doi.org/10.1038/s41565-024-01751-w},
	doi = {10.1038/s41565-024-01751-w},
	journal = {Nature Nanotechnology},
	author = {Kim, Jehyun and Dev, Himanshu and Kumar, Ravi and Ilin, Alexey and Haug, André and Bhardwaj, Vishal and Hong, Changki and Watanabe, Kenji and Taniguchi, Takashi and Stern, Ady and Ronen, Yuval},
	month = aug,
	year = {2024},
}

@article{neder_interference_2007,
	title = {Interference between two indistinguishable electrons from independent sources},
	volume = {448},
	copyright = {2007 Springer Nature Limited},
	issn = {1476-4687},
	url = {https://www.nature.com/articles/nature05955},
	doi = {10.1038/nature05955},
	language = {en},
	number = {7151},
	urldate = {2024-05-07},
	journal = {Nature},
	author = {Neder, I. and Ofek, N. and Chung, Y. and Heiblum, M. and Mahalu, D. and Umansky, V.},
	month = jul,
	year = {2007},
	note = {Publisher: Nature Publishing Group},
	pages = {333--337},
}

@article{Kundu_2023,
   title={Anyonic interference and braiding phase in a Mach-Zehnder interferometer},
   volume={19},
   ISSN={1745-2481},
   url={http://dx.doi.org/10.1038/s41567-022-01899-z},
   DOI={10.1038/s41567-022-01899-z},
   number={4},
   journal={Nature Physics},
   publisher={Springer Science and Business Media LLC},
   author={Kundu, Hemanta Kumar and Biswas, Sourav and Ofek, Nissim and Umansky, Vladimir and Heiblum, Moty},
   year={2023},
   month=jan, pages={515–521} }

@article{Halperin_2011,
   title={Theory of the Fabry-Pérot quantum Hall interferometer},
   volume={83},
   ISSN={1550-235X},
   url={http://dx.doi.org/10.1103/PhysRevB.83.155440},
   DOI={10.1103/physrevb.83.155440},
   number={15},
   journal={Physical Review B},
   publisher={American Physical Society (APS)},
   author={Halperin, Bertrand I. and Stern, Ady and Neder, Izhar and Rosenow, Bernd},
   year={2011},
   month=apr }

@article{PhysRevB.88.235415,
  title = {Hanbury Brown and Twiss correlations in quantum Hall systems},
  author = {Campagnano, Gabriele and Zilberberg, Oded and Gornyi, Igor V. and Gefen, Yuval},
  journal = {Phys. Rev. B},
  volume = {88},
  issue = {23},
  pages = {235415},
  numpages = {30},
  year = {2013},
  month = {Dec},
  publisher = {American Physical Society},
  doi = {10.1103/PhysRevB.88.235415},
  url = {https://link.aps.org/doi/10.1103/PhysRevB.88.235415}
}

@article{PhysRevLett.109.106802,
  title = {Hanbury Brown--Twiss Interference of Anyons},
  author = {Campagnano, Gabriele and Zilberberg, Oded and Gornyi, Igor V. and Feldman, Dmitri E. and Potter, Andrew C. and Gefen, Yuval},
  journal = {Phys. Rev. Lett.},
  volume = {109},
  issue = {10},
  pages = {106802},
  numpages = {5},
  year = {2012},
  month = {Sep},
  publisher = {American Physical Society},
  doi = {10.1103/PhysRevLett.109.106802},
  url = {https://link.aps.org/doi/10.1103/PhysRevLett.109.106802}
}

@article{PhysRevLett.116.156802,
  title = {Current Correlations from a Mesoscopic Anyon Collider},
  author = {Rosenow, Bernd and Levkivskyi, Ivan P. and Halperin, Bertrand I.},
  journal = {Phys. Rev. Lett.},
  volume = {116},
  issue = {15},
  pages = {156802},
  numpages = {5},
  year = {2016},
  month = {Apr},
  publisher = {American Physical Society},
  doi = {10.1103/PhysRevLett.116.156802},
  url = {https://link.aps.org/doi/10.1103/PhysRevLett.116.156802}
}

@article{PhysRevB.76.085333,
  title = {Shot noise in an anyonic Mach-Zehnder interferometer},
  author = {Feldman, D. E. and Gefen, Yuval and Kitaev, Alexei and Law, K. T. and Stern, Ady},
  journal = {Phys. Rev. B},
  volume = {76},
  issue = {8},
  pages = {085333},
  numpages = {9},
  year = {2007},
  month = {Aug},
  publisher = {American Physical Society},
  doi = {10.1103/PhysRevB.76.085333},
  url = {https://link.aps.org/doi/10.1103/PhysRevB.76.085333}
}

@article{PhysRevB.74.045319,
  title = {Electronic Mach-Zehnder interferometer as a tool to probe fractional statistics},
  author = {Law, K. T. and Feldman, D. E. and Gefen, Yuval},
  journal = {Phys. Rev. B},
  volume = {74},
  issue = {4},
  pages = {045319},
  numpages = {16},
  year = {2006},
  month = {Jul},
  publisher = {American Physical Society},
  doi = {10.1103/PhysRevB.74.045319},
  url = {https://link.aps.org/doi/10.1103/PhysRevB.74.045319}
}

@article{PhysRevB.55.2331,
  title = {Two point-contact interferometer for quantum Hall systems},
  author = {de C. Chamon, C. and Freed, D. E. and Kivelson, S. A. and Sondhi, S. L. and Wen, X. G.},
  journal = {Phys. Rev. B},
  volume = {55},
  issue = {4},
  pages = {2331--2343},
  numpages = {0},
  year = {1997},
  month = {Jan},
  publisher = {American Physical Society},
  doi = {10.1103/PhysRevB.55.2331},
  url = {https://link.aps.org/doi/10.1103/PhysRevB.55.2331}
}

@misc{girvin1999quantum,
      title={The Quantum Hall Effect: Novel Excitations and Broken Symmetries}, 
      author={Steven M. Girvin},
      year={1999},
      eprint={cond-mat/9907002},
      archivePrefix={arXiv},
      primaryClass={cond-mat.mes-hall}
}

@article{PhysRevLett.48.1559,
  title = {Two-Dimensional Magnetotransport in the Extreme Quantum Limit},
  author = {Tsui, D. C. and Stormer, H. L. and Gossard, A. C.},
  journal = {Phys. Rev. Lett.},
  volume = {48},
  issue = {22},
  pages = {1559--1562},
  numpages = {0},
  year = {1982},
  month = {May},
  publisher = {American Physical Society},
  doi = {10.1103/PhysRevLett.48.1559},
  url = {https://link.aps.org/doi/10.1103/PhysRevLett.48.1559}
}

@article{RevModPhys.80.1083,
  title = {Non-Abelian anyons and topological quantum computation},
  author = {Nayak, Chetan and Simon, Steven H. and Stern, Ady and Freedman, Michael and Das Sarma, Sankar},
  journal = {Rev. Mod. Phys.},
  volume = {80},
  issue = {3},
  pages = {1083--1159},
  numpages = {0},
  year = {2008},
  month = {Sep},
  publisher = {American Physical Society},
  doi = {10.1103/RevModPhys.80.1083},
  url = {https://link.aps.org/doi/10.1103/RevModPhys.80.1083}
}

@book{Sakurai_Napolitano_2020, 
place={Cambridge}, 
edition={3}, 
title={Modern Quantum Mechanics}, 
publisher={Cambridge University Press}, 
author={Sakurai, J. J. and Napolitano, Jim}, 
year={2020}}

@article{PhysRevX.13.011031,
  title = {Comparing Fractional Quantum Hall Laughlin and Jain Topological Orders with the Anyon Collider},
  author = {Ruelle, M. and Frigerio, E. and Berroir, J.-M. and Pla\ifmmode \mbox{\c{c}}\else \c{c}\fi{}ais, B. and Rech, J. and Cavanna, A. and Gennser, U. and Jin, Y. and F\`eve, G.},
  journal = {Phys. Rev. X},
  volume = {13},
  issue = {1},
  pages = {011031},
  numpages = {18},
  year = {2023},
  month = {Mar},
  publisher = {American Physical Society},
  doi = {10.1103/PhysRevX.13.011031},
  url = {https://link.aps.org/doi/10.1103/PhysRevX.13.011031}
}

@article{
doi:10.1126/science.aaz5601,
author = {H. Bartolomei  and M. Kumar  and R. Bisognin  and A. Marguerite  and J.-M. Berroir  and E. Bocquillon  and B. Plaçais  and A. Cavanna  and Q. Dong  and U. Gennser  and Y. Jin  and G. Fève },
title = {Fractional statistics in anyon collisions},
journal = {Science},
volume = {368},
number = {6487},
pages = {173-177},
year = {2020},
doi = {10.1126/science.aaz5601},
URL = {https://www.science.org/doi/abs/10.1126/science.aaz5601}
}

@article{PhysRevX.13.041012,
  title = {Fabry-P\'erot Interferometry at the $\ensuremath{\nu}=2/5$ Fractional Quantum Hall State},
  author = {Nakamura, J. and Liang, S. and Gardner, G. C. and Manfra, M. J.},
  journal = {Phys. Rev. X},
  volume = {13},
  issue = {4},
  pages = {041012},
  numpages = {11},
  year = {2023},
  month = {Oct},
  publisher = {American Physical Society},
  doi = {10.1103/PhysRevX.13.041012},
  url = {https://link.aps.org/doi/10.1103/PhysRevX.13.041012}
}

@article{nakamura_direct_2020,
	title = {Direct observation of anyonic braiding statistics},
	volume = {16},
	issn = {1745-2481},
	url = {https://doi.org/10.1038/s41567-020-1019-1},
	doi = {10.1038/s41567-020-1019-1},
	number = {9},
	journal = {Nature Physics},
	author = {Nakamura, J. and Liang, S. and Gardner, G. C. and Manfra, M. J.},
	month = sep,
	year = {2020},
	pages = {931--936},
}

@article{PhysRevLett.53.722,
  title = {Fractional Statistics and the Quantum Hall Effect},
  author = {Arovas, Daniel and Schrieffer, J. R. and Wilczek, Frank},
  journal = {Phys. Rev. Lett.},
  volume = {53},
  issue = {7},
  pages = {722--723},
  numpages = {0},
  year = {1984},
  month = {Aug},
  publisher = {American Physical Society},
  doi = {10.1103/PhysRevLett.53.722},
  url = {https://link.aps.org/doi/10.1103/PhysRevLett.53.722}
}

@article{PhysRevLett.49.957,
  title = {Quantum Mechanics of Fractional-Spin Particles},
  author = {Wilczek, Frank},
  journal = {Phys. Rev. Lett.},
  volume = {49},
  issue = {14},
  pages = {957--959},
  numpages = {0},
  year = {1982},
  month = {Oct},
  publisher = {American Physical Society},
  doi = {10.1103/PhysRevLett.49.957},
  url = {https://link.aps.org/doi/10.1103/PhysRevLett.49.957}
}

@book{Khare2005,
   author = {Avinash Khare},
   doi = {doi:10.1142/5752},
   isbn = {978-981-256-160-2},
   month = {2},
   note = {doi:10.1142/5752},
   pages = {320},
   publisher = {WORLD SCIENTIFIC},
   title = {Fractional Statistics and Quantum Theory},
   url = {https://doi.org/10.1142/5752},
   year = {2005}
}

@book{Fetter,
  address = {Boston},
  author = {Fetter, A. L. and Walecka, J. D.},
  publisher = {McGraw-Hill},
  title = {Quantum Theory of Many-Particle Systems},
  year = 1971
}

@article{Blanter_2000,
   title={Shot noise in mesoscopic conductors},
   volume={336},
   ISSN={0370-1573},
   url={http://dx.doi.org/10.1016/S0370-1573(99)00123-4},
   DOI={10.1016/s0370-1573(99)00123-4},
   number={1–2},
   journal={Physics Reports},
   publisher={Elsevier BV},
   author={Blanter, Ya.M. and Büttiker, M.},
   year={2000},
   month=sep, pages={1–166} }

@article{td98-5ltj,
  title = {Quantum Statistics and Self-Interference in Extended Colliders},
  author = {Samal, Sai Satyam and Vishveshwara, Smitha and Gefen, Yuval and V\"ayrynen, Jukka I.},
  journal = {Phys. Rev. Lett.},
  volume = {136},
  issue = {7},
  pages = {076301},
  numpages = {7},
  year = {2026},
  month = {Feb},
  publisher = {American Physical Society},
  doi = {10.1103/td98-5ltj},
  url = {https://link.aps.org/doi/10.1103/td98-5ltj}
}

@article{PhysRevLett.94.166802,
  title = {Topologically Protected Qubits from a Possible Non-Abelian Fractional Quantum Hall State},
  author = {Das Sarma, Sankar and Freedman, Michael and Nayak, Chetan},
  journal = {Phys. Rev. Lett.},
  volume = {94},
  issue = {16},
  pages = {166802},
  numpages = {4},
  year = {2005},
  month = {Apr},
  publisher = {American Physical Society},
  doi = {10.1103/PhysRevLett.94.166802},
  url = {https://link.aps.org/doi/10.1103/PhysRevLett.94.166802}
}

@book{gordanQM,
author = {Baym, G.},
title = {Lectures On Quantum Mechanics (1st ed.)},
publisher = {CRC Press},
year = {1969},
doi = {10.1201/9780429499265},
address = {},
edition   = {},
URL = {https://taylorfrancis.com/books/mono/10.1201/9780429499265/lectures-quantum-mechanics-gordon-baym}
}

@article{PhysRevB.99.045430,
  title = {Engineering statistical transmutation of identical quantum particles},
  author = {Barbarino, Simone and Fazio, Rosario and Vedral, Vlatko and Gefen, Yuval},
  journal = {Phys. Rev. B},
  volume = {99},
  issue = {4},
  pages = {045430},
  numpages = {7},
  year = {2019},
  month = {Jan},
  publisher = {American Physical Society},
  doi = {10.1103/PhysRevB.99.045430},
  url = {https://link.aps.org/doi/10.1103/PhysRevB.99.045430}
}

@book{Nazarov_Blanter_2009, place={Cambridge}, title={Quantum Transport: Introduction to Nanoscience}, publisher={Cambridge University Press}, author={Nazarov, Yuli V. and Blanter, Yaroslav M.}, year={2009}}

\end{document}